\documentclass[twocolumn, aps, prx,  superscriptaddress, 10pt]{revtex4-2}

\usepackage{mathrsfs}
\usepackage{amsmath}
\usepackage{amsthm}
\usepackage{amsfonts,amssymb} 
\usepackage{physics}

\usepackage{xcolor}
\usepackage[normalem]{ulem}

\usepackage{tcolorbox}
\usepackage{graphicx}

\usepackage{subcaption}

\usepackage{mathtools}
\newcommand{\FZ} {\text{\usefont{U}{euf}{m}{n}Z}}
\newcommand\Ribbon {\hat{F}}

\newcommand{\Strans}[1]{\mathbf{S}\left[ #1 \right]}
\newcommand{\Ttrans}[1]{\mathbf{T}\left[ #1 \right]}
\usepackage{dsfont}
\renewcommand{\mathbb}{\mathds}

\usepackage{tikz}
\usetikzlibrary{arrows,arrows.meta,decorations.markings,decorations.pathreplacing,
decorations.pathmorphing,calc}

\usepackage{booktabs}
\usepackage{makecell}

\usepackage[hypertexnames=true,
            colorlinks,
            linkcolor=blue,
            anchorcolor=blue,
            citecolor=blue]{hyperref}

\theoremstyle{plain}
\newtheorem{theorem}{Theorem}[section]
\newtheorem{theoremph}[theorem]{Theorem$^{\mathrm{ph}}$}
\newtheorem{lemma}[theorem]{Lemma}
\newtheorem{conjecture}{Conjecture}[section]

\theoremstyle{definition}
\newtheorem{definition}{Definition}[section]
\newtheorem{remark}{Remark}[section]

\begin{document}


\title{Gapped Boundaries of Kitaev's Quantum Double Models: A Lattice Realization of Anyon Condensation from Lagrangian Algebras}



\author{Mu Li}
\thanks{These authors contributed equally to this work.}
\affiliation{Shenzhen Institute for Quantum Science and Engineering, 
Southern University of Science and Technology, Shenzhen, 518055, China}
\affiliation{International Quantum Academy, Shenzhen 518048, China}
\affiliation{Guangdong Provincial Key Laboratory of Quantum Science and Engineering,
Southern University of Science and Technology, Shenzhen, 518055, China}

\author{Xiao-Han Yang}
\thanks{These authors contributed equally to this work.}
\affiliation{Hefei National Laboratory, University of Science and Technology of China, Hefei 230088, China}
\affiliation{Hefei National Research Center for Physical Sciences at the Microscale and School of Physical Sciences, University of Science and Technology of China, Hefei 230026, China}

\author{Xiao-Yu Dong}
\email{dongxyphys@ustc.edu.cn}
\affiliation{Hefei National Laboratory, University of Science and Technology of China, Hefei 230088, China}


\date{\today}
\begin{abstract}    
The macroscopic theory of anyon condensation, rooted in the categorical structure of topological excitations, provides a complete classification of gapped boundaries in topologically ordered systems, where distinct boundaries correspond to the condensation of different Lagrangian algebras. However, an intrinsic and direct understanding of anyon condensation in lattice models, grounded in the framework of Lagrangian algebras, remains undeveloped. In this paper, we propose a systematic framework for constructing all gapped boundaries of Kitaev's quantum double models directly from the data of Lagrangian algebras. Central to our approach is the observation that bulk interactions in the quantum double models admit two complementary interpretations: the anyon-creating picture and the anyon-probing picture. Generalizing this insight to the boundary, we derive the consistency condition for boundary ribbon operators that respect the mathematical axiomatic structure of Lagrangian algebras. Solving these conditions yields explicit expressions for the local boundary interactions required to realize gapped boundaries. We also provide two families of solutions that cover a broad range of cases. Our construction provides a microscopic characterization of the bulk-to-boundary anyon condensation dynamics via the action of ribbon operators. Moreover, all these boundary terms are supported within a common effective Hilbert space, making further studies on pure boundary phase transitions natural and convenient. Given the broad applicability of anyon condensation theory, we believe that our approach can be generalized to planar topological codes, extended string-net models, or higher-dimensional topologically ordered systems.  
\end{abstract}


\maketitle
\tableofcontents

\section{Introduction}

The concept of topological order emerged from seminal studies on the fractional quantum Hall effects~\cite{MOORE1991362, PhysRevB.41.9377, wenTopologicalOrdersEdge1995, PhysRevLett.66.802, BLOK1992615}, where conventional symmetry-breaking descriptions fail to characterize different phases. In the infrared regime, the universal properties of topologically ordered phases are effectively described by the Chern-Simons field theories~\cite{witten1989quantum, PhysRevLett.65.1502, PhysRevB.46.2290}. Topological excitations of a topologically ordered phase are described by a category $\mathscr{C}$, which is a unitary modular tensor category (UMTC) in 2+1D. In the ultraviolet limit, topologically ordered phases have been realized in explicit microscopic exactly solvable lattice models, including Kitaev's quantum double models~\cite{KITAEV20032} and Levin-Wen's string-net models~\cite{levin2005string, lin2021generalized}.

On lattices with spatial boundaries, topologically ordered phases~\cite{levin2005string} exhibit a range of remarkable properties, most notably the holographic duality~\cite{wen1994chiral, cappelli1997modular, MOORE1991362}. A boundary of a topologically ordered system is well-defined only when the interactions near the boundary are specified in a manner consistent with the bulk Hamiltonian. Depending on the nature of the low-energy excitation spectrum, a boundary is classified as gapped or gapless. Mathematically, a boundary of a topological order $\mathscr{C}$ is described by a category $\mathcal{B}$. For any 1+1D gapped boundary of a 2+1D topological order, $\mathcal{B}$ is a unitary fusion category (UFC). A general bulk-boundary relation states that the bulk $\mathscr{C}$ is the Drinfeld center $\FZ_1(\mathcal{B})$ of its boundary $\mathcal{B}$, i.e., $\mathscr{C} \simeq \FZ_1(\mathcal{B})$~\cite{kong_boundary_2017,kong_gapless_2020,kong_gapless2_2021}, which is also known as the holographic principle. This relation determines the unique one-dimensional higher bulk topological order for a given boundary.

A given bulk topological order can admit multiple distinct gapped boundaries. Without altering the properties of the bulk, different gapped boundaries can be viewed as different `gapped-boundary phases' of the bulk-boundary quantum system, and the phase transitions between them that only change the properties of the boundary are called the `pure boundary phase transitions'~\cite{Chetan_PureEdgeTrans, ChenPurePhaseTrans2020, lu_boundary_2023}.

As a special class of quantum phase transitions, pure boundary phase transitions are particularly notable for admitting a precise mathematical characterization of their critical points. Within the framework of topological Wick rotation, the critical point of a pure boundary phase transition corresponds to a gappable nonchiral gapless boundary. The macroscopic observables of these boundaries are rigorously described by an enriched fusion category~\cite{kong_gapless_2020,kong_gapless2_2021}. While this categorical description is mathematically precise and physically intuitive, it remains largely abstract at a macroscopic level. The correspondence between the macroscopic categorical descriptions and the computable physical observables remains poorly understood. This gap between macroscopic description and microscopic realization motivates the construction of explicit boundary terms of lattice models, as they provide a bridge between abstract categorical descriptions and tangible physical systems.

In Ref.~\cite{ChenPurePhaseTrans2020}, the authors constructed two gapped boundaries of the toric code model using Majorana zero modes and studied the critical point of the phase transition between them through topological Wick rotation~\cite{kong_gapless_2020,kong_gapless2_2021}. Subsequently, their construction was generalized to the $\mathbb{Z}_N$ quantum double~\cite{lu_boundary_2023}. However, the study of pure boundary phase transitions for a general topological order is still lacking. One of the most significant reasons is that the existing microscopic realizations of gapped boundaries of topological orders are subject to certain limitations, making them unsuitable for investigating pure boundary phase transitions.


In the framework of a 2+1D $\mathcal{G}$-string-net model, where the input data is specified by a UFC $\mathcal{G}$, the 1+1D gapped boundaries of the bulk topological order $\FZ_1(\mathcal{G})$ can be systematically characterized through three distinct yet mathematically equivalent data:
\begin{enumerate}
    \item Module categories of $\mathcal{G}$;
    \item Frobenius algebras in $\mathcal{G}$;
    \item Lagrangian algebras in $\FZ_1(\mathcal{G})$.
\end{enumerate}

Concrete constructions of lattice models with boundaries based on the first two types of data have been developed in previous works~\cite{kitaev2012models, hu2017boundary}. However, both approaches face notable limitations in the context of studying the pure boundary phase transitions. In the first approach~\cite{kitaev2012models}, distinct gapped boundaries are realized within different lattice Hilbert spaces, making it unnatural and unconventional to study pure boundary phase transitions. In the second approach~\cite{hu2017boundary}, different Frobenius algebras may realize identical gapped boundaries, introducing redundancy and unnecessary complexity in the analysis of pure boundary phase transitions.



These limitations arise from two primary factors. First, both types of data are defined in terms of the input data $\mathcal{G}$ rather than the output topological data $\FZ_1(\mathcal{G})$. This factor will be addressed later. Second, the local Hilbert space of the string-net model is insufficient to fully accommodate the required degrees of freedom. Under the constraints imposed by the size of the local Hilbert space and the artificial dependence of plaquette operators on the fusion rules, the introduction of auxiliary spaces are essential for the proper definition of simple topological excitations~\cite{buerschaperMappingQDSN2009}. When a boundary is introduced on the lattice, the corresponding effective boundary Hilbert space lacks the capacity to simultaneously support all distinct gapped boundary phases. Consequently, in the first approach~\cite{kitaev2012models}, different boundary Hilbert spaces are required to realize different boundaries. Alternatively, if the boundary Hilbert space is artificially enlarged through a duplication process, as in the second approach~\cite{hu2017boundary}, it inevitably introduces unnecessary redundancy. Therefore, we shift our focus from the string-net models to the quantum double models.

The relationship between the quantum double models and the string-net models is multifaceted. At a superficial level, there is an overlap between these two frameworks: for a finite Abelian group $A$, a quantum double model with input $A$ can be viewed as a string-net model with input $\mathcal{G} \simeq \text{Rep}_A$, where $\text{Rep}_A$ denotes the category of $A$-representations. At a deeper level, a quantum double model based on a non-Abelian group (or more generally, a $C^*$-Hopf algebra) is equivalent, via a Fourier transform, to a so-called extended string-net model~\cite{buerschaperMappingQDSN2009}. Here, the term `extended' refers to the canonical enlargement of the local Hilbert space through a fiber functor $\omega: \mathcal{G} \to \text{Vec}$. This equivalence can be interpreted as a manifestation of the electric-magnetic duality~\cite{buerschaperEM2013}, which is further extended to models with gapped boundaries~\cite{wang2020electric}. In this context, the local Hilbert space of a quantum double model is larger than that of the corresponding string-net model, making it more suitable for studying pure boundary phase transitions.

The study of the microscopic construction of gapped boundaries in quantum double models can be traced back to the smooth and rough boundaries of the toric code model ($\mathbb{Z}_2$ quantum double)~\cite{KITAEV20032}. For general $G$-quantum double models, gapped boundaries can be systematically constructed using the representation theory of the quantum double algebra~\cite{beigi_quantum_2011}. While this construction provides valuable insights, its implementation relies on the algebraic properties of the input data $G$, which imposes certain limitations. In particular, the physical interpretation of topological excitations on the boundary and the processes of bulk-to-boundary anyon condensation remain predominantly algebraic rather than geometrically intuitive. These features underscore the need for a complementary approach that could provide more geometric and physically motivated interpretations and also extend beyond the limitations imposed by the input data.

In this work, we aim to utilize the Lagrangian algebras in $\FZ_1(\text{Vec}_G)$ to construct all gapped boundaries of 2+1D $G$-quantum double models, where $\FZ_1(\text{Vec}_G)$ is the category describing the bulk topological order and $\text{Vec}_G$ is category of $G$-graded vector spaces.


{The mathematical notion of a Lagrangian algebra was originated from the study of 1+1D rational conformal field theories~\cite{kong2009cardy,Muger_2010}, and was later formulated slightly more generally and officially named in Ref.~\cite{DGNO2013}.} Over time, Lagrangian algebras have evolved from their abstract mathematical origins into essential tools for understanding the physical realization of boundaries of topologically ordered systems, particularly through their intricate relationship with anyon condensation.

Anyon condensation provides a physical mechanism and mathematical framework for characterizing transitions between phases with distinct topological orders~\cite{SlingerlandAnyonCond,BurnellAnyonCondense2018,kong_2014_anyoncondense}. Central to this approach is the concept of a condensable algebra in the initial phase, which determines both the topological order of the resulting phase and the structure of the gapped domain wall separating them. In the special case where the condensation drives the system into the trivial phase $\text{Vec}$, this framework yields a complete classification of gapped boundaries of the initial phase. The corresponding condensable algebra is precisely a Lagrangian algebras. This perspective has been applied to lattice models with a comparatively simple class of gapped boundaries, where the topological boundary excitations and defects are analyzed algebraically~\cite{cong2017defects,CongAlgebraicTheories2017}.

A Lagrangian algebra corresponds to a certain type of topological excitation in the bulk that condenses to the vacuum state on a gapped boundary. Different choices of Lagrangian algebra lead to distinct boundaries, serving as a macroscopic observable that classifies and distinguishes gapped boundaries. As such, the construction based on the Lagrangian algebras offers a more physically intuitive framework for investigating pure boundary phase transitions compared to previously discussed other approaches and holds greater potential for establishing connections with experimental systems. Since the theory of anyon condensation applies universally to all 2+1D topologically ordered systems and also admits natural generalizations to higher dimensions~\cite{kong2024higher}, our construction can, in principle,  be extended to a wider class of systems, such as extended string-net models and higher-dimensional models. A detailed exploration of these generalizations is left for future work.

This work also carries broader significance. Through the framework of topological Wick rotation~\cite{kong_gapless_2020,kong_gapless2_2021}, gapped boundaries of 2+1D topological orders are in one-to-one correspondence with 1+1D gapped phases with symmetry, where the topological order in the bulk serves as the categorical symmetry~\cite{kongAlgebraicHigherSymmetry2020,chatterjee_2023_symTO} or the symTO/symTFT~\cite{freed_2024_symTFT} of the latter. In several specific cases, the phase diagrams of such phase transitions have been calculated~\cite{chatterjee_2023_symTO,chatterje_HoloPhaseTrans_2023,chatterjee_NonInvertibleTrans_2024}. However, similar to the study of pure boundary phase transitions, the investigation of phase transitions in 1+1D gapped phases with symmetry still lacks a universal and systematic microscopic framework. Within the context of topological Wick rotation, our construction provides valuable insights for systematically constructing 1+1D gapped lattice models with symmetry, thereby establishing a robust microscopic foundation for exploring phase transitions in these systems.


This work also advances the study of quantum error-correcting codes with open boundaries~\cite{Bombin2006,kesselring2018boundaries}. Quantum double models with open boundaries naturally generalize the planar surface code of $Z_2$ qubits~\cite{kitaev1998Z2bdy}, and their planar geometry facilitates experimental realization. Different boundary conditions yield distinct code properties, including the number of encoded logical qubits and the structure of logical operators. Recent developments include algorithmic methods to systematically construct gapped boundaries in topological Pauli stabilizer codes~\cite{Liang2024Operator}, which lead to the construction of planar quantum low-density parity-check codes with open boundaries~\cite{Liang2025Planar}, as well as analyses of anyon condensation in color codes and the design of fault-tolerant logical operators~\cite{Kesselring2024Color}. Experimental realizations of planar codes are easier when all gapped boundaries reside on the same lattice structure and share the same local Hilbert spaces, which is precisely the setting of our construction. Our work extends these directions by generalizing the framework of planar topological codes to non-Abelian cases.


%

This paper is organized as follows. In Section~\ref{Intro_QD}, we review some basics of Kitaev's quantum double models, including the bulk Hamiltonians, ribbon operators, and the definition of sites.

Section~\ref{sec:CreatAndProb} establishes the correspondence between anyonic excitations and ribbon operators in Kitaev's quantum double models within two distinct physical frameworks: the anyon-creating picture and the anyon-probing picture. Additionally, we introduce the concept of internal degrees of freedom (DOFs) for anyons.

Section~\ref{sec:ModularTrans} constructs the operator-state correspondence between ribbon operators and quantum states on the torus. Furthermore, it elucidates the duality between anyon-creating operators and anyon-probing operators through the $S$-transformation and clarifies its relationship to anyon braiding. This correspondence plays a pivotal role in the construction of gapped boundaries.

Section~\ref{sec:QDbdy} constitutes the central contribution of this work. The analysis commences with a review of the macroscopic Lagrangian algebra framework for gapped boundary construction. In \S\ref{subsec:zigzag}, we define the zig-zag lattice configuration and the effective Hilbert space for boundary states. \S\ref{subsec:bdyProbAndCreat} analyzes the physical processes of anyon probing and anyon creation at the boundary and formally presents the interaction terms for gapped boundaries. In \S\ref{subsec:DefCondition}, we derive the consistency conditions for the boundary interaction terms based on the algebraic properties of Lagrangian algebras. The results are summarized in the physical Theorem~\ref{thmph:bdyTerms}. In the subsequent subsection \S\ref{subsec:ConsructiveExtence}, we present Theorem~\ref{thm:SSBterms} and Theorem~\ref{thm:AbelianBoundaryTerms} as two families of constructively derived solutions to Theorem~\ref{thmph:bdyTerms}. Subsection~\S\ref{subsec:GroundDynamics} concludes this section with a microscopic analysis of bulk-to-boundary anyon condensation dynamics.

Section~\ref{sec:examples} validates our theoretical framework through three paradigmatic examples, supported by explicit computational demonstrations. Comparative analyses with existing methodologies highlight the operational efficiency and practical advantages of our approach.

Section~\ref{sec:Summary} gives the summary and outlook.


\section{Basics of the quantum double models}
\label{Intro_QD}

\subsection{Bulk Hamiltonian}
\begin{figure}
    \centering
    \includegraphics{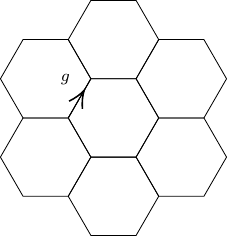}
    \caption{Kitaev's quantum double models defined on a honeycomb lattice.}
    \label{Fig:honeycomb}
\end{figure}

We consider 2+1D Kitaev's quantum double models~\cite{KITAEV20032} defined on a honeycomb lattice as shown in Fig.~\ref{Fig:honeycomb}. Given a finite group $G$, a local Hilbert space $\mathcal{H}_{\mathrm{loc}}=\mathbf{span}\{\ket{g}\}_{g\in G}$ is attached to each edge of the lattice. We can make a convention about the direction of each edge by an arrow, and a label $g$ for an edge represents the local physical state $|g\rangle$. The state $|g\rangle$ can also be represented by a reversed arrow with a label $g^{-1}$, i.e., reversing the arrow inverts the group element.

For each vertex $\alpha$ of the lattice, a vertex operator is defined as:
\begin{equation}
    \includegraphics{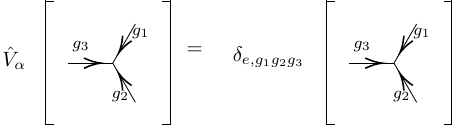},
\end{equation}
where $e$ is the identity element of $G$. For each plaquette $\beta$, the total plaquette operator has the form:
\begin{equation}
    \hat{P}_{\beta}=\frac{1}{|G|}\sum_{h\in G}\hat{P}_{\beta}( h),
\end{equation}
where $|G|$ is the rank of group $G$, and each term $\hat{P}_{\beta}(h)$ is defined as:
\begin{equation}
    \includegraphics{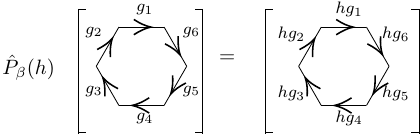}.
\end{equation}

The Hamiltonian of a Kitaev's quantum double model is:
\begin{eqnarray}
    H_{{\mathrm{QD}}}=-\sum_{{\mathrm{vertices}\ \alpha}}\hat{V}_{\alpha}-\sum_{{\mathrm{plaquettes}\ \beta}}\hat{P}_{\beta}.
\end{eqnarray}
Here $\hat{V}_{\alpha}$ and $\hat{P}_{\beta}$ ($\forall \alpha,\beta$) are projectors and commute with each other. The ground states are the common eigenvectors of all $\hat{V}_{\alpha}$ and $\hat{P}_{\beta}$ with eigenvalues $+1$.

\subsection{Ribbon operators and sites}\label{subsec:RibAndSite}
In the quantum double model with input data $G$, point-like topological excitations (i.e., anyons) form a UMTC $\mathrm{Rep}(D(G))\simeq \FZ_1(\mathrm{Vec}_G)$, where $D(G)$ is the quantum double of group $G$ and $\FZ_1(\mathrm{Vec}_G)$ is the Drinfeld center of the category $\mathrm{Vec}_G$~\cite{kitaev2006anyons,bombin_family_2008,EGNO_2015,kong_2022_invitation}. Anyons are created and moved by ribbon operators, which act along oriented ribbon configurations (called paths in the following) on the lattice. There are two primary types of ribbon operators: the charge-like ribbon operators and the flux-like ribbon operators. 

A charge-like ribbon operator $\hat{Y}^g(\text{Path})$ is labeled by a group element $g\in G$ and defined on any directed path as:
\begin{equation}
\begin{array}{c}
    \hat{Y}^g(\text{Path})=\delta_{g , x_Nx_{N-1}...x_3x_2x_1}.\\[10pt]
    \includegraphics{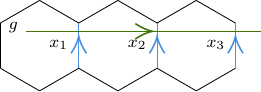}
\end{array}
\end{equation}

A flux-like ribbon operator $\hat{Z}^h(\mathrm{Path})$ is also labeled by a group element $h\in G$ and defined on any directed path as:
\begin{equation}
\begin{array}{c}
    \begin{aligned} \hat{Z}^h(\mathrm{Path})= &\text{ pull string initially labeled }h \text{ along the}\\ &\text{ path and fuse into the left edge.} \end{aligned}\\[10pt]
    \includegraphics{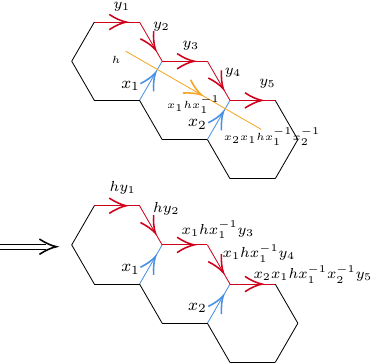}
\end{array}
\end{equation}

The charge-like and flux-like ribbon operators along the same path commute with each other, i.e., $\hat{Y}^{g}{(\mathrm{Path})}\hat{Z}^{h}{(\mathrm{Path})} = \hat{Z}^{h}{(\mathrm{Path})}\hat{Y}^{g}{(\mathrm{Path})}$.
  
A general ribbon operator has the form:
\begin{equation}
    \Ribbon^{g,h}{(\mathrm{Path})} = \hat{Y}^g{(\mathrm{Path})} \hat{Z}^h{(\mathrm{Path})}.
\end{equation}
 The ribbon operators on the same path satisfy the following multiplication rule:
\begin{equation}
    \Ribbon^{g_1,h_1}(\mathrm{Path})\Ribbon^{g_2,h_2}(\mathrm{Path}) = \delta_{g_1,g_2}\Ribbon^{g_1,h_1h_2}(\mathrm{Path}).
\end{equation}
The Hermitian conjugation~\cite{beigi_quantum_2011} of a ribbon operator is:
\begin{equation}
    (\Ribbon^{g,h})^{\dagger}{(\mathrm{Path})} = \Ribbon^{g,h^{-1}}{(\mathrm{Path})}.
\end{equation}

The ribbon operators have two key properties. First, they commute with all vertex and plaquette operators, except at the endpoints of their paths. This property is straightforward to prove. Second, the ribbon operators satisfy the pulling-through property shown in Eq.~(\ref{eq:PullingThrough1}) and Eq.~(\ref{eq:PullingThrough2}):

\begin{equation}
\label{eq:PullingThrough1}
\begin{array}{c}
    \includegraphics{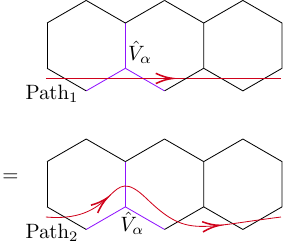}\\[10pt]
    \Ribbon^{g,h}{(\mathrm{Path}_1)}\hat{V}_\alpha = \Ribbon^{g,h}{(\mathrm{Path}_2)}\hat{V}_\alpha,
\end{array}
\end{equation}

\begin{equation}
\label{eq:PullingThrough2}
\begin{array}{c}
    \includegraphics{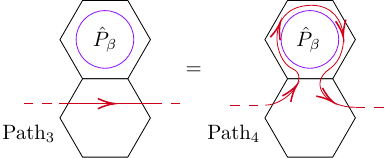}\\[10pt]
    \Ribbon^{g,h}{(\mathrm{Path}_3)}\hat{P}_\beta = \Ribbon^{g,h}{(\mathrm{Path}_4)}\hat{P}_\beta.
\end{array}
\end{equation}

The $0$-eigenstates of a plaquette operator are referred to as charge defects, which can be created and moved by charge-like ribbon operators. Similarly, the $0$-eigenstates of a vertex operator are termed flux defects, and they can be created and moved by flux-like ribbon operators. An anyon, as a local excitation of the Hamiltonian, typically manifests as a composite defect combining both charge and flux defects. To precisely locate an anyon on the lattice, it is necessary to formalize the concept of a ``site''.

\begin{figure}
    \centering
    \includegraphics{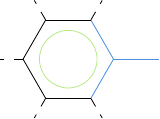}
    \caption{In the quantum double model, a site in the bulk is defined as a combination of a plaquette and an adjacent vertex. The illustrated site consists of the green plaquette and the blue vertex.}
    \label{fig:BulkSite}
\end{figure}

\begin{figure}
  \centering
  \begin{subfigure}[b]{0.48\columnwidth}
    \centering
    \includegraphics{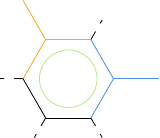}
    \subcaption{ }
  \end{subfigure}
  \hfill
  \begin{subfigure}[b]{0.48\columnwidth}
    \centering
    \includegraphics{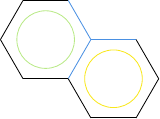}
    \subcaption{ }
  \end{subfigure}
  \caption{Two distinct sites may share a common vertex or plaquette: (a) Two sites sharing the common green plaquette; (b) Two sites sharing the common blue vertex.}
  \label{fig:OverlapSites}
\end{figure}

A site is defined as a combination of a plaquette and an adjacent vertex, as illustrated in Fig.~\ref{fig:BulkSite}. Note that each plaquette is adjacent to multiple nearest-neighbor vertices, and conversely, each vertex interacts with a number of neighboring plaquettes. As a result, two distinct sites may share a common vertex or plaquette, as depicted in Fig.~\ref{fig:OverlapSites}.

\begin{figure}
    \centering
    \includegraphics{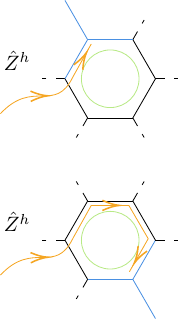}
    \caption{The action of the flux ribbon operator must terminate at the vertex part of the respective sites at the endpoints of the path. In these two illustrated examples, the endpoint of the operator $\hat{Z}^h$ acts exclusively on the edges parallel to the orange line. The number of edges at the endpoint of $\hat{Z}^h$ varies depending on the position of the vertex component of the end site.}
    \label{fig:FluxEndAction}
\end{figure}

Consider a ribbon operator $\Ribbon^{g,h}(\mathrm{Path}_{ij})$ defined along a path from site $j$ to site $i$. At the endpoints $j$ or $i$, the action of the flux-like component $\hat{Z}^h$ must terminate at the vertex part of the respective site as illustrated in Fig.~\ref{fig:FluxEndAction}.

\section{Creating and probing anyons in the bulk}
\label{sec:CreatAndProb}
The macroscopic description of gapped boundaries via anyon condensation is grounded in the physical picture of topological excitations. Topological excitations are excited states that are created from the ground state exclusively by non-local operators. Two excited states belong to the same type of topological excitation (or called topological sector) if and only if they can be connected by local operators.

In this section, we provide a detailed interpretation of topological excitations within Kitaev's quantum double model. Ribbon operators, which are intrinsically linked to topological excitations, will play a pivotal role as an essential tool in the subsequent construction of boundaries.

\subsection{Creating and moving anyons}
The ribbon operator $\Ribbon^{g,h}{(\mathrm{Path})}$ creates a pair of anyons, where one anyon lies at the ending site of the path and its dual anyon lies at the starting site. However, in general, these anyons are not simple, which means that they are direct sums of some simple anyons. A simple anyon cannot be decomposed further. In a quantum double model, a simple anyon is characterized by a paired index $[C,R]$. Here, the index $C$ denotes a conjugate class of the input group $G$. The index $R$ represents an irreducible representation of $Z(r_C)$, which is the centralizer of a selected representative element $r_C$ for each class $C$. To create simple anyons, the ribbon operators $\Ribbon^{g,h}{(\mathrm{Path})}$ should be superposed in the following way:
\begin{equation}
\label{eq:CreatingRibbon}
\begin{array}{c}
  \begin{tikzpicture}
    \tikzset{-<-/.style={decoration={markings,mark=at position .5 with {\arrow{>}}},postaction={decorate}}}
    \draw[-<-] (5,0) node[above]{mp} -- (0,0) node[above]{nq};
    
    \fill (5,0) circle (2pt);
    \fill (0,0) circle (2pt);
  \end{tikzpicture}\\[10pt]
  \hat{M}^{[C,R]}_{nq,mp}(\mathrm{Path}) = \sum_{z\in Z(r_C)}\rho^{R}_{nm}(z){\Ribbon}^{qzp^{-1},pr_Cp^{-1}}(\mathrm{Path}).
\end{array}
\end{equation}
The operator $\hat{M}^{[C,R]}_{nq,mp}(\mathrm{Path})$ creates a simple anyon labeled by $[C,R]$ and internal DOF $nq$ at the ending site of the path and its dual anyon with internal DOF $mp$ at the starting site of the path. For any element $c$ in the conjugacy class $C$, we select a unique group element $p$ such that $p r_C p^{-1} = c$ to represent $c$, thus the class $C$ can be denoted as $\left\{ p \right\}^C$. The $p,q$ in the subindex of $\hat{M}^{[C,R]}_{nq,mp}(\mathrm{Path})$ are two elements in $\left\{ p \right\}^C$. The coefficient $\rho^{R}_{nm}$ is the $(n,m)$ matrix element of the irreducible representation $R$.

The dual anyon type corresponding to $[C,R]$ is given by $[C^{-1},\bar{R}]$, where $C^{-1}$ represents the conjugacy class of $r_C^{-1}$, and $\bar{R}$ denotes the complex conjugate representation of $R$. Note that $Z(r_C^{-1}) = Z(r_C)$, which ensures consistency in the definition. The Hermitian conjugate of the operator $\hat{M}^{[C,R]}_{nq,mp}$ precisely corresponds to the dual anyon-creating operator defined on the same path:

\begin{eqnarray}
    \left( \hat{M}^{[C,R]}_{nq,mp} \right)^{\dagger} &=& \sum_{z \in Z(r_C)} \bar{\rho}_{nm}^{R}(z) \, {\Ribbon}^{qzp^{-1}, pr_C^{-1}p^{-1}} \nonumber \\
    &=& \sum_{z \in Z(r_C)} \rho_{nm}^{\bar{R}}(z) \, {\Ribbon}^{qzp^{-1}, pr_C^{-1}p^{-1}} \nonumber \\
    &=& \hat{M}^{[C^{-1},\bar{R}]}_{nq,mp}.
\end{eqnarray}

\begin{figure}
  \centering
  \includegraphics{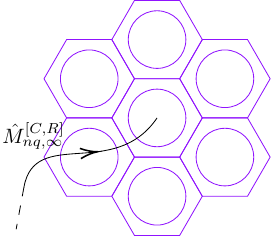}
  \caption{A non-local operator $\hat{M}^{[C,R]}_{nq,\infty}$ creates a topological excitation $\ket{[C,R];nq}$ at its ending site. The purple vertices and plaquettes represent local eigenstates of the vertex operators and plaquette operators with eigenvalues equal to $1$, respectively.}
  \label{fig:LocalTopoExc}
\end{figure}

Considering a half-infinite path, we could look at the anyon at the ending site of the path locally. As shown in Fig.~\ref{fig:LocalTopoExc}, one anyon with state $\ket{[C,R];nq}$ can be created by the operator $\hat{M}^{[C,R]}_{nq,\infty}$ acting on $\ket{\Omega}$, which is the unique ground state of the quantum double model on a infinite plane. This relationship is expressed as:
\begin{equation}
    \ket{[C,R];nq} = \hat{M}^{[C,R]}_{nq,\infty} \ket{\Omega},
\end{equation}
where the subscript $\infty$ indicates that the starting site of the path is infinitely far from its ending site, and the internal DOF is arbitrary. Due to the pulling-through property, the operator $\hat{M}^{[C,R]}_{nq,\infty}$, defined on any half-infinite path with the same ending site, will always produce the same local state $\ket{[C,R];nq}$.

Considering two paths $\mathrm{Path}_{ii^{\prime}}$ and $\mathrm{Path}_{i^{\prime}j}$ that are connected end-to-end at site $i^{\prime}$, the anyon-creating operators defined on these paths are concatenated according to the following rule:
\begin{eqnarray}
&&\hat{M}^{[C,R]}_{nq,mp}(\mathrm{Path}_{ii^{\prime}}*\mathrm{Path}_{i^{\prime}j}) \nonumber\\
&&= \sum_{n^{\prime}q^{\prime}}\hat{M}^{[C,R]}_{nq,n^{\prime}q^{\prime}}(\mathrm{Path}_{ii^{\prime}})\hat{M}^{[C,R]}_{n^{\prime}q^{\prime},mp}(\mathrm{Path}_{i^{\prime}j}).
\end{eqnarray}
As a result, $\hat{M}^{[C,R]}_{nq,mp}$ also plays the role of moving the location of anyons.

\subsection{Probing of anyon types}\label{subsec:LocalProbing}
\begin{figure}
  \centering
  \includegraphics{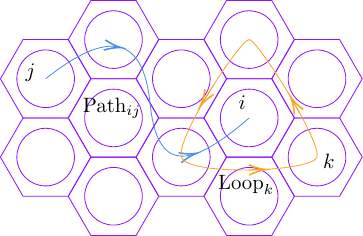}
  \caption{The orange path $\mathrm{Loop}_{k}$ is a loop with the same starting and ending site $k$. The blue path $\mathrm{Path}_{ij}$ starts at site $j$ and ends at site $i$.}
  \label{fig:OpenClosedPath}
\end{figure}

In addition to creating anyons, we can take another dual view of ribbon operators. Consider a path $\mathrm{Path}_{ij}$ and a loop $\mathrm{Loop}_{k}$ as shown in Fig.~\ref{fig:OpenClosedPath}, where $\mathrm{Loop}_{k}$ is a path whose starting and ending points are the same site $k$ and surrounds the ending site $i$ of $\mathrm{Path}_{ij}$. Our goal here is to identify the topological excitations enclosed by $\mathrm{Loop}_k$.

We consider the algebra generated by all the ribbon operators supported on $\mathrm{Loop}_{k}$, which commute with all the $\hat{V}_{\alpha}$ and $\hat{P}_{\beta}$ operators except those at the site $k$. These ribbon operators can themselves create excitations at the endpoints; however, we temporarily ignore this feature and disregard the commutation properties at the endpoints. The idempotent decomposition of this algebra yields the projectors associated with the simple anyonic excitations enclosed by $\mathrm{Loop}_k$.

The algebra generated by all the ribbon operators on $\mathrm{Loop}_{k}$ is:
\begin{equation}
  \mathfrak{C}_{\mathrm{Loop}_{k}} =  \mathbf{gen}\left\{ \mathrm{Ribbon\ operators\ } {\Ribbon}^{g,h}(\mathrm{Loop}_{k})\right\}.
\end{equation}
Due to the completeness of representation matrix elements, the algebra $\mathfrak{C}_{\mathrm{Loop}_{k}}$ can also be written as:
\begin{eqnarray}
\mathfrak{C}_{\mathrm{Loop}_{k}} = \mathbf{gen}\left\{ \hat{P}^{[C,R]}_{mp,nq}(\mathrm{Loop}_{k})\right\},
\end{eqnarray}
where 
\begin{eqnarray}
\hat{P}^{[C,R]}_{mp,nq}= \frac{|C|d_R}{|G|}\sum_{z\in Z(r_C)}\bar{\rho}^{R}_{mn}(z){\Ribbon}^{pr_Cp^{-1},pzq^{-1}},
\end{eqnarray}
in which $|C|$ is the rank of class $C$, $d_R$ is the dimension of the irreducible representation $R$, and $\bar{\rho}^{R}_{mn}$ is the complex conjugation of $\rho^{R}_{mn}$. Since the multiplication and linear combinations of ribbon operators defined on a specific path are independent of the path itself, the algebras $\mathfrak{C}_{\mathrm{Path}}$ defined on different paths are isomorphic. In this context, we often omit the subscript and denote the algebra as $\mathfrak{C}$.

The idempotent decomposition of the algebra $\mathfrak{C}$ is given by the diagonal terms:
\begin{eqnarray} 
\hat{P}^{[C,R]}_{nq}(\mathrm{Loop}_{k}) &\equiv& \hat{P}^{[C,R]}_{nq,nq}(\mathrm{Loop}_{k}).
\end{eqnarray}
The $\{\hat{P}^{[C,R]}_{nq}\}$ forms a complete set of mutually orthogonal projection operators. The orthogonality and normalization of $\{\hat{P}^{[C,R]}_{nq}\}$ are proven in Appendix~\ref{apdx:AlgCompute}. 

We name the operator $\hat{P}^{[C,R]}_{nq}(\mathrm{Loop}_{k})$ as an anyon-probing operator, since it can detect the excited states created by $\hat{M}^{[C,R]}_{nq,mp}(\mathrm{Path}_{ij})$ at its ending site $i$, as was demonstrated in~\cite{bombin_family_2008}: 
\begin{eqnarray}
\label{Detect_LDoF}
&&\hat{P}^{[C,R]}_{nq}(\mathrm{Loop}_{k})\hat{M}^{[C',R']}_{n'q',m'p'}(\mathrm{Path}_{ij})\ket{\Omega} \nonumber\\
&&= \delta_{[C,R],[C',R']}\delta_{n,n'}\delta_{q,q'}\hat{M}^{[C',R']}_{n'q',m'p'}(\mathrm{Path}_{ij})\ket{\Omega}.
\end{eqnarray}

The off-diagonal operator $\hat{P}^{[C,R]}_{n^{\prime\prime} q^{\prime\prime},nq}(\mathrm{Loop}_{k})$ can change the internal DOF of the excited states at site $i$:
\begin{eqnarray}
\label{Change_LDoF}
&&\hat{P}^{[C,R]}_{n^{\prime\prime} q^{\prime\prime},nq}(\mathrm{Loop}_{k})\hat{M}^{[C',R']}_{n'q',m'p'}(\mathrm{Path}_{ij})\ket{\Omega}\nonumber\\
&&= \delta_{[C,R],[C',R']}\delta_{n,n'}\delta_{q,q'}\hat{M}^{[C',R']}_{n^{\prime\prime}q^{\prime\prime},m'p'}(\mathrm{Path}_{ij})\ket{\Omega}.
\end{eqnarray}

Due to the pulling-through property, the anyon-probing operator can be defined on any loop encircling one endpoint of the anyon-creating operator, rather than being restricted to the specific $\mathrm{Loop}_{k}$ in Fig.~\ref{fig:OpenClosedPath}, and the algebraic relations in Eq.~\eqref{Detect_LDoF} and Eq.~(\ref{Change_LDoF}) still hold.

Restricting to the subspace of local excited states at the ending site of $\hat{M}^{[C,R]}_{-,\infty}$ and considering a minimal loop encircling it, the diagonal $\hat{P}$-operators are projectors of the excited states and the off-diagonal $\hat{P}$-operators can change its internal DOF. In this sense, $\hat{P}$-operators can be written as:
\begin{equation}
  \hat{P}^{[C,R]}_{mp,nq} \overset{\mathrm{restricted}}{=} \ket{[C,R];mp}\bra{[C,R];nq}.
\end{equation}

The trace of $\hat{P}^{[C,R]}_{mp,nq}$ is the projector on the local subspace of topological sector $[C,R]$:
\begin{eqnarray}
      &&\hat{P}^{[C,R]} = \tr\left(\hat{P}^{[C,R]}_{nq,mp} \right) = \sum_{n,q}\hat{P}^{[C,R]}_{nq,nq}\nonumber\\
  &&= \frac{|C|d_R}{|G|}\sum_{p\in\{p\}^{C}}\sum_{z\in Z(r_C)}\bar{\chi}^{R}(z){\Ribbon}^{pr_Cp^{-1},pzp^{-1}},
\end{eqnarray}
where $\bar{\chi}^R$ is the complex conjugation of the character of the representation $R$.

\subsection{Superposition of internal degrees of freedom}
We have introduced the index set $\{nq\}$ to represent the internal DOF of the topological excitations. In general, the internal state can be any superposition of them, for example, $\frac{1}{\sqrt{2}}(\ket{a;1}+\ket{a;2})$, where $1,2\in \{nq\}$ and $a$ is a simplified notation of an anyon type $[C,R]$. The probing operator that detects this superposed state is:
\begin{eqnarray}
      \hat{P}^{a}_{\frac{1}{\sqrt{2}}({1}+{2})} &\equiv& \frac{1}{2}\left(\ket{a;1}+\ket{a;2})(\bra{a;1}+\bra{a;2} \right) \nonumber\\
  &=& \frac{1}{2}\left( \hat{P}^{a}_{1,1}+\hat{P}^{a}_{1,2}+\hat{P}^{a}_{2,1}+\hat{P}^{a}_{2,2} \right).
\end{eqnarray}
The creating operators that carry this superposed state at one of its endpoints are:
\begin{eqnarray}
&&\hat{M}^{a}_{\frac{1}{\sqrt{2}}({1}+{2}),j} \equiv \frac{1}{\sqrt{2}}(\hat{M}^{a}_{1,j}+\hat{M}^{a}_{2,j}),\\
&&\hat{M}^{a}_{j,\frac{1}{\sqrt{2}}({1}+{2})} \equiv \frac{1}{\sqrt{2}}(\hat{M}^{a}_{j,1}+\hat{M}^{a}_{j,2}).
\end{eqnarray}
The following relation, which is similar to Eq.~(\ref{Detect_LDoF}), holds:
\begin{equation}
  \hat{P}^{a}_{\frac{1}{\sqrt{2}}({1}+{2})}\hat{M}^{a}_{\frac{1}{\sqrt{2}}({1}+{2}),j}\ket{\Omega} = \hat{M}^{a}_{\frac{1}{\sqrt{2}}({1}+{2}),j}\ket{\Omega}.
\end{equation}


\subsection{Anyon-creating and anyon-probing pictures}\label{subsec:twopictures}
Now, we consider the trace of the anyon-creating operator $\hat{M}^{[C,R]}_{nq,mp}(\mathrm{Loop}_{k})$:
\begin{eqnarray}
&&\hat{M}^{[C,R]}(\mathrm{Loop}_{k}) = \tr\left(\hat{M}^{[C,R]}_{nq,mp}(\mathrm{Loop}_{k}) \right)\nonumber\\
&&= \sum_{n,q}\hat{M}^{[C,R]}_{nq,nq}(\mathrm{Loop}_{k}) \nonumber\\
&&= \sum_{p\in\{p\}^{C}}\sum_{z\in Z(r_C)}\chi^{R}(z){\Ribbon}^{pzp^{-1},pr_Cp^{-1}}({\mathrm{Loop}_{k}}).
\end{eqnarray}
It is intuitive that the action of $\hat{M}^{[C,R]}(\mathrm{Loop}_{k})$ corresponds to a `dynamical process' involving the creation of a pair of anyon and dual-anyon, moving one of them around the loop, then annihilating the pair. It's easy to verify that $\hat{M}^{[C,R]}(\mathrm{Loop}_{k})$ commutes with every term in the Hamiltonian, which is consistent with the physical intuition outlined above.

When acting on the ground state $\ket{\Omega}$:
\begin{equation}
\label{eq:LoopMovingComponent}
  \hat{M}^{[C,R]}_{nq,mp}(\mathrm{Loop}_{k})\ket{\Omega} = \delta_{m,n}\delta_{p,q}\ket{\Omega},
\end{equation}
and its trace $\hat{M}^{[C,R]}({\mathrm{Loop}_{k}})$ gives:
\begin{equation}
  \hat{M}^{[C,R]}({\mathrm{Loop}_{k}})\ket{\Omega} = |C|d_R\ket{\Omega},
\end{equation}
Note that $|C|d_R = \mathsf{dim}([C,R])$ is the quantum dimension of anyon $[C,R]$. This is consistent with the graph calculus within the UMTC, where the process of creating a pair of anyons and then annihilating them gives the quantum dimension of the anyon.

\begin{figure}[ht]
  \centering

  \begin{subfigure}[b]{0.5\textwidth}
      \includegraphics{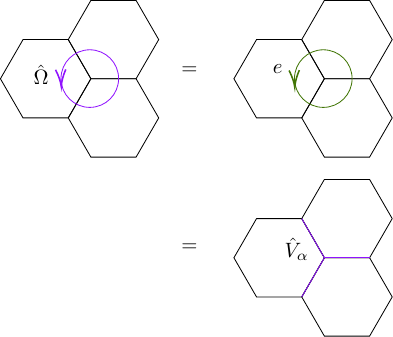}
      \caption{Minimal loop encircling a vertex.}
  \end{subfigure}
  
  \vspace{1em}

  \begin{subfigure}[b]{0.5\textwidth}
    \includegraphics{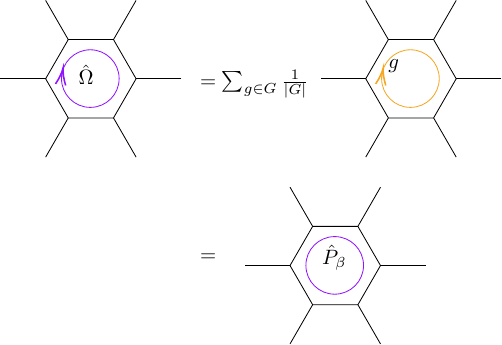}
    \caption{Minimal loop encircling a plaquette.}
  \end{subfigure}
  
  \caption{The $\Omega$-strand operator defined on a minimal loop that surrounds a vertex (plaquette) reduces to the vertex (plaquette) operator.}
  \label{fig:MinimalLoop}
\end{figure}

We can define the $\Omega$-strand operator as the weighted sum of anyon-creating operators:
\begin{eqnarray}
\label{eq:OmegaStrand}
\hat{\Omega} &=& \frac{1}{{\mathsf{dim}(\FZ_1(\mathrm{Vec}_G))}}\sum_{C,R}\mathsf{dim}([C,R])\hat{M}^{[C,R]}\nonumber\\
&=& \frac{1}{|G|}\sum_{C,R}\frac{|C|d_R}{|G|}\hat{M}^{[C,R]}= \left( \frac{1}{|G|}\sum_{g\in G} \hat{Z}^g \right)\hat Y^{e}.
\end{eqnarray}
Detailed calculations are provided in Appendix~\ref{apdx:AlgCompute}.
On the minimal loop that surrounds a vertex (plaquette), the $\Omega$-strand operator reduces to the vertex (plaquette) operator, as illustrated in Fig.~\ref{fig:MinimalLoop}.

Thus, we can rewrite the Hamiltonian of the quantum double model as a summation of $\Omega$-strands:
\begin{equation}
\begin{aligned}
  H_{\mathrm{QD}} =& -\sum_{{\mathrm{vertices}\ \alpha}}\hat{V}_{\alpha}-\sum_{{\mathrm{plaquettes}\ \beta}}\hat{P}_{\beta}\\
  =& -\sum_{{\mathrm{vertices}\ \alpha}} \hat{\Omega}_{\alpha} - \sum_{{\mathrm{plaquettes}\ \beta}}\hat{\Omega}_{\beta}.
\end{aligned}
\end{equation}
It is evident that the ground state of $H_{QD}$ is an eigenstate of all $\Omega$-strands with eigenvalues equal to $1$. This is precisely why we use the notation $\ket{\Omega}$ to represent the ground state of the quantum double.

We refer to this formulation of the Hamiltonian as `anyon-creating picture' of the quantum double model. In this picture, the model acquires a particularly intuitive interpretation: the Hamiltonian is composed of all the minimal local dynamical processes that create a pairs of anyons and subsequently annihilate them.

The projector $\hat{P}^{[C_e,1]}$ on the trivial excitation $\mathbb{1}\equiv[C_e,1]$ is just the $\Omega$-strand operator, where $C_e$ is the trivial conjugate class $\{e\}$, and $1$ is the trivial representation of $Z(e) = G$:
\begin{equation}\label{eq:trvProbingOmega}
\hat{P}^{\mathbb{1}}\equiv \hat{P}^{[C_e,1]} = \left( \frac{1}{|G|}\sum_{g\in G} \hat{Z}^g \right)\hat Y^{e} =\hat{\Omega}.
\end{equation}
Thus, the Hamiltonian can be written as the summation of trivial excitation probing operators:
\begin{eqnarray}
  H_{\mathrm{QD}} = -\sum_{{\mathrm{vertices}\ \alpha}} \hat{P}^{\mathbb{1}}(\alpha) - \sum_{{\mathrm{plaquettes}\ \beta}}\hat{P}^{\mathbb{1}}(\beta) .
\end{eqnarray}

Using the normalization of probing operators,
\begin{equation}
  \hat{P}^{\mathbb{1}} = 1 - \sum_{[C,R]\neq \mathbb{1}} \hat{P}^{[C,R]},
\end{equation}
the Hamiltonian can also be written as the summation of probing operators for all non-trivial anyons:
\begin{eqnarray}
  H_{\mathrm{QD}} &=& \sum_{{\mathrm{vertices}\ \alpha}}\ \sum_{[C,R]\neq \mathbb{1}}\hat{P}^{[C,R]}(\alpha)\nonumber\\
   && + \sum_{{\mathrm{plaquettes}\ \beta}}\ \sum_{[C,R]\neq\mathbb{1}}\hat{P}^{[C,R]}(\beta)+\mathrm{Const}.
\end{eqnarray}
This form of the Hamiltonian is referred to as the `anyon-probing picture' of the quantum double model. Due to the orthogonality of the probing operators, the ground state and excited states are clearly distinguished in this picture.

In summary, we observe that the bulk interactions of a quantum double model can be interpreted in two distinct ways, framed within the anyon language:
\begin{enumerate}
    \item In the \textbf{anyon-creating} picture, the bulk interactions are represented as a weighted sum over all minimal permissible dynamical processes within the ground state.
    \item In the \textbf{anyon-probing} picture, the bulk interactions are characterized as probing operators of the trivial excitation with negative coefficients or probing operators of all non-trivial simple anyons with positive coefficients. Both of them indicate that the ground state has no non-trivial anyons.
\end{enumerate}
The boundary interactions should be constructed according to the same underlying intuitive principles.

\section{$S$ and $T$ transformations}
\label{sec:ModularTrans}
In the previous section, we introduced two complementary pictures, i.e., the anyon-creating and anyon-probing pictures, each with a well-defined physical interpretation. In fact, these two pictures are not isolated but are dual to each other.

We define the $S$-transformation of ribbon operators as (\romannumeral 1) interchanging the upperscripts $g$ and $h$; (\romannumeral 2) taking the inverse of $g$:
\begin{equation}\label{eq:Def_S_hat}
    \hat{F}^{g,h}(\mathrm{Path})\mapsto \mathbf{S} [{\Ribbon}^{g,h}(\mathrm{Path})]\equiv {\Ribbon}^{h,g^{-1}}(\mathrm{Path})
\end{equation}
\begin{equation}
    \mathbf{S}^{-1}[{\Ribbon}^{g,h}(\mathrm{Path})]\equiv {\Ribbon}^{h^{-1},g}(\mathrm{Path})
\end{equation}
The $S$-transformation is linear but does not preserve the multiplication between ribbon operators:
\begin{equation}
    \mathbf{S} [{\Ribbon}^{a,b}]\mathbf{S} [{\Ribbon}^{c,d}]\neq \mathbf{S} [{\Ribbon}^{a,b}{\Ribbon}^{c,d}].
\end{equation}
From this definition, it can be directly derived that:
\begin{equation}
  \label{eq:LoopProbeToMove}
  \hat{P}^{[C,R]}_{k,l}({\mathrm{Loop}}) = \frac{|C|d_{R}}{|G|} \Strans{\hat{M}^{[C,R]}_{l,k}({\mathrm{Loop}})},
\end{equation}
where $k,l$ are some internal DOFs of $[C,R]$.

We also define the $T$-transformation of ribbon operators as:
\begin{equation}\label{eq:Def_T_Trans}
    \hat{F}^{g,h}(\mathrm{Path})\mapsto \mathbf{T} [{\Ribbon}^{g,h}(\mathrm{Path})]\equiv {\Ribbon}^{g,gh}(\mathrm{Path})
\end{equation}

When restricted to the set of mutually commuting group elements $C \equiv \{(g, h) \mid gh = hg\}$, the operators $\mathbf{S}_C$ and $\mathbf{T}_C$ generate the modular group $\operatorname{SL}_2(\mathbb{Z})$:
\begin{equation}
    \mathbf{S}_{C}^4 =1,\quad (\mathbf{S}_{C}\mathbf{T}_{C})^3=\mathbf{S}_{C}^2
\end{equation}
Indeed, the $S$ and $T$ transformations are fundamentally tied to modular invariance, a subject that will be addressed in greater detail later in this work.

\subsection{Operator-state correspondence}\label{subec:OpStaCorr}
To elucidate the physical implications of the aforementioned $S$- and $T$-transformation, we now establish the correspondence between the ribbon operators and the states on the torus.

\begin{figure}
  \centering
  \includegraphics{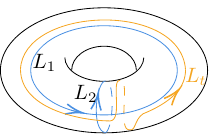}
  \caption{The simplest lattice decomposition of the torus. The path $L_1$ represents the longitudinal direction, $L_2$ denotes the meridional direction, and $L_t$ corresponds to a twisted path.}
  \label{fig:TorusDecomposition}
\end{figure}

Consider a $G$-quantum double model defined on a torus. Since the manifold of the degenerate ground states does not depend on the lattice decomposition of the torus, we choose the simplest one as shown in Fig.~\ref{fig:TorusDecomposition}, where two non-contractible loops are denoted as $L_1$ and $L_2$.
A state $\ket{g,h}$ on the torus is defined as:
\begin{equation}\label{eq:Torus_State}
  \ket{g,h} = \vcenter{\hbox{\includegraphics{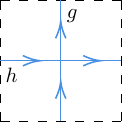}}},
\end{equation}
where the top and bottom, left and right dashed lines are identified, respectively. The set $\{|g,h\rangle,\forall g,h\in G\}$ forms the basis for states on the torus. There exists a correspondence between states on the torus and ribbon operators, in the sense that the linear space spanned by all ribbon operators defined on any fixed path is isomorphic to the linear space of states on the torus (see Appendix~\ref{apdx:Anyon_Basis}):
\begin{equation}
    \mathfrak{C}\simeq \{|g,h\rangle\}_{\mathrm{Torus}}
\end{equation}
This correspondence is realized by:
\begin{equation}
    \hat{F}^{g,h} \Leftrightarrow \ket{g,h},
\end{equation}

Under this established mapping, we specifically obtain the following relations:
\begin{eqnarray}
    \hat{M}^{[C,R]}_{nq,mp} &\Leftrightarrow&  \ket{C,R;nq,mp}_{L_1}\\
    \hat{M}^{[C,R]} &\Leftrightarrow&  \ket{C,R}_{L_1}
\end{eqnarray}

Here, states on the right-hand side are the anyon basis defined in Appendix~\ref{apdx:Anyon_Basis}. We will also employ, for instance, the notation $\ket{{M}^{[C,R]}_{nq,mp}}\equiv \ket{C,R;nq,mp}_{L_1}$ to represent the torus state corresponding to the ribbon operator $\hat{M}^{[C,R]}_{nq,mp}$.

Moreover, the $S$ and $T$ transformations of ribbon operators correspond to unitary operators on the torus states:
\begin{eqnarray}
    \Strans{-} &\Leftrightarrow& \hat{S}\label{S-S correspondence},\\
    \Ttrans{-} &\Leftrightarrow& \hat{T}\label{T-T correspondence},
\end{eqnarray}
where $\hat{S}$ and $\hat{T}$ are defined as:
\begin{eqnarray}
    \hat{S}\ket{g,h} \equiv \ket{h,g^{-1}},&&\quad \hat{T}\ket{g,h} \equiv \ket{g,gh},\\
    \ket{\mathbf{S}[\hat{F}^{g,h}]} = \hat{S}\ket{g,h},&&\quad \ket{\mathbf{T}[{\hat{F}^{g,h}}]} = \hat{T}\ket{g,h}.
\end{eqnarray}
This correspondence is also explained in detail in Appendix~\ref{apdx:Anyon_Basis}.

\subsection{Transformation between creating and probing operators}\label{subsec:GraphCalculus}
It has been established in the context of TQFT~\cite{freedman2008picture,dittrich2017quantum} and lattice models~\cite{bombin_family_2008,delcamp_fusion_2017} that the ground states of the quantum double models on the torus are indeed TQFT states. Thus, some topological invariants such as the overlap of states can be calculated using TQFT diagrams or graph calculus of tensor category.  

By finding the irreducible central idempotents of the tube algebra, a set of basis of the ground-state subspace on the torus is obtained~\cite{simon2023topological}:
\begin{equation}\label{eq:Any_Bas}
  \ket{C,R}_{L_1} = \sum_{p\in\{p\}^{C}}\sum_{z\in Z(r_C)}\chi^{R}(z)\ket{pzp^{-1},pr_Cp^{-1}},
\end{equation}
where $C,R$ are anyon labels and $L_1$ denotes the loop $L_1$ in Fig.~\ref{fig:TorusDecomposition}. We label the states on torus using the same labels $C,R$ as the operator $\hat{M}^{[C,R]}(L_1)$ we defined in \S\ref{subsec:twopictures} because of the operator-state correspondence introduced in \S\ref{subec:OpStaCorr}. The details of the correspondence are provided in Eq.~(\ref{eq:L1GSAnyonloop}):
\begin{equation}
    \ket{C,R}_{L_1}=\hat{M}^{[C,R]}(L_1)\ket{C_e,1}_{L_1}.
\end{equation}
Such ground states on the torus possess a physical interpretation: each ground state corresponds to a process that an anyon of type $[C, R]$ and its dual are created and propagate around the torus along the loop $L_1$ in opposite directions, then finally they meet and annihilate with each other~\cite{preskill1999lecture,simon2023topological}. These ground states can be represented graphically as:
\begin{equation}\label{eq:TorusketMa}
  \ket{a}_{L_1} = \vcenter{\hbox{\includegraphics{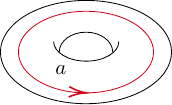}}},
\end{equation}
where we abbreviate the anyon label $[C,R]$ with $a$. In the context of TQFT, $\ket{a}_{L_1}$ is a state represented by such a solid torus $D^2 \times S^1$, with an anyon tube line of type $a$ dragged around the handle inside its bulk. It is well-known that the ground-state subspace of a topologically ordered system on the torus is generated by such physical processes~\cite{preskill1999lecture}.

Another set of basis of the torus ground-state subspace has the form:
\begin{equation}\label{eq:Any_Bas2}
  \ket{C,R}_{L_2} = \sum_{p\in\{p\}^{C}}\sum_{z\in Z(r_C)}\bar{\chi}^{R}(z)\ket{pr_Cp^{-1},pzp^{-1}},
\end{equation}
As demonstrated in Appendix~\ref{apdx:Anyon_Basis}, this set of basis admits a physical interpretation associated with anyon loops along $L_2$.

Using the definition of unitary operator $\hat{S}$, we get
\begin{equation}
\hat{S}\ket{C,R}_{L_1}=\ket{C,R}_{L_2}.
\end{equation}
It is elaborated that the operator $\hat{S}$ can be understood as interchanging the meridian and longitude of the torus. Under this operation, the topology of $D^2 \times S^1$ is transformed into $S^1 \times D^2$, illustrated as:
\begin{equation}
    D^2 \times S^1 \xrightarrow{\circlearrowleft} S^1 \times D^2.
\end{equation}
Therefore, the $L_2$ anyon basis can be represented graphically as an anyon tube of type $a$ encircling the handle within the bulk of $S^1 \times D^2$:
\begin{eqnarray}
    \ket{a}_{L_1}&\xrightarrow{\hat{S}}&\ket{a}_{L_2}\\[10pt]
    \vcenter{\hbox{\includegraphics{TorusketMa.pdf}}}&\xrightarrow{\circlearrowleft}&\vcenter{\hbox{\includegraphics{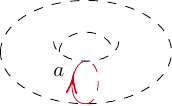}}}
\end{eqnarray}
Here, a dashed torus is used to indicate that it represents the boundary of $S^1 \times D^2$, as opposed to the boundary of $D^2 \times S^1$ in Eq.~(\ref{eq:TorusketMa}).

Consider the inner product:
\begin{equation}\label{eq:S_element}
{}_{L_1}\langle C',R'|C,R\rangle_{L_2}= {}_{L_1}\langle C',R'|\hat{S}|C,R\rangle_{L_1}.
\end{equation}
It corresponds to the gluing of two 3-manifolds by the $T^2$ face, as described by the following topological equivalence:
\begin{equation}
    D^2 \times S^1 \cup_{T^2} S^1 \times D^2 = S^3.
\end{equation}
Since the result is $S^3$, calculating the inner product is equivalent to evaluating the topological path integral for the localized anyon propagation in 2+1D spacetime. This computation is further equivalent to performing graph calculus within the framework of UMTC, yielding:
\begin{equation}
    \vcenter{\hbox{\includegraphics{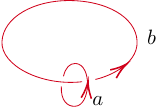}}} = \vcenter{\hbox{\includegraphics{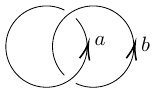}}}. 
\end{equation}
Therefore, we have
\begin{eqnarray}
{}_{L_1}\langle C',R'|\hat{S}|C,R\rangle_{L_1}&=&  \vcenter{\hbox{\includegraphics{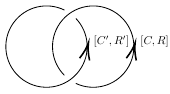}}}\nonumber \\
    &=&|G| S_{[C',R'],[C,R]}.
\end{eqnarray}
Here, $S_{[C',R'],[C,R]}$ is the $S$-matrix of the UMTC $\mathcal{Z}_1(\mathrm{Vec}_G)$, and the factor $|G|$ appears because:
\begin{eqnarray}\label{eq:ortho_CR}
&&{}_{L_1}\langle C',R'|C,R\rangle_{L_1}\nonumber\\
&&=\delta_{C,C'}\delta_{R,R'} \sum_{p\in\{p\}^{C}}\sum_{z\in Z(r_C)}\chi^{R}(z)\bar{\chi}^{R}(z)\nonumber \\
&&= \delta_{C,C'}\delta_{R,R'}|C|\sum_{z\in Z(r_C)}\chi^{R}(z)\bar{\chi}^{R}(z) \nonumber\\
&&= \delta_{C,C'}\delta_{R,R'}|C|\frac{|G|}{|C|}= \delta_{C,C'}\delta_{R,R'}|G|.
\end{eqnarray}
This gives the linear transformation between the two sets of anyon basis in the torus ground-state subspace:
\begin{equation}
    \ket{C,R}_{L_2} =\hat{S}\ket{C,R}_{L_1}=\sum_{C',R'}S_{[C',R'],[C,R]}\ket{C',R'}_{L_1}.
\end{equation}
According to correspondence between $\Strans{-}$ and $\hat{S}$ in Eq.~(\ref{S-S correspondence}), the identical linear relation holds for ribbon operators: 
\begin{equation}\label{Strans}
    \Strans{\hat{M}^{[C,R]}}=\sum_{C',R'}S_{[C',R'],[C,R]}\hat{M}^{[C',R']}.
\end{equation}
Substituting the Eq.~(\ref{Strans}) into Eq.~(\ref{eq:LoopProbeToMove}), we conclude that the probing operators and creating operators can be transformed to each other by $S$-matrix:
\begin{eqnarray}
        \hat{P}^{[C,R]} &=& \frac{\mathsf{dim}([C,R])}{|G|}\Strans{\hat{M}^{[C,R]}}\nonumber\\
        &=&\frac{\mathsf{dim}([C,R])}{|G|}\sum_{[C',R']}S_{{[C,R]},[C',R']} \hat{M}^{[C',R']},
\end{eqnarray}
and
\begin{eqnarray}\label{eq:LoopProbeToMoveSum}
\hat{M}^{[C,R]} &=& |G|\sum_{[C',R']} \bar{S}_{[C,R],[C',R']}\Strans{\hat{M}^{[C',R']}}\nonumber\\
 &=& |G|\sum_{[C',R']} \bar{S}_{[C,R],[C',R']}\frac{1}{|C'|d_{R'}}\hat{P}^{[C',R']}
\end{eqnarray}
The correctness of Eq.~(\ref{eq:LoopProbeToMoveSum}) can also be verified by the direct calculations in Appendix~\ref{apdx:Smatrix}.

As demonstrated in the analysis presented in this section, the duality between anyon-creating and anyon-probing operators induced by the $S$-transformation is closely tied to the braiding operation of anyons. 

Next, we turn our attention to analyzing the physical interpretation of the $T$-transformation. As elucidated in Eq.~(\ref{eq:ThreeTyptT}), the geometric intuition underlying the T-transformation is particularly clear:
\begin{eqnarray}
        \hat{T} \ket{a}_{L_2} &=& \ket{a}_{L_t}\\
    \vcenter{\hbox{\includegraphics{TorusKetSMS.pdf}}} &\xrightarrow{\mathrm{Twist}}&\vcenter{\hbox{\includegraphics{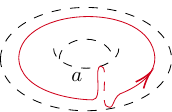}}}.
\end{eqnarray}
This operation precisely corresponds to the twist of the anyon, or equivalently, its self-statistics.

At this stage, the physical interpretation of Eq.~(\ref{eq:Def_S_hat}) and Eq.~(\ref{eq:Def_T_Trans}) becomes evident. These operations precisely correspond to the modular transformations on the torus, as elucidated through the operator-state correspondence.

The duality induced by the $S$-transformation, along with the geometric intuition of the $S$ and $T$ transformations, is expected to play a pivotal role in the construction of gapped boundaries.

\section{Boundaries of quantum double models}
\label{sec:QDbdy}
\subsection{Anyon condensation and Lagrangian algebras}
\label{subsec:AnyonCondense}
We briefly review some relevant results from category theory for constructing gapped boundaries via Lagrangian algebras. In a (2+1)D topologically ordered system with a (1+1)D gapped boundary, bulk topological excitations form a unitary modular tensor category, while boundary excitations are described by a unitary fusion category. These two categories are related through a bulk-boundary relation, as formalized in Theorem~\ref{thm:TOBdy}~\cite{kong_boundary_2017}.
\begin{theorem}[Boundary theory of 2+1D topological order]
\label{thm:TOBdy}
    A 2+1D topological order with a 1+1D gapped boundary is described by a triple $(\mathscr{C},\mathcal{B},F)$.
    \begin{enumerate}
        \item $\mathscr{C}$ is a UMTC formed by topological excitations in the bulk.
        \item $\mathcal{B}$ is a UFC formed by topological excitations on the boundary.
        \item $\mathscr{C}$ is braided equivalent to the Drinfeld center of $\mathcal{B}$: $\mathscr{C}\simeq \FZ_1(\mathcal{B})$.
        \item There exists a central functor $F:\mathscr{C}\to\mathcal{B}$, which describes the bulk-to-boundary map.
    \end{enumerate}
\end{theorem}

According to the mathematical theory of anyon condensation~\cite{kong_2014_anyoncondense}, a 1+1D gapped boundary of a 2+1D topological order $\mathscr{C}$ is uniquely determined by a Lagrangian algebra in $\mathscr{C}$ which condenses on the boundary.
\begin{definition}\label{def:LagAlg}
  A Lagrangian algebra in a UMTC $\mathscr{C}$ is an object $A$ in $\mathscr{C}$ with an associative multiplication $\mu_A: A\otimes A\to A$ such that:
  \begin{enumerate}
    \item $A$ is connected, i.e. $\mathrm{Hom}_{\mathscr{C}}(\mathbb{1},A)=\mathbb{C}$.
    \item $A$ is commutative, i.e. ${A}\otimes{A}\xrightarrow{c_{A,A}}{A}\otimes{A}\xrightarrow{\mu_A}{A}$ equals $A\otimes A\xrightarrow{\mu_A}{A}$, here $c_{{A,A}}$ is the braiding in $\mathscr{C}$.
    \item $A$ is separable, i.e. the multiplication $\mu_A$ admits a splitting $e_A:A\to A\otimes A$ as a $A$-$A$-bimodule map.
    \item $A$ is Lagrangian, i.e. the quantum dimensions of $A$ and $\mathscr{C}$ satisfy: $[\mathsf{dim}({A})]^2=\mathsf{dim}(\mathscr{C})$.
  \end{enumerate}
\end{definition}
For a $G$-quantum double model, the Lagrangian property in the last term equals to $\mathsf{dim}(A) = |G|$.


As elucidated in the mathematical framework of anyon condensation~\cite{kong_2014_anyoncondense}, the axiomatic definitions introduced here carry deep physical implications:
\begin{enumerate}
    \item The Lagrangian algebra characterizes the condensed vacuum state on the gapped boundary.
    \item The concept of connection signifies that the bulk vacuum can condense onto the boundary exclusively through a unique channel.
    \item Commutativity entails that $A$, as the boundary vacuum, exhibits both trivial braiding and twist. Specifically, the condensation process of two vacua $A \otimes A$ remains invariant regardless of whether one $A$-particle is moved around another along an arbitrary path before or after the condensation.
    \item Separability emphasizes that all internal degrees of freedom within $A$ are mutually independent and orthogonal, with no redundant or overlapping components.
    \item The Lagrangian property is intimately linked to modular invariants in the bulk phase on the torus, reflecting the $S$- and $T$-invariance of the condensed gapped boundary phase. This connection will be examined in greater detail in subsequent sections.
\end{enumerate}
In the following sections, our discussion of the Lagrangian algebra will not pay too much attention on its categorical axiomatic formulation. Instead, we will primarily focus on its intrinsic physical significance.

\begin{figure}
  \centering
  \includegraphics{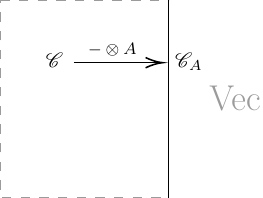}
  \caption{Anyon condensation from UMTC $\mathscr{C}$ to $\mathrm{Vec}$ through gapped boundary $\mathscr{C}_{A}$.}
  \label{fig:AnyonCondensation}
\end{figure}

\begin{theorem}[Anyon condensation in 2+1D]
\label{thm:anyoncondensation}
    Suppose a 2+1D topological order $\mathscr{C}$ condense to $\mathrm{Vec}$ through a 1+1D gapped boundary (as shown in Fig.~\ref{fig:AnyonCondensation}):
    \begin{enumerate}
      \item The vacuum particle on the gapped boundary is identified with a Lagrangian algebra $A$ in $\mathscr{C}$.
      \item The UFC that describes the excitations on the gapped boundary can be identified with $\mathscr{C}_{A}$, which denotes the category of right $A$-modules in $\mathscr{C}$.
      \item The bulk-to-boundary map is given by: $-\otimes A:\mathscr{C}\to\mathscr{C}_{A}$.
    \end{enumerate}
\end{theorem}

In particular, for any finite group $G$, the classification of Lagrangian algebras in $\FZ_1(\mathrm{Vec}_G)$ is already known~\cite{ostrik2002module}.
\begin{theorem}[Classification of Lagrangian algebras in $\FZ_1(\mathrm{Vec}_G)$]\label{thm:ClassLgrgAlg}
    For a finite group $G$, each Lagrangian algebra in $\FZ_1(\mathrm{Vec}_G)$ corresponds to a pair $(H,\omega)$. $H$ is a subgroup of $G$, up to conjugation. $\omega\in H^{2}(H,\mathbb{C}^{\times})$ where $H^{2}(H,\mathbb{C}^{\times})$ is the 2-cohomology group of $H$, and $\mathbb{C}^{\times}$ is the set of complex number without zero.
\end{theorem}

\subsection{Zig-Zag boundary of honeycomb lattice}
\label{subsec:zigzag}
\begin{figure}
    \centering
    \begin{subfigure}[t]{\columnwidth}
        \centering
        \includegraphics{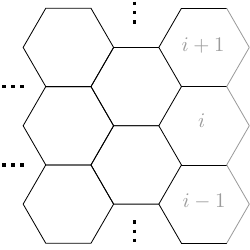}
        \caption{The honeycomb lattice exhibits a zig-zag boundary (gray color) and extends infinitely in all directions except to the right. Boundary sites are labeled sequentially by $i$, $i+1$, and so on, indicating their positions on the boundary.}
        \label{fig:boundedhoneycomb}
    \end{subfigure}
    
    \vspace{1em} 

    \begin{subfigure}[t]{\columnwidth}
            \centering
            \includegraphics{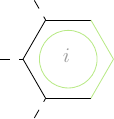}
            \caption{A boundary site on the zig-zag boundary. The illustrated site comprises the green plaquette and the green vertex. The green vertex consists of two edges, in contrast to the vertices in the bulk, which are intersections of three edges.}
            \label{fig:BdySite} 
    \end{subfigure}    
    \caption{The zig-zag bounded honeycomb lattice configuration.}
\end{figure}

We consider a honeycomb lattice with a zig-zag boundary, as depicted in Fig.~\ref{fig:boundedhoneycomb}. Each edge on the boundary hosts a local Hilbert space $\mathcal{H}_{\mathrm{loc}} = \mathrm{span}\{\ket{g}\}_{g \in G}$, identical to that of the bulk edges. A boundary site is defined as a composite of a vertex and a plaquette, arranged explicitly as shown in Fig.~\ref{fig:BdySite}. These boundary sites are constructed to be mutually disjoint and independent, ensuring that the position of a condensed anyon can be specified unambiguously. This structural clarity is one of the key advantages of employing a zigzag boundary in the lattice geometry.


\begin{figure}
    \centering
    \includegraphics{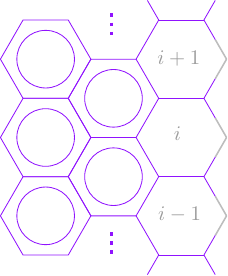}
    \caption{The bounded honeycomb lattice with the effective Hilbert space $\mathcal{H}^{\mathrm{ZigZag}}_{\mathrm{bdy}}$ after bulk interactions are introduced.}
    \label{fig:BulkFixHilbert}
\end{figure}
We introduce bulk vertex and plaquette operators on the purple vertices and plaquettes in Fig.~\ref{fig:BulkFixHilbert}. The bulk Hamiltonian is defined as:
\begin{equation}
    H_{\mathrm{bulk}} = -\sum_{\mathrm{bulk\ vertices}}\hat{V}_{\alpha}-\sum_{\mathrm{bulk\ plaquettes}}\hat{P}_{\beta}.
\end{equation}
Since the DOFs on the boundary are not fully constrained, the ground-state subspace of this Hamiltonian is highly degenerate. We denote the ground-state subspace of $H_{\mathrm{bulk}}$ as $\mathcal{H}^{\mathrm{ZigZag}}_{\mathrm{bdy}}$. 

\subsection{Anyon probing and creating on the boundary}
\label{subsec:bdyProbAndCreat}
\begin{figure}
  \centering
  \includegraphics{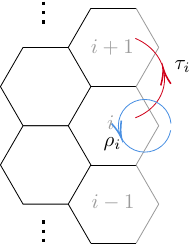}
  \caption{The minimal paths denoted by ${\rho_i}$ and ${\tau_i}$ are located on the boundary of the honeycomb lattice.}
  \label{fig:BdyPath}
\end{figure}

To lift the large degeneracy in $\mathcal{H}^{\mathrm{ZigZag}}_{\mathrm{bdy}}$ and obtain a gapped ground state, we need to introduce boundary interactions into the Hamiltonian. Similar to the bulk terms in Fig.~\ref{fig:MinimalLoop}, the boundary terms are supposed to be ribbon operators defined on the minimal paths on the boundary as illustrated in Fig.~\ref{fig:BdyPath}.

On the effective Hilbert space $\mathcal{H}^{\mathrm{ZigZag}}_{\mathrm{bdy}}$, the action of a ribbon operator on $\rho_i$ is defined as:
\begin{eqnarray}
&&\Ribbon^{k,g}(\rho_i) \left[\vcenter{\hbox{\includegraphics{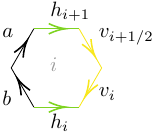}}} \right]\nonumber\\
&&= \left[\vcenter{\hbox{\includegraphics{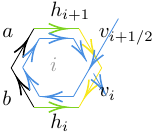}}} \right]\nonumber\\
&&= \delta_{k,v_{i}^{-1}v_{i+\frac{1}{2}}}\left[\vcenter{\hbox{\includegraphics{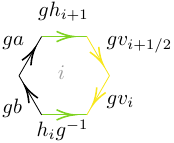}}} \right].\label{eq:BdyRibbon1}
\end{eqnarray}
And the action of a ribbon operator on $\tau_i$ is defined as:
\begin{eqnarray}
&&\Ribbon^{k,g}(\tau_i) \left[\vcenter{\hbox{\includegraphics{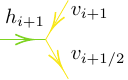}}} \right]\nonumber\\
&&= \left[\vcenter{\hbox{\includegraphics{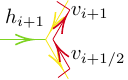}}} \right]\nonumber\\
&&= \delta_{k,v_{i+1}v_{i+1/2}^{-1}}\left[\vcenter{\hbox{\includegraphics{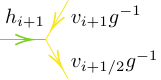}}} \right]\nonumber\\
&&= \delta_{k,h_{i+1}^{-1}}\left[\vcenter{\hbox{\includegraphics{BdyRibbonState6.pdf}}} \right].\label{eq:BdyRibbon2}
\end{eqnarray}
The third equality uses the vertex conservation law $v_{i+1}v_{i+1/2}^{-1}=h_{i+1}^{-1}$, which is satisfied within $\mathcal{H}^{\mathrm{ZigZag}}_{\mathrm{bdy}}$.

It is evident that all these boundary ribbon operators commute with both the vertex and plaquette operators in the bulk. A detailed analysis of the commutation relations for boundary terms will be presented in \S\ref{subsec:GroundDynamics}.

Consider a Lagrangian algebra $A$ in $\FZ_1(\mathrm{Vec}_G)$. In general, $A$ is a direct sum of some simple objects in $\FZ_1(\mathrm{Vec}_G)$:
\begin{equation}\label{eq:Asummands}
  A = \bigoplus_{a} K_a a,
\end{equation}
where $a$ is a simple object in $\FZ_1(\mathrm{Vec}_G)$ and $K_a\in\mathbb{N}$ is the summation coefficient. 

Recalling Theorem~\ref{thm:anyoncondensation}, at the categorical level, a bulk simple anyon $b$ transfers into a boundary topological excitation via the bulk-to-boundary map. It is important to note that the resulting boundary excitation is not necessarily simple:
\begin{equation}
    \mathrm{Irr}(\mathscr{C})\ni b \longrightarrow B\oplus C\oplus D\oplus\cdots\in \mathrm{Obj}(\mathscr{C}_A)
\end{equation}
At the lattice model level, this implies that the process by which anyons in the bulk are transported to and captured by the boundary is, in principle, non-unique, allowing for multiple distinct channels:
\begin{equation*}
    \includegraphics{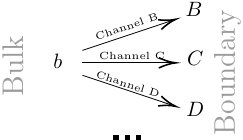}
\end{equation*}
In the context of anyon condensation, this phenomenon is described by the concepts of `partial condensation'~\cite{SlingerlandAnyonCond} and `anyon splitting'~\cite{Neupert_Bosoncondensation}. When anyons propagate to the boundary via distinct channels, they manifest a variety of behaviors, giving rise to different types of boundary topological excitations. We denote the specific channels through which a bulk anyon transforms into a trivial boundary excitation as \emph{condensable channels}.

A natural interpretation of the phenomenon of anyon splitting is that it reflects the presence of a hidden internal Hilbert space associated with simple anyons in the bulk~\cite{SlingerlandAnyonCond}. Within the bulk, different internal DOFs of the same simple anyon can be interconverted by local operators, rendering them indistinguishable at the macroscopic level. Nevertheless, during the process of anyon condensation, these internal DOFs can be differentiated based on their distinct condensation behaviors.

This relationship between internal DOFs and anyon splitting is especially evident in Kitaev's quantum double models. In these models, as shown in section\ref{sec:CreatAndProb}, the internal DOFs of bulk anyons are explicitly represented by orthogonal quantum states in the Hilbert space. Each bulk-to-boundary channel is associated with an internal DOF of the anyon, and {condensable channels} naturally yield \emph{condensable internal DOFs}. The coefficient $K_a$ in Eq.~(\ref{eq:Asummands}) represents the number of condensable channels associated with the anyon $a$. As a result, it is crucial to clearly distinguish between the semi-simple anyon $A$ in the bulk and the trivial excitation $\mathbb{1}_{\mathscr{C}_A}$ on the boundary. The latter corresponds to only a subset of the internal DOFs associated with the condensation. We will provide further elaboration on these concepts in \S\ref{subsec:GroundDynamics} through an explicit bulk-to-boundary operator, supplemented by physical intuition.

\begin{remark}
It is necessary to explore how to formalize the concepts of internal DOF of anyons and condensable internal DOFs within a general theoretical framework. In conventional string-net models, non-trivial topological excitations can only be defined with the assistance of auxiliary spaces. To accommodate all topological excitations within a unified lattice Hilbert space, it is necessary to introduce a functor $\omega: \mathcal{G} \to \text{Vec}$~\cite{buerschaperEM2013}, which in general is equipped with a weak separable Frobenius structure~\cite{PFEIFFER20093714}.

When constructing lattice models that realize topological order theories described by category theory, additional structures akin to $\omega$ are invariably required. In general, one topological excitation may correspond to multiple distinct quantum states, which can be locally converted into each other. In a model-dependent manner, this procedure leads to the concept of internal DOFs of anyons. When combined with the theory of anyon condensation, it furnish a canonical basis for understanding condensable internal DOFs and anyon splitting.
\end{remark}

\begin{figure}
    \centering
    \includegraphics{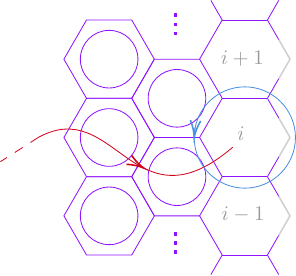}
    \caption{Anyon-probing process on the boundary. The blue circle $\mathrm{Loop}^{\mathrm{bdy}}_i$ surrounds a boundary site $i$. The red path $\mathrm{Path}_{i,\infty}$ ends at site $i$, while its starting site is infinitely far away.}
    \label{fig:BdyProbe}
\end{figure}

Let $k$ be a condensable internal DOF of $a$. The corresponding probing operator is $\hat{P}^{a}_{k}$. This operator is well-defined on the boundary loop $\mathrm{Loop}^{\mathrm{bdy}}_i$, which is the blue circle in Fig.~\ref{fig:BdyProbe}.

As a ribbon operator, $\hat{P}^{a}_{k}(\mathrm{Loop}^{\mathrm{bdy}}_i)$ commutes with $H_{\mathrm{bulk}}$. Thus, the subspace $\mathcal{H}^{\mathrm{ZigZag}}_{\mathrm{bdy}}$ can be partitioned into two parts based on the eigenvalues of $\hat{P}^{a}_{k}(\mathrm{Loop}^{\mathrm{bdy}}_i)$. Denoting the eigenstate of $\hat{P}^{a}_{k}(\mathrm{Loop}^{\mathrm{bdy}}_i)$ with eigenvalue $0$ in $\mathcal{H}^{\mathrm{ZigZag}}_{\mathrm{bdy}}$ as $\ket{i;a,k;0}$, we have:
\begin{equation}
    \hat{P}^{a}_{k}(\mathrm{Loop}^{\mathrm{bdy}}_i)\ket{i;a,k;0} = 0.
\end{equation}
Then, we consider the anyon-creating operator $\hat{M}^{a}_{k,-}(\mathrm{Path}_{i,\infty})$ defined on the red path $\mathrm{Path}_{i,\infty}$ in Fig.~\ref{fig:BdyProbe}. This operator creates an $a$ excitation with internal DOF $k$ at site $i$, and form the eigenstate of $\hat{P}^{a}_{k}(\mathrm{Loop}^{\mathrm{bdy}}_i)$ with eigenvalue $1$, which is denoted as $\ket{i;a,k;1}$:
\begin{eqnarray}
&&\hat{P}^{a}_{k}(\mathrm{Loop}^{\mathrm{bdy}}_i)\hat{M}^{a}_{k,-}(\mathrm{Path}_{i,\infty})\ket{i;a,k;0}\nonumber\\
&&= \hat{M}^{a}_{k,-}(\mathrm{Path}_{i,\infty})\ket{i;a,k;0} \equiv \ket{i;a,k;1}.\label{eq:BdylocalExc}
\end{eqnarray}

By introducing the term ``$-\hat{P}^{a}_{k}(\mathrm{Loop}^{\mathrm{bdy}}_i)$'' into the Hamiltonian, the local boundary state $\ket{i;a,k;1}$ becomes a ground state, whereas a similar state in the bulk remains an excited state. This observation is consistent with the macroscopic framework of anyon condensation theory, which predicts that certain bulk excitations condense into the ground state on the boundary.

\begin{figure}
  \centering
  \includegraphics{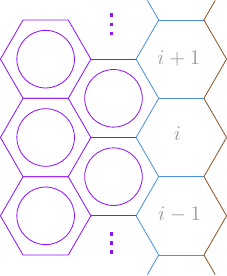}
  \caption{The bounded honeycomb lattice with the effective Hilbert space $\mathcal{H}^{A\text{-}\mathrm{Conf}}_{\mathrm{bdy}}$ after bulk interactions and boundary probing operators are introduced.}
  \label{fig:BdyConfineHilbert}
\end{figure}

The loop $\mathrm{Loop}^{\mathrm{bdy}}_i$ encircles multiple sites. In practice, it is sufficient to introduce probing operators acting $\rho_i$, which detect only local excitations at the boundary site $i$. To realize the gapped boundary corresponding to $A$ condensation, analogous to the bulk terms discussed in the probing picture in \S\ref{subsec:twopictures}, the boundary terms we introduce should include all probing operators that can detect every condensable internal DOFs within $A$. We can formally write the Hamiltonian with all boundary probing operators as:
\begin{gather}
    \hat{P}^{\mathbb{1}_{\mathscr{C}_A}} \equiv \sum_{a} \sum_{\substack{\mathrm{Condensable}\\ \mathrm{Internal\ DOF\ } k}} \hat{P}^{a}_k,\\
  H^{A\text{-}\mathrm{Conf}}_{\mathrm{bdy}} = H_{\mathrm{bulk}} - \sum_{i} \hat{P}^{\mathbb{1}_{\mathscr{C}_A}}(\rho_i).
\end{gather}
The superscript ``$A\text{-}\mathrm{Conf}$'' stands for ``$A\text{-}\mathrm{Confined}$'', the meaning of which will be explained later. We denote the ground subspace of $ H^{A\text{-}\mathrm{Conf}}_{\mathrm{bdy}}$ as $\mathcal{H}^{A\text{-}\mathrm{Conf}}_{\mathrm{bdy}}$, which is illustrated in Fig.~\ref{fig:BdyConfineHilbert}.

\begin{figure}
  \centering
  \includegraphics{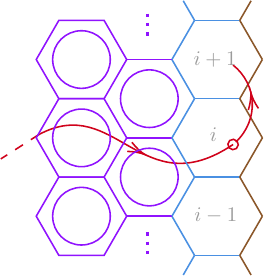}
  \caption{A condensable anyon from the bulk is moved to and confined at a specific boundary site $i$. A stabilizer that moves anyons between adjacent sites can deconfine the condensed anyon.}
  \label{fig:BdyDeconfine}
\end{figure}

While the inclusion of boundary probing terms facilitates anyon condensation at individual boundary sites, $\mathcal{H}^{A\text{-}\mathrm{Conf}}_{\mathrm{bdy}}$ remains highly degenerate. At every site on the boundary, analogous to Eq.~\ref{eq:BdylocalExc}, each condensable internal DOF of $A$ corresponds to a degenerate ground state. This results in a ground-state degeneracy that grows exponentially with the boundary length. The origin of this degeneracy is rather intuitive: when a condensable bulk anyon is moved to a specific boundary site $i$, it becomes confined to that site, unable to move freely along the boundary. This confinement prevents anyons at different sites from fusing or annihilating, leading to an undesired ground state degeneracy. To resolve this, it suffices to introduce boundary terms that move anyons between neighboring sites. These terms naturally correspond to the anyon-creating operators acting on $\tau_i$ as shown in Fig.~\ref{fig:BdyDeconfine}.

To deconfine all condensable internal DOFs, we introduce a complete $A$-creating operator on each minimal path $\tau_i$ as a boundary term. In accordance with the duality established in Section~\ref{sec:ModularTrans}, the $A$-creating operator is identified as the $S$-transformation of the $A$-probing operator. This construction leads to a formal Hamiltonian that describes the $A$-condensed gapped boundary:
\begin{gather}
      \hat{M}^{\mathbb{1}_{\mathscr{C}_A}} = \mathbf{S}^{-1}\left[\hat{P}^{\mathbb{1}_{\mathscr{C}_A}}\right] = \sum_{a} \sum_{\substack{\mathrm{Condensable}\\ \mathrm{Internal\ DOF\ } k}}\frac{|G|}{\mathsf{dim}(a)}{\hat{M}^{a}_{k,k}}\\
      H^{A}_{\mathrm{bdy}} =H_{\mathrm{bulk}} - \sum_{i}\hat{P}^{\mathbb{1}_{\mathscr{C}_A}}(\rho_i)-\sum_i\hat{M}^{\mathbb{1}_{\mathscr{C}_A}}(\tau_i) .\label{eq:FormalBdyHamiltonian}
\end{gather}
Here, we apply Eq.~(\ref{eq:LoopProbeToMove}) in a slightly generalized sense and observe that the $S$-transformation is linear, with the coefficient $\mathsf{dim}(A)/|G| = 1$.

\subsection{From Lagrangian algebras to consistency conditions}
\label{subsec:DefCondition}
Although Eq.~(\ref{eq:FormalBdyHamiltonian}) provides a formal expression for the Hamiltonian, a systematic method for identifying the condensable internal DOFs remains to be established. This task reduces to deriving a set of constraint equations that the boundary ribbon operators must satisfy, based on the properties of Lagrangian algebras. Solutions to these constraints determine the specific sets of condensable internal DOFs and, in turn, dictate the appropriate boundary terms. Each solution corresponds to a distinct type of gapped boundary.


We introduce the notation $\hat{A}$ as a shorthand for $\hat{P}^{\mathbb{1}_{\mathscr{C}_A}}$. Recall that:
\begin{equation}\label{eq:ProbeBdyterm}
  \hat{A} = \hat{P}^{\mathbb{1}_{\mathscr{C}_A}}= \sum_{a} \sum_{\substack{\mathrm{Condensable}\\ \mathrm{Internal\ DOF\ } k}} \hat{P}^{a}_k.
\end{equation}
The operator $\hat{A}$ must satisfy specific conditions imposed by the definition of the Lagrangian algebra.

First, the separability property of the Lagrangian algebra fundamentally reflects the stability of the boundary ground state, which implies that the condensable DOFs are mutually orthogonal. At the operator level, this implies that $\hat{A}$, being a sum of mutually orthogonal projectors, is itself a projector, which can probe the trivial boundary excitation:
\begin{equation}\label{eq:AIsProjector}
  \hat{A}^2 = \hat{A}.
\end{equation}



The connectivity property of $A$ indicates that, in its direct summation representation, the coefficient $K_{\mathbb{1}}$ corresponding to the trivial excitation equals $1$. This feature is reflected in the behavior of $\hat{A}$ when multiplied by the trivial anyon probing operator, identified as the omega loop operator $\hat{P}^{\mathbb{1}} = \hat{\Omega}$, as detailed in Eq.~(\ref{eq:trvProbingOmega}). Specifically, the following relations are satisfied:
\begin{align}
    \hat{P}^{a}_k \hat{\Omega} &= \delta_{a,\mathbb{1}} \hat{\Omega}, \\
    \hat{A} \hat{\Omega} &= \hat{\Omega}. \label{eq:connectionProperty}
\end{align}

Finally, the commutative property and Lagrangian property give that $\hat{A}$ is $S$ and $T$ invariant:
\begin{equation}
\label{eq:ModularInvariant}
  \mathbf{S}[\hat{A}] = \hat{A},\quad \mathbf{T}[\hat{A}] = \hat{A}.
\end{equation}
These relations can be obtained by graph calculus method as shown below.

Via the operator-state correspondence, the torus states corresponding to the ribbon operators $\hat{A}$ and $\mathbf{S}^{-1}[\hat{A}]$ are, respectively:
\begin{eqnarray}
\ket{A} &=&  \sum_{a} \sum_{\substack{\mathrm{Condensable}\\ \mathrm{Internal\ DOF\ } k}} \ket{P^{a}_{k}}\nonumber\\
&=&  \sum_{a} \sum_{\substack{\mathrm{Condensable}\\ \mathrm{Internal\ DOF\ } k}} \frac{\mathsf{dim}(a)}{|G|}\hat{S}\ket{{M}^{a}_{k,k}}\nonumber\\
&=&  \sum_{a} \sum_{\substack{\mathrm{Condensable}\\ \mathrm{Internal\ DOF\ } k}} \frac{\mathsf{dim}(a)}{|G|}\ket{a;k,k}_{L_2},
\end{eqnarray}
\begin{eqnarray}
      \ket{\mathbf{S}^{-1}[A]} &=&  \sum_{a} \sum_{\substack{\mathrm{Condensable}\\ \mathrm{Internal\ DOF\ } k}} \frac{\mathsf{dim}(a)}{|G|}\ket{{M}^{a}_{k,k}}\nonumber\\
  &=&  \sum_{a} \sum_{\substack{\mathrm{Condensable}\\ \mathrm{Internal\ DOF\ } k}} \frac{\mathsf{dim}(a)}{|G|}\ket{a;k,k}_{L_1}.
\end{eqnarray}
It is important to note that the two quantum states presented here are normalized follows from Eq.~(\ref{eq:TorusNormal1}), Eq.~(\ref{eq:TorusNormal2}) and the Lagrangian property:
\begin{equation}
    \bra{A}\ket{A} = \bra{\mathbf{S}^{-1}[A]}\ket{\mathbf{S}^{-1}[A]} = 1.
\end{equation}

Similar to the process in \S\ref{subsec:GraphCalculus}, we illustrate torus states as:
\begin{equation}
  \ket{A} = \vcenter{\hbox{\includegraphics{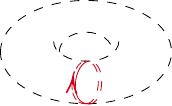}}},
\end{equation}
\begin{equation}
  \ket{\mathbf{S}^{-1}[A]}= \vcenter{\hbox{\includegraphics{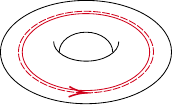}}}.
\end{equation}
Here we employ dashed-solid double lines to emphasize that $\hat{A}$ is the vacuum of the gapped boundary. Thus, we can calculate the inner product using graph calculus:
\begin{equation}
  \bra{\mathbf{S}^{-1}[A]}\ket{A} = \vcenter{\hbox{\includegraphics{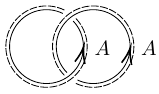}}}.
\end{equation}

Since $A$, depicted here as a dashed-solid double line, does not correspond to a pure superposition of bulk anyons, graphical calculus in this context actually constitutes an extension of that in UMTC. In addition to inheriting the graph calculus rules from $\FZ_1(\text{Vec}_G)$, we should also take into account the properties of $A$ as the vacuum of the gapped boundary. Within the framework of this extended graph calculus, the commutative property of the Lagrangian algebra can be understood as follows: the braiding operation performed on $A$ must be trivial. This leads to the following equality in graph calculus :
\begin{equation}\label{eq:Commutativity}
  \includegraphics{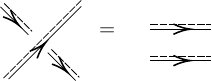}.
\end{equation}
Therefore, we have
\begin{equation}
\label{eq:doubleALoop}
  \vcenter{\hbox{\includegraphics{DashSmatrix.pdf}}} = \vcenter{\hbox{\includegraphics{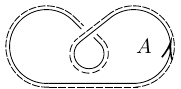}}} = \vcenter{\hbox{\includegraphics{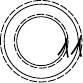}}}.
\end{equation}
In the glued $S^3$ spacetime, the $A$-loop in Eq.~(\ref{eq:doubleALoop}) should be interpreted as a summation of the propagation paths of each component of $A$ in the 2+1D spacetime.
\begin{equation}
\label{eq:Aloop}
  \vcenter{\hbox{\includegraphics{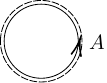}}} = \sum_{a}\sum_{\substack{\text{Condensable}\\ \text{Internal DOF } k}}\frac{\mathsf{dim}(a)}{|G|}\vcenter{\hbox{\includegraphics{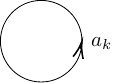}}}.
\end{equation}
As discussed in Refs.~\cite{delcamp_fusion_2017,dittrich2017quantum}, the value of the $a_k$-loop corresponds to the expectation value of the associated anyon-creating operator acting on the ground state of the quantum double model defined on the plane:
\begin{equation}
\label{eq:akLoopValue}
  \vcenter{\hbox{\includegraphics{akLoop.pdf}}} = \bra{\Omega}\hat{M}^{a}_{k,k}\ket{\Omega} = 1.
\end{equation}
The second equality follows from Eq.~(\ref{eq:LoopMovingComponent}), which allows the value of the A-loop to be computed as follows:
\begin{equation}\label{eq:LagrangianUnit1}
\begin{aligned}
      \vcenter{\hbox{\includegraphics{Dashloop.pdf}}} 
      &= \sum_{a} \sum_{\substack{\text{Condensable} \\ \text{Internal DOF } k}} \frac{\mathsf{dim}(a)}{|G|} \bra{\Omega} \hat{M}^{a}_{k,k} \ket{\Omega} \\
      &= \frac{\sum_a K_a \mathsf{dim}(a)}{|G|} = \frac{\mathsf{dim}(A)}{|G|} = 1.
\end{aligned}
\end{equation}
The last equality is precisely the Lagrangian property. Thus, the value of the double A-loop in Eq.~(\ref{eq:doubleALoop}) is given by $1 \times 1 = 1$. Consequently, we arrive at:
\begin{equation}
\bra{\mathbf{S}^{-1}[A]}\ket{A} = 1.
\end{equation}
Since both states are normalized, the inner product equals $1$ implies that the two states are identical: $\ket{A} = \ket{\mathbf{S}^{-1}[A]}$. By invoking the state-operator correspondence, this result leads to a self-duality relation for the operator $\hat{A}$:
\begin{equation}
\hat{A} = \mathbf{S}[\hat{A}].
\end{equation}

The self-duality of the operator $\hat{A}$ reveals its dual identity: it acts both as a boundary $A$-creating operator and as a boundary $A$-probing operator:
\begin{eqnarray}
    \label{eq:moveBdyterm}
   &&\hat{A} = \hat{P}^{\mathbb{1}_{\mathscr{C}_A}} = \mathbf{S}^{-1}[\hat{A}] = \hat{M}^{\mathbb{1}_{\mathscr{C}_A}}\nonumber\\
   &&= \sum_{a} \sum_{\substack{\mathrm{Condensable}\\ \mathrm{Internal\ DOF\ } k}}\frac{|G|}{\mathsf{dim}(a)}{\hat{M}^{a}_{k,k}}
\end{eqnarray}

In the form of the anyon-creating operator, the consistency conditions given by Eq.~(\ref{eq:AIsProjector}) and Eq.~(\ref{eq:connectionProperty}) can be reinterpreted as follows.

The Eq.~(\ref{eq:AIsProjector}) reflects that the fusion rule $A \otimes_A A = A$ in the UFC $\FZ_1(\mathrm{Vec}_G)_A$. A projector must be self-adjoint, satisfying $\hat{A}^{\dagger} = \hat{A}$, which reflects the fact that the dual excitation of $A$ is itself. Meanwhile, the connection property in Eq.~(\ref{eq:connectionProperty}) and the graphic calculus in Eq.~(\ref{eq:LagrangianUnit1}) are mutually corroborating, providing complementary perspectives on the underlying structure.

On the other hand, the analysis of the $T$-invariant proceeds in a manner analogous to the preceding discussion. Specifically, we consider the following expression for the torus state:
\begin{equation}
    \ket{\mathbf{T}[A]} = \hat{T}\ket{A} = \vcenter{\hbox{\includegraphics{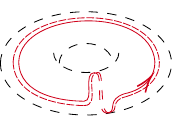}}}.
\end{equation}
The associated inner product is then computed as follows:
\begin{eqnarray}
    \bra{\mathbf{T}[A]}\ket{\mathbf{S}^{-1}[A]}
    &=& \vcenter{\hbox{\includegraphics{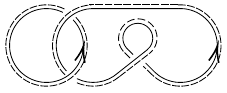}}} \nonumber\\
    &=& \vcenter{\hbox{\includegraphics{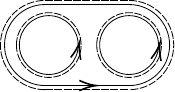}}} = 1.
\end{eqnarray}
This result leads to the conclusion that:
\begin{equation}
    \mathbf{T}[\hat{A}] = \mathbf{S}^{-1}[\hat{A}] = \hat{A}.
\end{equation}

\begin{remark}
    Although the Lagrangian condition in Definition~\ref{def:LagAlg} may seem to be brief, Ref.~\cite{kong2009cardy} establishes that a simple commutative separable algebra in a modular tensor category is modular invariant if and only if it is Lagrangian. Indeed, in our preceding computations, the $S$ and $T$ invariance of $\hat{A}$ emerges not solely from the commutativity condition in Eq.~(\ref{eq:Commutativity}), but also depends critically on the Lagrangian property in Eq.~(\ref{eq:LagrangianUnit1}).
\end{remark}

We summarize the consistency conditions as follows:
\begin{theoremph}
\label{thmph:bdyTerms}
    Consider a Kitaev's quantum double model with a finite group $G$ as the input data. The point-like topological excitations in this model form a unitary modular tensor category $\FZ_1(\mathrm{Vec}_G)$. A Lagrangian algebra $A$ in $\FZ_1(\mathrm{Vec}_G)$ corresponds uniquely, up to a unitary transformation, to a ribbon operator $\hat{A}$ that satisfies the following consistency conditions:
  \begin{center}
  \begin{tabular}{c@{\hspace{1cm}}c}
      1. $\hat{A}^2 = \hat{A}$ & 2. $\hat{A}\hat{\Omega} = \hat{\Omega}$\\
      3. $\mathbf{S}[\hat{A}] = \hat{A}$ & 4. $\mathbf{T}[\hat{A}] = \hat{A}$
  \end{tabular}
  \end{center}
  


The boundary interacting terms that realize the $A$-condensed gapped boundary are:
\begin{equation*}
  \left\{ \hat{A}(\rho_i),\hat{A}(\tau_i)\right\}_i.
\end{equation*}
The total Hamiltonian is:
\begin{equation*}
  H^{A}_{\mathrm{bdy}} = H_{\mathrm{bulk}} - \sum_i \hat{A}(\rho_i) - \sum_i \hat{A}(\tau_i).
\end{equation*}

\end{theoremph}

\subsection{From consistency conditions to Lagrangian algebras}\label{subsec:ConsructiveExtence}
In \S\ref{subsec:DefCondition}, we derived the consistency conditions that boundary terms must satisfy, grounded in the categorical axiomatic definition of Lagrangian algebras. In this section, we constructively present two families of solutions that encompass a wide range of possible solutions to these conditions. These constructively existent solutions not only highlight the self-consistency of our framework but also serve to demonstrate the sufficiency of Theorem~\ref{thmph:bdyTerms}.

Referring back to Theorem~\ref{thm:ClassLgrgAlg}, Lagrangian algebras in Kitaev's quantum double models are intrinsically linked to subgroups and their associated 2-cohomology groups. This connection serves as the foundation for identifying specific solutions. One such family of solutions is relatively straightforward to uncover:
\begin{theoremph}[Spontaneous Symmetry Breaking Boundary Terms]\label{thm:SSBterms}
    For a Kitaev's quantum double model with a finite group $G$ serving as the input data, the following ribbon operator satisfies the consistency constraints for boundary terms:
    \begin{equation}\label{eq:SSBterms}
        \hat{A}_{H} = \frac{1}{|H|} \sum_{h_1, h_2 \in H} \Ribbon^{h_1, h_2},
    \end{equation}
    where $H \subseteq G$ denotes a subgroup.
\end{theoremph}

The four consistency conditions stated in Theorem~\ref{thmph:bdyTerms} can be directly verified for these operators. Notably, if two subgroups differ only by a conjugation transformation, i.e., $H_1 = g H_2 g^{-1}$, then the corresponding operators $\hat{A}_{H_1}$ and $\hat{A}_{H_2}$ are related via a unitary transformation:
\begin{equation}
    \hat{A}_{H_1} =  \hat{Z}^{g} \hat{C}^{g} \hat{A}_{H_2} \hat{C}^{g^{-1}} \hat{Z}^{g^{-1}},
\end{equation}
where $\hat{C}^{g}$ is defined as follows:
\begin{equation}\label{ConjugateRibbon}
\begin{array}{c}
    \hat{C}^{g}(\mathrm{path}) = \text{Acting }g\text{ conjugately on orthogonal edges}.\\[10pt]
    \includegraphics{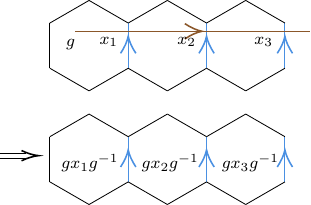}
\end{array}
\end{equation}

The term `Spontaneous Symmetry Breaking (SSB) Boundary Terms' derives its name from the underlying subgroup structure, which will be further shown in \S\ref{sec:examples}. These SSB terms account for a range of  scenarios in the classification of Lagrangian algebras corresponding to the identity element of the 2-cohomology group.

In light of the classification results of Lagrangian algebras and the validity of the ansatz Eq.~(\ref{eq:SSBterms}), We propose a reasonable conjecture that more general solutions will assume the following form:
\begin{equation}
    \hat{A}_{H}^{\mu} = \frac{1}{|H|} \sum_{h_1, h_2 \in H} f^{\mu}(h_1, h_2) \Ribbon^{h_1, h_2}.
\end{equation}
The coefficients $f^{\mu}(h_1, h_2)$ are expected to correspond to a 2-cocycle $\mu \in Z^2(H, \mathbb{C}^\times)$. Here, we work with $U(1)$ phases in 2-cocycle rather than group elements in the 2-cohomology, up to equivalence under modifications by a 1-coboundary.

To determine constraints on $f^{\mu}(h_1, h_2)$, we substitute this ansatz with undetermined coefficients into the consistency condition given by Eq.~(\ref{eq:AIsProjector}):
\begin{eqnarray}
    &&\hat{A}_{H}^{\mu} \hat{A}_{H}^{\mu}\nonumber\\
    &&= \frac{1}{|H|^2} \sum_{h_1, h_2 \in H} \sum_{h_3, h_4 \in H} f^{\mu}(h_1, h_2) f^{\mu}(h_3, h_4) \Ribbon^{h_1, h_2} \Ribbon^{h_3, h_4} \nonumber \\
    &&= \frac{1}{|H|^2} \sum_{h_1, h_2 \in H} \sum_{h_4 \in H} f^{\mu}(h_1, h_2) f^{\mu}(h_1, h_4) \Ribbon^{h_1, h_2 h_4} \nonumber \\
    &&= \frac{1}{|H|^2} \sum_{h_1, h \in H} \left( \sum_{h_2 \in H} f^{\mu}(h_1, h_2) f^{\mu}(h_1, h_2^{-1} h) \right) \Ribbon^{h_1, h}. \nonumber\\
    \label{eq:IntermediateStep}
\end{eqnarray}

For the idempotency condition $\hat{A}_{H}^{\mu} \hat{A}_{H}^{\mu} = \hat{A}_{H}^{\mu}$ to hold, it is necessary that:
\begin{equation}\label{eq:WeakBicharacter}
    \sum_{h_2 \in H} f^{\mu}(h_1, h_2) f^{\mu}(h_1, h_2^{-1} h) = |H| f^{\mu}(h_1, h).
\end{equation}

To identify specific solutions, we impose a stronger condition:
\begin{equation}\label{eq:StrongBicharacter1}
    f^{\mu}(h_1, h_2) f^{\mu}(h_1, h_3) = f^{\mu}(h_1, h_2h_3),\ \forall h_i \in H.
\end{equation}
It is straightforward to verify that Eq.~(\ref{eq:StrongBicharacter1}) implies Eq.~(\ref{eq:WeakBicharacter}). From the perspective of the representation theory of group-theoretical associative algebras, these two conditions are equivalent when the group under consideration is Abelian~\cite{BantayAlgebraicAspectsOrbifold1994}.

Next, we examine the consistency condition in Eq.~(\ref{eq:ModularInvariant}), which imposes the following requirement:
\begin{equation}\label{eq:Strans_of_fab}
    f^{\mu}(h_1, h_2) = S[f^{\mu}(h_1, h_2)] \equiv f^{\mu}(h_2, h_1^{-1}).
\end{equation}
The combination of Eq.~(\ref{eq:StrongBicharacter1}) with Eq.~(\ref{eq:Strans_of_fab}) naturally leads to:
\begin{equation}\label{eq:StrongBicharacter2}
    f^{\mu}(h_2, h_1) f^{\mu}(h_3, h_1) = f^{\mu}(h_2h_3,h_1),\ \forall h_1 \in H.
\end{equation}

The conditions expressed in Eq.~(\ref{eq:StrongBicharacter1}) and Eq.~(\ref{eq:StrongBicharacter2}) mathematically defines a bicharacter over the group $H$.

\begin{definition}[Alternating Bicharacters]\label{def:Bicharacters}
    Let $Q$ be an abelian group. An alternating bicharacter on $Q$ is a bivariate mapping 
    \[
    B: Q \times Q \to \mathbb{C}^\times,
    \]
    satisfying the following properties:
    \begin{subequations}
    \begin{align}
        &B(q_1 q_2, q) = B(q_1, q) B(q_2, q), \\
        &B(q, q_1 q_2) = B(q, q_1) B(q, q_2), \\
        &B(q, q) = 1, \quad \forall q, q_1, q_2 \in Q.
    \end{align}
    \end{subequations}
    The set of all alternating bicharacters forms a group, denoted by $\Lambda^2(Q)$.
\end{definition}

Interestingly, at least for abelian groups, there is a well-known rigorous correspondence between alternating bicharacters and 2-cohomology classes, as described in Proposition 2.6 of Ref.~\cite{TambaraRepTensorCat2000} (see also subsection 2.4 of Ref.~\cite{Naidu_2008}). This correspondence is encapsulated in the following theorem.

\begin{theorem}[Isomorphism Between Bicharacters and 2-Cohomology]\label{thm:BiCoIsomorphism}
    For an abelian group $Q$, the 2-cohomology group $H^2(Q, \mathbb{C}^\times)$ is isomorphic to the group of alternating bicharacters $\Lambda^2(Q)$. The isomorphism is defined as:
    \begin{align}
        \operatorname{alt}: H^2(Q, \mathbb{C}^\times) &\to \Lambda^2(Q), \\
        \{\mu\} &\mapsto \operatorname{alt}_\mu,
    \end{align}
    where $\{\mu\}$ denotes an element in the 2-cohomology group represented by a 2-cocycle $\mu \in Z^2(Q, \mathbb{C}^\times)$, and:
    \begin{equation}
        \operatorname{alt}_\mu(q_1, q_2) \equiv \frac{\mu(q_1, q_2)}{\mu(q_2, q_1)}.
    \end{equation}
\end{theorem}

Recall the defining property of a 2-cocycle:
\begin{equation}\label{eq:defining2cocycle}
    \mu(q_1, q_2) \mu(q_1 q_2, q_3) = \mu(q_1, q_2 q_3) \mu(q_2, q_3).
\end{equation}
It can be directly verified that $\operatorname{alt}_\mu$ satisfies the conditions outlined in Definition~\ref{def:Bicharacters}. 

From this point onward, we restrict our attention to the case where the subgroup $H$ is Abelian. Let's simply set $f^{\mu}(h_1, h_2) = \operatorname{alt}_{\mu}(h_1, h_2)$.

Essentially, we observe that $\operatorname{alt}_{\mu}(h_1, h_2)$ automatically satisfies the consistency condition in Eq.~(\ref{eq:connectionProperty}). Apparently, the coefficient $\operatorname{alt}_\mu(e, h)= 1$ follows from $\mu(e, h) = \mu(h, e) = 1$. Therefore:
\begin{align}
    \hat{A}^{\mu}_{H}\hat{\Omega}=& \left[\frac{1}{|H|} \sum_{h_1,h_2 \in H} \operatorname{alt}_\mu(h_1, h_2)\Ribbon^{h_1, h_2}\right]\left[\frac{1}{|G|} \sum_{g \in G} \Ribbon^{e,g}\right] \nonumber \\
    =& \frac{1}{|H|} \sum_{h_2 \in H} \hat{Y}^{e}\left[\frac{1}{|G|} \sum_{g \in G} \hat{Z}^{h_2g}\right]
    = \hat{Y}^{e}\left[\frac{1}{|G|} \sum_{g \in G} \hat{Z}^{g}\right] \nonumber\\
    =& \hat{\Omega}.
\end{align}
Here, the third equality arises from the rearrangement property of groups.

Furthermore, $\operatorname{alt}_\mu(h_1, h_2)$ is modular invariant:
\begin{eqnarray}\label{eq:modularInvarOfalt}
    &&\operatorname{alt}_\mu(h_1, h_2) = S[\operatorname{alt}_\mu(h_1, h_2)] \equiv \operatorname{alt}_\mu(h_2, h_1^{-1}),\label{eq:modularInvarOfalt1}\\
    &&\operatorname{alt}_\mu(h_1, h_2) = T[\operatorname{alt}_\mu(h_1, h_2)] \equiv \operatorname{alt}_\mu(h_1, h_1h_2),\label{eq:modularInvarOfalt2}
\end{eqnarray}
which are proved in Appendix~\ref{apdx:AlgCompute} and imply that $\hat{A}^{\mu}_{H}$ is modular invariant.

Thus, for the case that the subgroup is Abelian, we obtain a complete set of solutions to the consistency equations: 
\begin{theoremph}[Complete Set of Solutions for Abelian Boundary Terms]\label{thm:AbelianBoundaryTerms}
    For a Kitaev's quantum double model with a finite group $G$ as input, each of its gapped boundaries is labeled by a pair $(H\subseteq G,\omega\in H^2(H, \mathbb{C}^\times))$.
    
    We refer to such boundaries as {gapped Abelian boundaries} when the corresponding subgroup $H$ is Abelian. All Abelian boundary terms satisfying the consistency conditions take the following form:
    \begin{equation*}
        \hat{A}_{H}^{\mu} = \frac{1}{|H|} \sum_{h_1, h_2 \in H} \operatorname{alt}_\mu(h_1, h_2) \Ribbon^{h_1, h_2},
    \end{equation*}
    where $\mu$ is in the 2-cocycle $Z^2(H, \mathbb{C}^\times)$, and:
    \begin{equation*}
        \operatorname{alt}_\mu \equiv \frac{\mu(h_1, h_2)}{\mu(h_2, h_1)}
    \end{equation*}
\end{theoremph}
Clearly, when the input group $G$ is Abelian, physical Theorem~\ref{thm:AbelianBoundaryTerms} fully recovers the classification of Lagrangian algebras presented in mathematical Theorem~\ref{thm:ClassLgrgAlg}.

In the context of non-Abelian subgroups, in addition to the SSB terms presented in Theorem~\ref{thm:SSBterms}, there should also exist boundary terms associated with non-trivial elements of the 2-cohomology group. However, the absence of suitable algebraic tools currently prevents an explicit construction of these terms. In light of this, we propose the following physically motivated conjecture:

\begin{conjecture}
    For any finite group $G$, all bivariate maps $B: G \times G \to \mathbb{C}^\times$ satisfying the following conditions are classified by the pair $(H\subseteq G,\omega\in H^2(H, \mathbb{C}^\times))$.
    \begin{gather}
        \sum_{g^\prime\in G}B(g_1,g^\prime)B(g_1,g^{\prime-1}g_2) = |G|B(g_1,g_2),\label{eq:conjecture1}\\
        B(e,g) = B(g,e) = 1,\label{eq:conjecture2}\\
        B(g_1,g_2) = B(g_2,g_1^{-1}),\ B(g_1,g_2) = B(g_1,g_1g_2).\label{eq:conjecture3}
    \end{gather}
\end{conjecture}
\begin{remark}
    Readers familiar with the representation theory of group-theoretical associative algebras will recognize that our conjecture defines certain bicharacters on a group that satisfies modular invariance.
    
    In a related context, similar objects indeed appear in the classification of group-theoretical modular invariants~\cite{DavydovModularInvariantsGrouptheoretical2009a}. The characters of Lagrangian algebras defined in that context satisfy Eqs.~(\ref{eq:conjecture2}) and (\ref{eq:conjecture3}) but fail to satisfy Eq.~(\ref{eq:conjecture1}) in non-Abelian cases. This discrepancy arises because the characters of Lagrangian algebras are generally reducible. A key underlying reason is the introduction of the novel concept of condensable internal DOFs.
    
    The actual situation in our conjecture might be slightly more complex. Fundamentally, the bivariate maps we consider do not satisfy strong conjugate invariance. Applying a conjugate transformation to a bivariate map generates another map that is algebraically isomorphic, rather than being identical. We suggest that the object we seek may correspond to a specific restriction or reduction of the characters of Lagrangian algebras in the bulk when projected onto the boundary. Consequently, the development of new algebraic tools appears necessary, although this lies beyond the scope of the present work.
\end{remark}
\begin{remark}
    Under the assumption that the conjecture can be rigorously proven, our work suggests two significant implications: On the one hand, by considering the physical implications of Lagrangian algebras, it reproduces the classification results of these algebras; on the other hand, it furnishes a lattice-based framework to clarify the connection between Lagrangian algebras and modular invariants. Specifically, it employs a physical construction to establish the equivalence between the distinct mathematical characterizations of modular invariants discussed in \cite{kong2009cardy} and \cite{DGNO2013}.
\end{remark}

\subsection{Dynamics of anyon condensation}\label{subsec:GroundDynamics}
Utilizing the physical interpretations of $\hat{A}$, we can analyze the commutation relations among the boundary terms and shed light on the microscopic dynamical processes underlying anyon condensation on the lattice.

When viewed as an $A$-creating operator, $\hat{A}(\mathrm{Path})$ can be interpreted as the spacetime trajectory of $A$. This perspective naturally leads to the following relations:
\begin{eqnarray}
  \hat{A}(\rho_{i})\hat{A}(\tau_{i}) &=& \vcenter{\hbox{\includegraphics{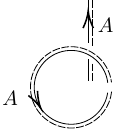}}}, \\
  \hat{A}(\tau_{i})\hat{A}(\rho_{i}) &=& \vcenter{\hbox{\includegraphics{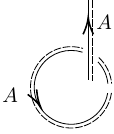}}}.
\end{eqnarray}

From Eq.~(\ref{eq:Commutativity}), it follows that:
\begin{equation}
  \vcenter{\hbox{\includegraphics{Termcommutativity1.pdf}}} = \vcenter{\hbox{\includegraphics{Termcommutativity2.pdf}}}.
\end{equation}
This equality establishes the commutative relation:
\begin{equation}
  [\hat{A}(\rho_{i}), \hat{A}(\tau_{i})] = 0,
\end{equation}
which can be extended to:
\begin{equation}
  [\hat{A}(\rho_{i}), \hat{A}(\tau_{j})] = 0, \quad \forall i, j.
\end{equation}
In summary, all local terms in $H^{A}_{\mathrm{bdy}}$ commute with one another. Therefore, the constructed $H^{A}_{\mathrm{bdy}}$ is a commuting projector Hamiltonian, which is necessarily gapped. {In Appendix~\ref{apdx:SSB_GSD}, we explicitly compute the gapped ground states corresponding to the SSB boundary terms introduced in Theorem~\ref{thm:SSBterms}, along with the global ground state degeneracy (GSD) associated with the SSB pattern.}

\begin{remark}
    In the context of quantum error correcting codes, general planar codes are typically not at a RG fixed point. For such bulk models, boundary condensation terms alone may not always suffice to eliminate all accidental degeneracies, occasionally requiring additional stabilizers, as discussed in Ref.~\cite{Liang2024Operator}.
    
    In contrast, our analysis centers on Kitaev's quantum double model, which resides precisely at an RG fixed point, where the lattice Hilbert space is ``tight'' with respect to the underlying topological data. The completeness of the Lagrangian algebra—specifically, the condition $(\dim A)^2 = \dim \mathcal{C}$—ensures that the condensation is ``maximal'' and complete at the topological level. The only remaining question is whether the explicit truncation of the lattice to introduce a boundary introduces extraneous degrees of freedom due to discretization effects. This aspect encompasses the boundary confinement, which we have already eliminated via the $\hat{A}(\tau)$ terms as discussed in Fig.~\ref{fig:BdyDeconfine}, as well as the global ground state degeneracy associated with the SSB pattern analyzed in Appendix~\ref{apdx:SSB_GSD}. Apart from these contributions, there are no additional sources of degeneracy.
\end{remark}

\begin{figure}
  \centering
  \begin{subfigure}[b]{0.48\columnwidth}
    \centering
    \includegraphics{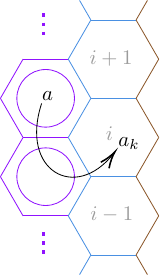}
    \subcaption{ }
    \label{fig:BdyCondense}
  \end{subfigure}
  \hfill
  \begin{subfigure}[b]{0.48\columnwidth}
    \centering
    \includegraphics{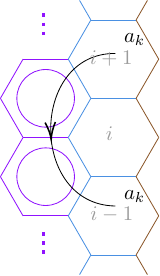}
    \subcaption{ }
    \label{fig:BdyDynamics}
  \end{subfigure}
  \caption{Dynamical processes in the effective Hilbert space $\mathcal{H}^{A\text{-}\mathrm{Conf}}_{\mathrm{bdy}}$. (a) A topological excitation in the bulk condenses to the boundary through the condensable channel $a_k$ via anyon-creating operator; (b) A dynamical process that creates a pair of anyons in the bulk, then condenses them to the gapped boundary.}
  \label{fig:BdyProcesses}
\end{figure}
In the anyon-probing picture, consider a path $\mathrm{Path}_{ij}$ whose starting site $j$ is in the bulk and ending site $i$ is on the boundary, as shown in Fig.~\ref{fig:BdyCondense}. Let $a$ be a summand of the Lagrangian algebra $A$, and $k$ be a condensable internal DOF of $a$. The corresponding $\hat{P}^a_k$ component exists in boundary terms. Within the subspace $\mathcal{H}^{A\text{-}\mathrm{Conf}}_{\mathrm{bdy}}$, the anyon-creating operator $\hat{M}^{a}_{k,k'}(\mathrm{Path}_{ij})$ commutes with all bulk terms except those associated with the bulk site $j$. At the boundary site $i$, the anyon created by this operator fuses with the trivial $A$-excitation, and the fusion process must respect the fusion rule $A \otimes_A A = A$ within the category $\mathscr{C}_A$. As a result, the action of $\hat{M}^{a}_{k,k'}(\mathrm{Path}_{ij})$ leaves the boundary state invariant. This implies that the Hilbert subspace $\mathcal{H}^{A\text{-}\mathrm{Conf}}_{\mathrm{bdy}}\backslash(\mathrm{Site}_j)$, defined by excluding the bulk site $j$, remains invariant under the action of $\hat{M}^{a}_{k,k'}(\mathrm{Path}_{ij})$.

Therefore, $\hat{M}^{a}_{k,-}(\mathrm{Path}_{ij})$, with a free bulk index, can be interpreted as a bulk-to-boundary map that transforms the bulk excitation $a$ into the trivial boundary excitation $A$ through the condensable channel $a_k$. For every condensable channel of $a$, there is a similar creating operator with free starting point in the bulk that carries the complete condensable anyon. This correspondence reveals that the coefficient $K_a$ in the direct sum decomposition of $A$ quantifies the number of independent condensable internal DOFs associated with $a$. Evidently, this number is bounded above by $\mathsf{dim}(a)$.
 
In the anyon-creating picture, consider a path $\mathrm{Path}_{i-1,i+1}$ that starts at boundary site $i+1$ and ends at boundary site $i-1$, as illustrated in Fig.~\ref{fig:BdyDynamics}. The anyon-creating operator $\hat{M}^{a}_{k,k}(\mathrm{Path}_{i-1,i+1})$ describes a dynamical process in which a pair of anyons is created in the bulk and subsequently condenses onto the gapped boundary, provided that $k$ is a condensable internal DOF. This dynamical process can also be interpreted as moving a condensed anyon from one boundary site to another. Such a mechanism prevents condensed anyons from being confined to fixed boundary sites. When this operator is defined on a minimal path $\tau_i$, it becomes component of the boundary term given in Eq.~(\ref{eq:moveBdyterm}).
 
Ultimately, these two forms of boundary terms can be understood as natural generalizations of the bulk terms discussed in \S\ref{subsec:twopictures}, mirroring their structure and functional roles in a closely analogous manner.
\medskip

Finally, we discuss gapped domain walls and general anyon condensations. There is a fundamental and intuitive tool in the study of topological orders named ``\textit{folding-trick}''. As illustrated in Fig.~\ref{fig:folding_trick}, a gapped domain wall between two topological orders $\mathcal{C}$ and $\mathcal{D}$ can be mathematically mapped to a gapped boundary of the folded system $\mathcal{C} \boxtimes \overline{\mathcal{D}}$.

\begin{figure}
    \centering
    \begin{subfigure}[b]{\columnwidth}
        \centering
        \begin{tikzpicture}[scale=0.7]
            \draw[thick] (0,0) -- (0,2);
            \fill[gray!20] (-1.2,0) rectangle (0,2);
            \fill[gray!20] (0,0) rectangle (1.2,2);
            \node[font=\scriptsize] at (-0.6, 1) {$\mathcal{C}$};
            \node[font=\scriptsize] at (0.6, 1) {$\mathcal{D}$};
            
            \draw[->, thick] (1.5,1) -- (2.5,1) node[midway, above, font=\tiny] {Fold};
            
            \draw[thick] (3,0) -- (3,2);
            \fill[gray!40] (3,0) rectangle (4.5,2);
            \node[font=\scriptsize] at (3.75, 1) {$\mathcal{C} \boxtimes \overline{\mathcal{D}}$};
            \node[right, font=\tiny] at (4.5, 1) {Vac.};
            \draw[thick] (4.5,0) -- (4.5,2);
        \end{tikzpicture}
        \caption{Folding trick: fold $\mathcal{C}$ and $\mathcal{D}$ into a single phase, so that the domain wall becomes a boundary.}
        \label{fig:folding_trick}
    \end{subfigure}
    
    \vspace{1em} 
    
    \begin{subfigure}[b]{\columnwidth}
        \centering
        \begin{tikzpicture}[scale=0.7]
            \tikzstyle{bulk}=[fill=gray!20]
            \tikzstyle{boundary}=[thick]
            \tikzstyle{label}=[font=\scriptsize]
            \tikzstyle{tiny_label}=[font=\tiny]

            \begin{scope}[local bounding box=fig1]
                \fill[bulk] (-1.2, -1.5) rectangle (0, 1.5);
                \draw[boundary] (0, -1.5) -- (0, 1.5); 
                \node[label] at (-0.6, 0) {$\mathcal{C}$};
                \node[tiny_label, below] at (0, -1.5) {Bdy $\mathcal{B}_1$};

                \node[tiny_label] at (0.6, 0) {Vac.};

                \fill[bulk] (1.2, -1.5) rectangle (2.4, 1.5);
                \draw[boundary] (1.2, -1.5) -- (1.2, 1.5); 
                \node[label] at (1.8, 0) {$\mathcal{D}$};
                \node[tiny_label, below] at (1.2, -1.5) {Bdy $\mathcal{B}_2$};
                
                \node[label, above] at (0.6, 1.6) {1. Separated};
            \end{scope}


            \begin{scope}[shift={(6,0)}, local bounding box=fig2]
                
                \coordinate (J) at (0, 0);       
                \coordinate (Top) at (0, 2);     
                \coordinate (BL) at (-0.8, -2);  
                \coordinate (BR) at (0.8, -2);   
                
                \fill[bulk] (-2, 2) -- (Top) -- (J) -- (BL) -- (-2, -2) -- cycle;
                
                \fill[bulk] (2, 2) -- (Top) -- (J) -- (BR) -- (2, -2) -- cycle;

                \draw[boundary, ultra thick] (J) -- (Top);
                
                \draw[boundary] (J) -- (BL);
                \draw[boundary] (J) -- (BR);

                \node[label] at (-1, 1) {$\mathcal{C}$};
                \node[label] at (1, 1) {$\mathcal{D}$};
                \node[tiny_label] at (0, -1.5) {Vac.};

                \draw[<-, thin] (0.1, 1) -- (1.5, 1.5) node[right, tiny_label, align=left] {Domain Wall\\ $\mathcal{B}_1\boxtimes \mathcal{B}_2$};
                
                \fill[red] (J) circle (2pt); 
                \draw[<-, thin] (0.1, 0) -- (1.5, 0) node[right, tiny_label, align=left, text=red] {\textbf{Boundary}\\\textbf{Junction} \\ ($E_1$ condens-\\-able algebra)};
                
                \draw[<-, thin] (-0.4, -1) -- (-1.5, -1.5) node[left, tiny_label] {Bdy};

                \node[label, above] at (0, 2.1) {2. The Junction};
            \end{scope}
        \end{tikzpicture}
        \caption{Merging two patches $\mathcal{C}$ and $\mathcal{D}$ through a boundary junction, forming a domain wall above and separating boundaries below.}
        \label{fig:Boundary_junction}
    \end{subfigure}
    \caption{Illustration of folding-trick and boundary junction.}
\end{figure}
    
This equivalence implies that our construction of boundaries based on Lagrangian algebras naturally extends to the construction of gapped domain walls.

Consequently, the boundary junction—which is the 0d domain wall between distinct domain walls—falls within the framework of anyon condensation. Let us illustrate the boundary junction in a more dynamical manner as depicted in Fig.~\ref{fig:Boundary_junction}. The two separated paths can be brought together such that, on the upper side of the branched structure, the two boundaries merge into a single domain wall, while remaining distinct on the lower side. A boundary junction forms at the branching node. According to the folding trick, the two boundaries and the domain wall can be described by Lagrangian algebras in $\mathcal{C}$, $\mathcal{D}$, and $\mathcal{C}\boxtimes\overline{\mathcal{D}}$, respectively, and thus all can be realized on the lattice via our construction.

The boundary junction is fundamentally a 0d domain wall separating two instances of $\mathcal{B}_1\boxtimes \mathcal{B}_2$, and its classification is determined by the $E_1$ condensable algebras \cite{xu20252moritaequivalentcondensablealgebras}. Extending our construction to lower dimension would therefore facilitate a systematic lattice realization of boundary junctions. Some concrete examples can be found in \cite{xu20252moritaequivalentcondensablealgebras, huang2025hybridlatticesurgerynonclifford}.

\section{Examples}
\label{sec:examples}
Having established the general lattice construction for gapped boundaries of quantum double models via Lagrangian algebras, we now demonstrate its operational power through three representative examples: the $\mathbb{Z}_2$, $\mathbb{Z}_2 \times \mathbb{Z}_2$, and $S_3$ quantum doubles. These carefully selected examples serve complementary purposes:
\begin{enumerate}
    \item The $\mathbb{Z}_2$ case serves as a bridge to the well-known physics of toric code, offering a simple and intuitive demonstration of the effectiveness of our new construction.
    \item The $\mathbb{Z}_2 \times \mathbb{Z}_2$ example illustrates the normal form of our construction for groups with nontrivial 2-cohomology groups.
    \item The $S_3$ quantum double exemplifies the capacity of our framework to handle non-Abelian bulk topological orders.
\end{enumerate}

\subsection{Two $\mathds{Z}_2$ gapped boundaries}\label{subsec:ExampleZ2}
\begin{figure}
    \centering
    \includegraphics{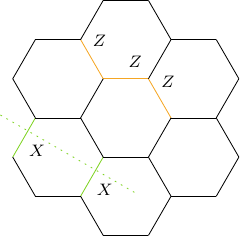}
    \caption{Ribbon operators in $\mathbb{Z}_2$ quantum double. $\hat{M}^{e}$ acts on orange bonds parallel to the path; $\hat{M}^{m}$ acts on green bonds perpendicular to the path.}
    \label{fig:Z2BdyLattice}
\end{figure}

Considering the $\mathds{Z}_2$-quantum double model defined on a lattice with a spin-1/2 residing on each edge. We write the group $\mathbb{Z}_2=\{1,-1\}$, and also denote its two irreducible representations by $1$ and $-1$, respectively. The anyon types in the $\mathds{Z}_2$-quantum double model are enumerated as follows:
\begin{equation}
\begin{aligned}
    &\mathbb{1} = [\{1\},1],\quad & e = [\{-1\},1],\\
    &m = [\{1\},-1],\quad & f = [\{-1\},-1].
\end{aligned}
\end{equation}
The four corresponding anyon-creating operators are:
\begin{equation}
\begin{aligned}
    &\hat{M}^{\mathbb{1}} = Id,\quad & \hat{M}^{e} = \bigotimes_{\substack{\mathrm{Parallel}\\\mathrm{bonds\ }k}} \hat{Z}_k,\\
    &\hat{M}^{m} = \bigotimes_{\substack{\mathrm{Vertical}\\\mathrm{bonds\ }l}} \hat{X}_l,\quad & \hat{M}^{f} = \hat{M}^{e}\hat{M}^{m}.
\end{aligned}
\end{equation}
Here $\hat{X}$ and $\hat{Z}$ are Pauli matrix. The actions of these operators are illustrated in Fig.~\ref{fig:Z2BdyLattice}.

There are two types of gapped boundaries: the smooth boundary corresponds to $A_s = \mathbb{1}\oplus m$ condensation and the rough boundary corresponds to $A_r = \mathbb{1}\oplus e$ condensation. The corresponding classification indices and boundary ribbon operators are summarized in Table~\ref{tab:Z2Bdy}.
\begin{table}
  \caption{Local boundary terms corresponding to two gapped boundaries of $\mathds{Z}_2$ quantum double.}
  \label{tab:Z2Bdy}
  \begin{tabular}{l @{\hspace{1.5em}} c @{\hspace{2em}} c}
    \toprule
    Lagrangian Algebra & $(H,\omega)$ & Boundary Term \\
    \midrule
    $A_s = \mathbb{1} \oplus m$ & $(\{1\}, e)$ & $\hat{Y}^{1}\hat{Z}^{1}$ \\
    $A_r = \mathbb{1} \oplus e$ & $(\mathbb{Z}_2, e)$ & $\frac{1}{2}(\hat{Y}^{1} + \hat{Y}^{-1})(\hat{Z}^{1}+\hat{Z}^{-1})$ \\
    \bottomrule
  \end{tabular}
\end{table}

\begin{figure}
    \centering
    \includegraphics{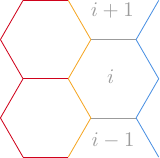}
    \caption{A segment of the lattice that comprises the zig-zag boundary, a transition region, and the bulk. Edges in different regions are color-coded for clarity.}
    \label{fig:TransitionUnit}
\end{figure}

Figure~\ref{fig:TransitionUnit} depicts a segment of the lattice near the boundary, where edges belonging to different regions are distinguished by color. These color-coded edges are used in the following to streamline our notation.

For a Kitaev's quantum double model based on an Abelian group, each anyon possesses only one internal DOF. Using Eq.~(\ref{eq:moveBdyterm}), we can obtain the boundary terms for both types of the boundary. For the smooth boundary, they are:
\begin{align}
  \hat{A}_{s}(\rho_i)  =& \frac{1}{2}\left( 1+\vcenter{\hbox{\includegraphics{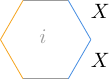}}} \right)\label{eq:Z2SmoothTerm1}, \\
  \hat{A}_{s}(\tau_i)  =& \frac{1}{2}\left( 1+\vcenter{\hbox{\includegraphics{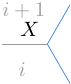}}} \right)\label{eq:Z2SmoothTerm2}.
\end{align}
And for the rough boundary, they are:
\begin{align}
  \hat{A}_{r}(\rho_i)  =& \frac{1}{2}\left( 1+\vcenter{\hbox{\includegraphics{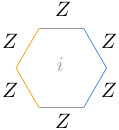}}} \right) \label{eq:Z2RoughTerm1}, \\
  \hat{A}_{r}(\tau_i)  =& \frac{1}{2}\left( 1+\vcenter{\hbox{\includegraphics{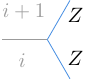}}} \right)\label{eq:Z2RoughTerm2}.
\end{align}

Next, we establish the equivalence between the results of our construction and the well-known boundary formulation presented in Ref.~\cite{kitaev1998Z2bdy}.

We begin by analyzing the smooth boundary. In the ground-state subspace, the boundary terms in Eq.~(\ref{eq:Z2SmoothTerm2}) fix each gray edge to the eigenstate of $X$ with eigenvalue $1$. Then, the vertex operators that involve the gray edges can be effectively rewritten as:
\begin{gather}
    \includegraphics{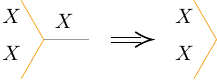} \label{eq:VerifySmooth1},\\
    \includegraphics{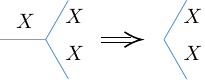} \label{eq:VerifySmooth2}.
\end{gather}
As a result, within the gound-state subspace, the blue zig-zag boundary becomes effectively decoupled from the bulk. The residual interactions on the blue zig-zag boundary, given by Eqs.~\eqref{eq:Z2SmoothTerm1} and \eqref{eq:VerifySmooth2}, correspond exactly to the fixed-point Hamiltonian of an Ising chain in its spontaneously symmetry-broken phase.

After decoupling the blue and gray edges from the main system, we observe that the interactions in Eq.~(\ref{eq:VerifySmooth1}) constrain the two orange edges in each pair to occupy identical local states. This observation enables a deformation of the lattice in which each pair of orange edges is merged into a single edge, resulting in the following transformation of the interacting terms:
\begin{gather}
    \includegraphics{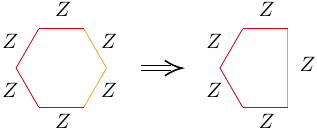} \label{eq:VerifySmooth3},\\
    \includegraphics{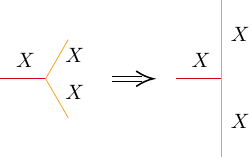} \label{eq:VerifySmooth4}.
\end{gather}
It is evident that Eqs.~\eqref{eq:VerifySmooth3} and \eqref{eq:VerifySmooth4} coincide precisely with the well-known smooth boundary terms presented in Ref.~\cite{kitaev1998Z2bdy}.

We now turn our attention to the rough boundary described by Eqs.~(\ref{eq:Z2RoughTerm1}) and (\ref{eq:Z2RoughTerm2}). Within the ground-state subspace, the interaction term in Eq.~(\ref{eq:Z2RoughTerm2}), together with the vertex operator on the same vertex, restricts the configuration of the three associated edges to only two allowed states. Denoting the $-1$-eigenstate of the $X$ operator with purple lines and the $+1$-eigenstate with black lines, these two states are graphically represented as follows:
\begin{align}
    \ket{-}_{i,i+1} &= \vcenter{\hbox{\includegraphics{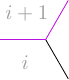}}} + \vcenter{\hbox{\includegraphics{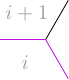}}}, \\
    \ket{+}_{i,i+1} &= \vcenter{\hbox{\includegraphics{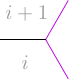}}} + \vcenter{\hbox{\includegraphics{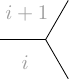}}}.
\end{align}
It is straightforward to verify that the action of the boundary term in Eq.~(\ref{eq:Z2RoughTerm1}) on these effective states is given by:
\begin{align}
    \hat{A}_{r}(\rho_i) \ket{+}_{i,i+1} &= \ket{-}_{i,i+1}, \\
    \hat{A}_{r}(\rho_i) \ket{-}_{i,i+1} &= \ket{+}_{i,i+1}, \\
    \hat{A}_{r}(\rho_i) \ket{+}_{i-1,i} &= \ket{-}_{i-1,i}, \\
    \hat{A}_{r}(\rho_i) \ket{-}_{i-1,i} &= \ket{+}_{i-1,i},
\end{align}
Hence, the blue zig-zag edges contribute no additional DOF. We may therefore consolidate the DOFs of the three edges of a vertex onto its corresponding gray edge, yielding:
\begin{align}
    \ket{-} &\sim \vcenter{\hbox{\includegraphics{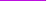}}}, \\
    \ket{+} &\sim \vcenter{\hbox{\includegraphics{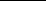}}}.
\end{align}
The boundary term in Eq.~(\ref{eq:Z2RoughTerm1}) then reduces to:
\begin{equation}
    \includegraphics{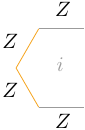}
\end{equation}
This expression is fully consistent with the established form of the rough boundary in Ref.~\cite{kitaev1998Z2bdy}.

We have successfully demonstrated the emergence of the 1d lower symmetric phases within the boundary states of topological orders. The ground state degeneracy at the boundary exhibits a precise correspondence with the SSB patterns of the respective subgroups encoded in the classification indices of Lagrangian algebras. Up to an electromagnetic duality transformation, this finding is in exact agreement with the classification scheme based on enriched fusion category for 1+1D Ising phases~\cite{kong1DGapped2022}.

Furthermore, a straightforward generalization of the proof presented in the $Z_2$ examples demonstrates that, for a generic SSB gapped boundary, there is a global GSD $d_{K} = |K\backslash G / K|$, which corresponds to the number of double cosets of the subgroup $K$. This precisely reflects the phenomenon of `boundary SSB pattern' associated with the topological holography, which can not be characterized by other boundary construction schemes. We provide a detailed proof of this result in the Appendix~\ref{apdx:SSB_GSD}.


\subsection{Six $\mathds{Z}_2\times\mathds{Z}_2$ gapped boundaries}
We employ $(\pm 1,\pm 1)$ to express group elements of $\mathds{Z}_2\times\mathds{Z}_2$, and treat the local Hilbert space on each edge of the lattice as two coupled spin-1/2. Four irreducible representations of $\mathbb{Z}_2\times \mathbb{Z}_2$ are shown in Tab.~\ref{tab:z2z2_reps}.
\begin{table}
  \caption{Irreducible representations of $\mathbb{Z}_2\times \mathbb{Z}_2$ group elements.}
  \label{tab:z2z2_reps}
  \begin{tabular}{ccccc}
    \toprule
    $\mathbb{Z}_2\times \mathbb{Z}_2$ & $I$ & $O$ & $E$ & $D$ \\
    \midrule
    $(1,1)$     & $1$ & $1$ & $1$ & $1$ \\
    $(-1,1)$    & $1$ & $-1$ & $1$ & $-1$ \\
    $(1,-1)$    & $1$ & $1$ & $-1$ & $-1$ \\
    $(-1,-1)$   & $1$ & $-1$ & $-1$ & $1$ \\
    \bottomrule
  \end{tabular}
\end{table}

\begin{table}
  \caption{$16$ types of anyon in a $\mathds{Z}_2\times\mathds{Z}_2$ quantum double model.}
  \label{tab:z2z2_anyon}
  \begin{tabular}{c|cccc}
    \toprule
    $[C,R]$ & $I$ & $O$ & $E$ & $D$ \\
    \midrule
    $\{(1,1)\}$     & $\mathbb{1}$ & $m_1$ & $m_2$ & $m_0$ \\
    $\{(-1,1)\}$    & $e_1$ & $f_{11}$ & $f_{12}$ & $f_{10}$ \\
    $\{(1,-1)\}$    & $e_2$ & $f_{21}$ & $f_{22}$ & $f_{20}$ \\
    $\{(-1,-1)\}$   & $e_0$ & $f_{01}$ & $f_{02}$ & $f_{00}$ \\
    \bottomrule
  \end{tabular}
\end{table}

The $\mathds{Z}_2\times\mathds{Z}_2$-quantum double model has $16$ types of anyons as listed in Tab.~\ref{tab:z2z2_anyon}. There are six Lagrangian algebras in $\FZ_1({\mathrm{Vec}}_{\mathds{Z}_2\times\mathds{Z}_2})$, as shown in Table.~\ref{tab:z2z2bdyterms}, where we utilize $\pm 1$ to represent elements in $H^2(\mathds{Z}_2 \times \mathds{Z}_2,\mathds{C}^\times)=\mathbb{Z}_2$.

\begin{table*}
\centering
  \caption{Local boundary terms corresponding to six gapped boundaries of $\mathds{Z}_2 \times \mathds{Z}_2$ quantum double.}
  \label{tab:z2z2bdyterms}
  \begin{tabular}{l @{\hspace{1.5em}} l @{\hspace{2em}} c}
    \toprule
    Lagrangian Algebra & $(H,\omega)$ & Boundary Term \\
    \midrule
        $A_b = \mathbb{1}\oplus m_1\oplus m_2\oplus m_0$ & $(\{(1,1)\},e)$ & $\hat{Z}^{(1,1)}\hat{Y}^{(1,1)}$\\[10pt]
        
        $A_o = \mathbb{1}\oplus e_2\oplus m_1\oplus f_{21}$  & $(\{(1,1),(1,-1)\},e)$ & { $\displaystyle \frac{1}{2}\bigl[(\hat{Y}^{(1,1)}+\hat{Y}^{(1,-1)})(\hat{Z}^{(1,1)}+\hat{Z}^{(1,-1)} )\bigr]$ } \\[10pt]
        
        $A_e = \mathbb{1}\oplus e_1\oplus m_2\oplus f_{12}$  & $(\{(1,1),(-1,1)\},e)$ & { $\displaystyle \frac{1}{2}\bigl[(\hat{Y}^{(1,1)}+\hat{Y}^{(-1,1)})(\hat{Z}^{(1,1)}+\hat{Z}^{(-1,1)} )\bigr]$ } \\[10pt]
        
        $A_d = \mathbb{1}\oplus e_0\oplus m_0\oplus f_{00}$  & $(\{(1,1),(-1,-1)\},e)$ & { $\displaystyle \frac{1}{2}\bigl[(\hat{Y}^{(1,1)}+\hat{Y}^{(-1,-1)})(\hat{Z}^{(1,1)}+\hat{Z}^{(-1,-1)} )\bigr]$ } \\[10pt]

        $A_s = \mathbb{1}\oplus e_1\oplus e_2\oplus e_0$  & $(\mathds{Z}_2 \times \mathds{Z}_2,1)$ & { $\displaystyle \begin{aligned}[t] \frac{1}{4}\bigl[ &(\hat{Z}^{(1,1)}+\hat{Z}^{(-1,1)}+\hat{Z}^{(1,-1)}+\hat{Z}^{(-1,-1)}) \\ &(\hat{Y}^{(1,1)}+\hat{Y}^{(-1,1)}+\hat{Y}^{(1,-1)}+\hat{Y}^{(-1,-1)})\bigr] \end{aligned}$ }\\[10pt]
        
        $A_c = \mathbb{1}\oplus f_{12}\oplus f_{21}\oplus f_{00}$  & $(\mathds{Z}_2 \times \mathds{Z}_2,-1)$ & { $\displaystyle \begin{aligned}[t] &\frac{1}{4}\bigl[ (\hat{Y}^{(1,1)}+\hat{Y}^{(-1,1)}+\hat{Y}^{(1,-1)}+\hat{Y}^{(-1,-1)})\hat{Z}^{(1,1)} \\ &+(\hat{Y}^{(1,1)}-\hat{Y}^{(-1,1)}+\hat{Y}^{(1,-1)}-\hat{Y}^{(-1,-1)})\hat{Z}^{(1,-1)} \\&+(\hat{Y}^{(1,1)}+\hat{Y}^{(-1,1)}-\hat{Y}^{(1,-1)}-\hat{Y}^{(-1,-1)})\hat{Z}^{(-1,1)} \\&+(\hat{Y}^{(1,1)}-\hat{Y}^{(-1,1)}-\hat{Y}^{(1,-1)}+\hat{Y}^{(-1,-1)})\hat{Z}^{(-1,-1)} \bigr] \end{aligned}$ }\\
    \bottomrule
  \end{tabular}
\end{table*}

The gapped boundary terms of the $\mathbb{Z}_2 \times \mathbb{Z}_2$-quantum double model, expressed in terms of ribbon operators, are summarized in the third column of Tab.~\ref{tab:z2z2bdyterms}. These terms can be obtained either via the direct sum decomposition of the Lagrangian algebra, as detailed in Eq.~(\ref{eq:ProbeBdyterm}) or Eq.~(\ref{eq:moveBdyterm}), or alternatively, through bicharacters as described in Theorem~\ref{thm:AbelianBoundaryTerms}.

As an illustrative example, the boundary terms associated with $A_c$ can be expressed as follows:
\begin{equation}
    \includegraphics{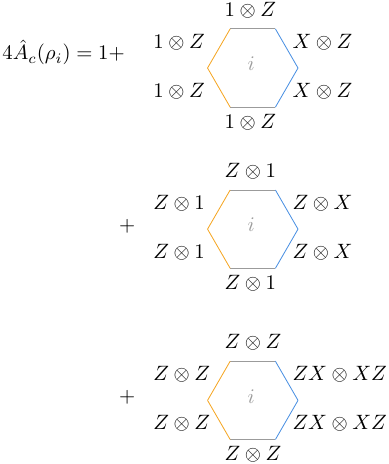}
\end{equation}
\begin{equation}
  \includegraphics{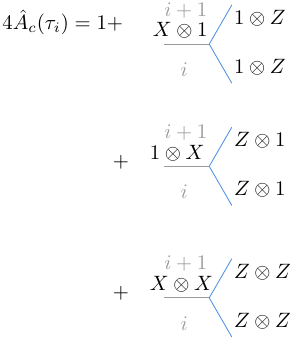}
\end{equation}

Once more, this result is consistent with the classification of $\mathbb{Z}_2 \times \mathbb{Z}_2$ phases presented in Ref.~\cite{xuCategoricalDescriptions1dimensional2022}, particularly the correspondence in which the $A_c$ term maps to the non-trivial SPT phase of the $\mathbb{Z}_2 \times \mathbb{Z}_2$ symmetric quantum chain.

\subsection{Four ${S}_3$ gapped boundaries}
The $S_3$ group is generated by the elements $e$, $r$, and $x$. Here $e$ is the identity element, $r^3 = e$, $x^2 = e$, and the relations $xr = r^2x$ and $rx = xr^2$ hold.

The conjugate classes and their centralizers are shown in Tab.~\ref{tab:S3conj_classes}. All irreducible representations of the three centralizers are shown in Tab.~\ref{tab:combined_S3reps}.
\begin{table}
  \caption{Conjugate classes and their centralizers in $S_3$ group}
  \label{tab:S3conj_classes}
  \begin{tabular}{lcl}
    \toprule
    \makecell{Conjugate\\ Class}       & \makecell{Representative\\ Element} & Centralizer \\
    \midrule
    $C_e = \{ e \}$         & $e$            & $S_3$ \\
    $C_r = \{ r,r^2 \}$     & $r$            & $\mathbb{Z}_3 = \{e,r,r^2\}$ \\
    $C_x = \{ x,xr,xr^2 \}$ & $x$            & $\mathbb{Z}_2 = \{e,x\}$ \\
    \bottomrule
  \end{tabular}
\end{table}

\begin{table}[t]
\caption{Irreducible representations of the $S_3$ group and its subgroups $\mathbb{Z}_3$ and $\mathbb{Z}_2$. Here we define $\omega = e^{i2\pi/3}$.
\label{tab:combined_S3reps}}
\renewcommand{\arraystretch}{1.4} 
\setlength{\tabcolsep}{6pt}       
\begin{tabular}{l c c c @{\hspace{3em}} l c c c}
\toprule
\multicolumn{4}{c}{\textbf{$S_3$ Group}} & \multicolumn{4}{c}{\textbf{Subgroups ($\mathbb{Z}_3, \mathbb{Z}_2$)}} \\
\cmidrule(r{2em}){1-4} \cmidrule{5-8} 
$S_3$ & $I$ & $S$ & $V$ & $\mathbb{Z}_3^{r}$ & 1 & $\omega$ & $\omega^{2}$ \\
\cmidrule(r{2em}){1-4} \cmidrule{5-8} 
$e$    & $1$ & $1$  & $\begin{bmatrix}1&0 \\ 0&1\end{bmatrix}$ & 
$\mathbb{Z}_3$ ($e$) & $1$ & $1$ & $1$ \\

$r$    & $1$ & $1$  & $\begin{bmatrix}\omega & 0 \\ 0 & \omega^{2}\end{bmatrix}$ & 
$\mathbb{Z}_3$ ($r$) & $1$ & $\omega$ & $\omega^{2}$ \\

$r^2$  & $1$ & $1$  & $\begin{bmatrix}\omega^{2} & 0 \\ 0 & \omega\end{bmatrix}$ & 
$\mathbb{Z}_3$ ($r^2$) & $1$ & $\omega^{2}$ & $\omega$ \\

\cmidrule{5-8} 
$x$    & $1$ & $-1$ & $\begin{bmatrix}0 & 1 \\ 1 & 0\end{bmatrix}$ & 
$\mathbb{Z}^{x}_2$ & $+$ & $-$ \\

$xr$   & $1$ & $-1$ & $\begin{bmatrix}0 & \omega^{2} \\ \omega & 0\end{bmatrix}$ & 
$\mathbb{Z}_2$ ($e$) & $1$ & $1$ &  \\

$xr^2$ & $1$ & $-1$ & $\begin{bmatrix}0 & \omega \\ \omega^{2} & 0\end{bmatrix}$ & 
$\mathbb{Z}_2$ ($x$) & $1$ & $-1$ &  \\
\bottomrule
\end{tabular}
\end{table}

Next, we introduce the anyon labels as listed in Table~\ref{tab:anyon_dimensions}, along with their corresponding quantum dimensions and the conventions for the internal DOFs.

\begin{table}
  \caption{Anyon labels, quantum dimensions and conventions for the internal DOFs of $S_3$ quantum double}
  \label{tab:anyon_dimensions}
  \begin{tabular}{lccl}
    \toprule
    \makecell{Group\\ Label}      & \makecell{Anyon\\ Label} & \makecell{Quantum\\ Dimension} & \makecell{Internal\\ DOFs} \\
    \midrule
    $[C_e,I]$        & $\mathbb{1}$         & 1 & $\ket{\mathbb{1};1} = \ket{[C_e,I];(e,1)}$\\[1ex]
    $[C_e,S]$        & $a_1$         & 1 & $\ket{s_1;1} = \ket{[C_e,S];(e,1)}$\\[4ex]
    $[C_e,V]$        & $a_2$       & 2 & {$\displaystyle \begin{aligned}\ket{s_2;1} &= \ket{[C_e,V];(e,1)}\\ \ket{s_2;2} &= \ket{[C_e,V];(e,2)}\end{aligned}$}\\[4ex]
    $[C_r,I]$        & $b$         & 2 & {$\displaystyle \begin{aligned}\ket{b;1} &= \ket{[C_r,I];(e,1)}\\ \ket{b;2} &= \ket{[C_r,I];(x,1)}\end{aligned}$}\\[4ex]
    $[C_r,\omega]$   & $b_1$       & 2 & {$\displaystyle \begin{aligned}\ket{b_1;1} &= \ket{[C_r,\omega];(e,1)}\\ \ket{b_1;2} &= \ket{[C_r,\omega];(x,1)}\end{aligned}$}\\[4ex]
    $[C_r,\omega^2]$ & $b_2$       & 2 & {$\displaystyle \begin{aligned}\ket{b_2;1} &= \ket{[C_r,\omega^{2}];(e,1)}\\ \ket{b_2;2} &= \ket{[C_r,\omega^{2}];(x,1)}\end{aligned}$}\\[6ex]
    $[C_x,+]$        & $c$         & 3 & {$\displaystyle \begin{aligned}\ket{c;1} &= \ket{[C_x,+];e,1}\\ \ket{c;2} &= \ket{[C_x,+];r,1}\\ \ket{c;3} &= \ket{[C_x,+];r^{2},1} \end{aligned}$}\\[6ex]
    $[C_x,-]$        & $c_1$       & 3 & {$\displaystyle \begin{aligned}\ket{c_1;1} &= \ket{[C_x,-];e,1}\\ \ket{c_1;2} &= \ket{[C_x,-];r,1}\\ \ket{c_1;3} &= \ket{[C_x,-];r^{2},1}\end{aligned}$}\\
    \bottomrule
  \end{tabular}
\end{table}

We list all the anyon-creating operators in the $S_3$ quantum double model in Table~\ref{tab:anyon_operators}.

\begin{table*}[t]
\caption{Anyon-creating operators in the $S_3$ quantum double model. \label{tab:anyon_operators}}

  \begin{tabular}{l c l}
  \toprule
  Anyon Type & Dim & Anyon Move Operators\\
  \midrule
  $\mathds{1} = [C_e,I]$ & 1 & 
  $\displaystyle \hat{M}_{1,1}^{\mathds{1}} = \hat{Y}^{e}+\hat{Y}^{r}+\hat{Y}^{r^{2}}+\hat{Y}^{x}+\hat{Y}^{xr}+\hat{Y}^{xr^2}$ \\[1.5ex]

  $a_1 = [C_e,S]$ & 1 & 
  $\displaystyle \hat{M}_{1,1}^{a_1} = \hat{Y}^{e}+\hat{Y}^{r}+\hat{Y}^{r^{2}}-\hat{Y}^{x}-\hat{Y}^{xr}-\hat{Y}^{xr^2}$ \\[2.5ex]

  $a_2 = [C_e,V]$ & 2 & 
  $\begin{aligned}
  \hat{M}_{1,1}^{a_2} &= \hat{Y}^{e}+e^{i{2\pi}/{3}}\hat{Y}^{r}+e^{-i{2\pi}/{3}}\hat{Y}^{r^{2}}, &
  \hat{M}_{1,2}^{a_2} &= \hat{Y}^{x}+e^{-i{2\pi}/{3}}\hat{Y}^{xr}+e^{i{2\pi}/{3}}\hat{Y}^{xr^{2}} \\
  \hat{M}_{2,1}^{a_2} &= \hat{Y}^{x}+e^{i{2\pi}/{3}}\hat{Y}^{xr}+e^{-i{2\pi}/{3}}\hat{Y}^{xr^{2}}, &
  \hat{M}_{2,2}^{a_2} &= \hat{Y}^{e}+e^{-i{2\pi}/{3}}\hat{Y}^{r}+e^{i{2\pi}/{3}}\hat{Y}^{r^{2}}
  \end{aligned}$ \\[5.5ex]

  $b = [C_r,I]$ & 2 & 
  $\begin{aligned}
  \hat{M}_{1,1}^{b} &= \hat{Z}^r(\hat{Y}^{e}+\hat{Y}^{r}+\hat{Y}^{r^{2}}), &
  \hat{M}_{1,2}^{b} &= \hat{Z}^{r^2}(\hat{Y}^{x}+\hat{Y}^{xr^2}+\hat{Y}^{xr}) \\
  \hat{M}_{2,1}^{b} &= \hat{Z}^r(\hat{Y}^{x}+\hat{Y}^{xr}+\hat{Y}^{xr^{2}}), &
  \hat{M}_{2,2}^{b} &= \hat{Z}^{r^2}(\hat{Y}^{e}+\hat{Y}^{r^2}+\hat{Y}^{r})
  \end{aligned}$ \\[5.5ex]

  $b_1 = [C_r,\omega]$ & 2 & 
  $\begin{aligned}
  \hat{M}_{1,1}^{b_1} &= \hat{Z}^r(\hat{Y}^{e}+e^{i{2\pi}/{3}}\hat{Y}^{r}+e^{-i{2\pi}/{3}}\hat{Y}^{r^{2}}), &
  \hat{M}_{1,2}^{b_1} &= \hat{Z}^{r^2}(\hat{Y}^{x}+e^{i{2\pi}/{3}}\hat{Y}^{xr^2}+e^{-i{2\pi}/{3}}\hat{Y}^{xr}) \\
  \hat{M}_{2,1}^{b_1} &= \hat{Z}^r(\hat{Y}^{x}+e^{i{2\pi}/{3}}\hat{Y}^{xr}+e^{-i{2\pi}/{3}}\hat{Y}^{xr^{2}}), &
  \hat{M}_{2,2}^{b_1} &= \hat{Z}^{r^2}(\hat{Y}^{e}+e^{i{2\pi}/{3}}\hat{Y}^{r^2}+e^{-i{2\pi}/{3}}\hat{Y}^{r})
  \end{aligned}$ \\[5.5ex]

  $b_2 = [C_r,\omega^2]$ & 2 & 
  $\begin{aligned}
  \hat{M}_{1,1}^{b_2} &= \hat{Z}^r(\hat{Y}^{e}+e^{-i{2\pi}/{3}}\hat{Y}^{r}+e^{i{2\pi}/{3}}\hat{Y}^{r^{2}}), &
  \hat{M}_{1,2}^{b_2} &= \hat{Z}^{r^2}(\hat{Y}^{x}+e^{-i{2\pi}/{3}}\hat{Y}^{xr^2}+e^{i{2\pi}/{3}}\hat{Y}^{xr}) \\
  \hat{M}_{2,1}^{b_2} &= \hat{Z}^r(\hat{Y}^{x}+e^{-i{2\pi}/{3}}\hat{Y}^{xr}+e^{i{2\pi}/{3}}\hat{Y}^{xr^{2}}), &
  \hat{M}_{2,2}^{b_2} &= \hat{Z}^{r^2}(\hat{Y}^{e}+e^{-i{2\pi}/{3}}\hat{Y}^{r^2}+e^{i{2\pi}/{3}}\hat{Y}^{r})
  \end{aligned}$ \\[5.5ex]

  $c = [C_x,+]$ & 3 & 
  $\begin{aligned}
  \hat{M}_{1,1}^{c} &= \hat Z^x(\hat Y^e + \hat Y^x), &
  \hat{M}_{1,2}^{c} &= \hat Z^{xr}(\hat Y^{r^2} + \hat Y^{xr^2}), &
  \hat{M}_{1,3}^{c} &= \hat Z^{xr^2}(\hat Y^{r} + \hat Y^{xr}) \\
  \hat{M}_{2,1}^{c} &= \hat Z^x(\hat Y^r + \hat Y^{xr^2}), &
  \hat{M}_{2,2}^{c} &= \hat Z^{xr}(\hat Y^{e} + \hat Y^{xr}), &
  \hat{M}_{2,3}^{c} &= \hat Z^{xr^2}(\hat Y^{r^2} + \hat Y^{x}) \\
  \hat{M}_{3,1}^{c} &= \hat Z^x(\hat Y^{r^2} + \hat Y^{xr}), &
  \hat{M}_{3,2}^{c} &= \hat Z^{xr}(\hat Y^{r} + \hat Y^{x}), &
  \hat{M}_{3,3}^{c} &= \hat Z^{xr^2}(\hat Y^{e} + \hat Y^{xr^2})
  \end{aligned}$ \\[7.5ex]

  $c_1 = [C_x,-]$ & 3 & 
  $\begin{aligned}
  \hat{M}_{1,1}^{c_1} &= \hat Z^x(\hat Y^e - \hat Y^x), &
  \hat{M}_{1,2}^{c_1} &= \hat Z^{xr}(\hat Y^{r^2} - \hat Y^{xr^2}), &
  \hat{M}_{1,3}^{c_1} &= \hat Z^{xr^2}(\hat Y^{r} - \hat Y^{xr}) \\
  \hat{M}_{2,1}^{c_1} &= \hat Z^x(\hat Y^r - \hat Y^{xr^2}), &
  \hat{M}_{2,2}^{c_1} &= \hat Z^{xr}(\hat Y^{e} - \hat Y^{xr}), &
  \hat{M}_{2,3}^{c_1} &= \hat Z^{xr^2}(\hat Y^{r^2} - \hat Y^{x}) \\
  \hat{M}_{3,1}^{c_1} &= \hat Z^x(\hat Y^{r^2} - \hat Y^{xr}), &
  \hat{M}_{3,2}^{c_1} &= \hat Z^{xr}(\hat Y^{r} - \hat Y^{x}), &
  \hat{M}_{3,3}^{c_1} &= \hat Z^{xr^2}(\hat Y^{e} - \hat Y^{xr^2})
  \end{aligned}$ \\
  \bottomrule
  \end{tabular}

\end{table*}

Since all 2-cohomology groups of the $S_3$ group and its subgroups are trivial, the boundary terms can be directly determined by applying Theorem~\ref{thm:SSBterms}. Here, we explicitly demonstrate the computation of the boundary terms by utilizing Eq.~(\ref{eq:moveBdyterm}) facilitated by direct sum decompositions of the Lagrangian algebras.

There are four Lagrangian algebras in $\FZ_1({\mathrm{Vec}}_{S_3})$:
\begin{equation}
\begin{aligned}
    &A_1 = \mathbb{1}\oplus a_1\oplus 2a_2 ,\quad &A_2 = \mathbb{1}\oplus a_1\oplus 2b, \\
    &A_3 = \mathbb{1}\oplus a_2\oplus c ,\quad &A_4 = \mathbb{1}\oplus b\oplus c.
\end{aligned}
\end{equation}

        
        
        

For $A_1 = \mathbb{1}\oplus a_1\oplus 2a_2$, there are two independent condensable internal DOFs of $a_2$, which are $\ket{a_2;1}$ and $\ket{a_2;2}$.
\begin{align}
  M_{11}^{a_2} = \hat{Z}^e(\hat{Y}^{e}+e^{i{2\pi}/{3}}\hat{Y}^{r}+e^{-i{2\pi}/{3}}\hat{Y}^{r^{2}}),\\
  M_{22}^{a_2} = \hat{Z}^e(\hat{Y}^{e}+e^{-i{2\pi}/{3}}\hat{Y}^{r}+e^{i{2\pi}/{3}}\hat{Y}^{r^{2}}),
\end{align}

Using Eq.~(\ref{eq:moveBdyterm}), it is easy to get boundary term:
\begin{eqnarray}
\hat{A}_1 &=& \frac{1}{6} \hat{M}^{\mathbb{1}} + \frac{1}{6}\hat{M}^{a_1} + \frac{2}{6}(\hat{M}^{a_2}_{11}+\hat{M}^{a_2}_{22})\nonumber\\
&=& \hat Z^e\hat Y^e.
\end{eqnarray}

For $A_2 = \mathbb{1}\oplus a_1\oplus 2 b$, there are two independent condensable internal DOFs of $b$, which are $\ket{b;1}$ and $\ket{b;2}$.
\begin{align}
  M_{11}^{b} = \hat{Z}^r(\hat{Y}^{e}+\hat{Y}^{r}+\hat{Y}^{r^{2}}),\\
  M_{22}^{b} = \hat{Z}^{r^2}(\hat{Y}^{e}+\hat{Y}^{r}+\hat{Y}^{r^{2}}),
\end{align}
Thus, we have
\begin{eqnarray}
      \hat{A}_2 &=& \frac{1}{6} \hat{M}^{\mathbb{1}} + \frac{1}{6}\hat{M}^{a_1} + \frac{2}{6}(\hat{M}^{b}_{11}+\hat{M}^{b}_{22})\nonumber\\
  &=& \frac{1}{3}\left[(\hat{Z}^{e}+\hat{Z}^{r}+\hat{Z}^{r^{2}})(\hat{Y}^e+\hat{Y}^r+\hat{Y}^{r^2})\right].
\end{eqnarray}

For $A_3 = \mathbb{1}\oplus a_2\oplus c$, condensable internal DOF of $a_2$ is $\frac{1}{\sqrt{2}}(\ket{a_2;1}+\ket{a_2;2})$ and condensable internal DOF of $c$ is $\ket{c;1}$.
\begin{eqnarray}
        &&\hat{M}^{a_2}_{\frac{1}{\sqrt{2}}(1+2),\frac{1}{\sqrt{2}}(1+2)} =  \frac{1}{2} \sum_{mn}\hat{M}^{a_2}_{mn}\nonumber\\
        &&=  \frac{1}{2}\hat{Z}^e(2\hat{Y}^{e}-\hat{Y}^{r}-\hat{Y}^{r^{2}}+2\hat{Y}^{x}-\hat{Y}^{xr}-\hat{Y}^{xr^2}).
\end{eqnarray}
\begin{equation}
  \hat{M}_{11}^{c} = \hat{Z}^x(\hat{Y}^e + \hat{Y}^x).
\end{equation}
Thus:
\begin{eqnarray}
      \hat{A}_3 &=& \frac{1}{6} \hat{M}^{\mathbb{1}} + \frac{2}{6}\hat{M}^{a_2}_{\frac{1}{\sqrt{2}}(1+2),\frac{1}{\sqrt{2}}(1+2)} + \frac{3}{6}\hat{M}^{c}_{11}\nonumber\\
  &=& \frac{1}{2}\left[(\hat{Z}^{e}+\hat{Z}^{x})(\hat{Y}^e+\hat{Y}^x)\right].
\end{eqnarray}

For $A_4 = \mathbb{1}\oplus b\oplus c$, condensable internal DOF of $b$ is $\frac{1}{\sqrt{2}}(\ket{b;1}+\ket{b;2})$ and condensable internal DOF of $c$ is $\frac{1}{\sqrt{3}}(\ket{c;1}+\ket{c;2}+\ket{c;3})$.
\begin{eqnarray}
&& \hat{M}^{b}_{\frac{1}{\sqrt{2}}(1+2),\frac{1}{\sqrt{2}}(1+2)} = \frac{1}{2} \sum_{mn}\hat{M}^{b}_{mn}\nonumber\\
&&= \frac{1}{2}(\hat{Z}^r+\hat{Z}^{r^2})(\hat{Y}^{e}+\hat{Y}^{r}+\hat{Y}^{r^{2}}+\hat{Y}^{x}+\hat{Y}^{xr}+\hat{Y}^{xr^2}),\nonumber\\
\end{eqnarray}
\begin{eqnarray}
&&\hat{M}^{c}_{\frac{1}{\sqrt{3}}(1+2+3),\frac{1}{\sqrt{3}}(1+2+3)} = \frac{1}{3} \sum_{mn}\hat{M}^{c}_{mn}\nonumber\\
&&=  \frac{1}{3}(\hat{Z}^{xr}+\hat{Z}^{xr^2})(\hat{Y}^{e}+\hat{Y}^{r}+\hat{Y}^{r^{2}}+\hat{Y}^{x}+\hat{Y}^{xr}+\hat{Y}^{xr^2}).\nonumber\\
\end{eqnarray}
Thus:
\begin{eqnarray}
\hat{A}_4 &=& \frac{1}{6} \hat{M}^{\mathbb{1}}+ \frac{2}{6}\hat{M}^{b}_{\frac{1}{\sqrt{2}}(1+2),\frac{1}{\sqrt{2}}(1+2)}\nonumber\\
  &&+ \frac{3}{6}\hat{M}^{c}_{\frac{1}{\sqrt{3}}(1+2+3),\frac{1}{\sqrt{3}}(1+2+3)}\nonumber\\
  &=& \frac{1}{6}\bigl[(\hat{Z}^{e}+\hat{Z}^{r}+\hat{Z}^{r^{2}}+\hat{Z}^{x}+\hat{Z}^{xr}+\hat{Z}^{xr^2})\nonumber\\
  &&(\hat{Y}^{e}+\hat{Y}^{r}+\hat{Y}^{r^{2}}+\hat{Y}^{x}+\hat{Y}^{xr}+\hat{Y}^{xr^2})\bigr] 
\end{eqnarray}

The complete results of gapped boundaries terms of $S_3$ quantum double are shown in Tab.~\ref{tab:S3bdyterms}. These terms are defined on the $\rho_i$ and $\tau_i$ paths, and their operator actions are given by Eqs.~(\ref{eq:AK_rhoi}) and (\ref{eq:AK_taui}), respectively. 
Here, we present the boundary terms of $A_3$ as well as the bulk-to-boundary map as a visual illustration:
\begin{eqnarray}
&& \hat{A}_{\mathbb{Z}^x_2}(\rho_i) \left[\vcenter{\hbox{\includegraphics{BdyRibbonState1.pdf}}} \right]\nonumber\\
&&= \frac{\delta(v_{i+\frac{1}{2}}\in v_{i}\mathbb{Z}^x_2) }{2} \left[\vcenter{\hbox{\includegraphics{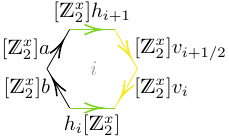}}} \right]\label{eq:AZ2_rhoi}
\end{eqnarray}
\begin{eqnarray}
&&\hat{A}_{\mathbb{Z}^x_2}(\tau_i) \left[\vcenter{\hbox{\includegraphics{BdyRibbonState4.pdf}}} \right]\nonumber\\
&&= \frac{\delta(h_{i+1}\in \mathbb{Z}^x_2 )}{2}  \left[\vcenter{\hbox{\includegraphics{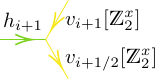}}} \right]\label{eq:AZ2_taui}.
\end{eqnarray}
Here we use the notation $[\mathbb{Z}_{2}^{x}] = e+x$. We can examine the bulk-to-boundary map in this example. Consider the following boundary-site basis state:
\begin{equation}
    \ket{a,h_{i+1},v_{i+1/2}} \equiv \left[\vcenter{\hbox{\includegraphics{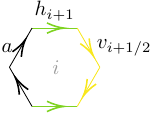}}} \right]
\end{equation}
For $A_3 = \mathbb{1}\oplus a_2 \oplus c$, $\ket{c;1}$ constitutes a condensable internal DOF, as demonstrated in Table~\ref{tab:S3bdyterms}. Consider the bulk-to-boundary map $\hat{M}^c_{1,1}$ defined on the representative shortest path:
\begin{eqnarray}
&&\hat{M}^c_{1,1}(p_s)\ket{a,h_{i+1},v_{i+1/2}} \nonumber\\
&&= \frac{\delta(a\in \mathbb{Z}^x_2 )}{2}  \left[\vcenter{\hbox{\includegraphics{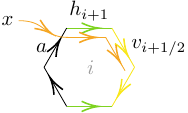}}}\right]\nonumber\\
&&= \frac{\delta(a\in \mathbb{Z}^x_2 )}{2} \ket{a,xh_{i+1},xv_{i+1/2}}.
\end{eqnarray}
It is straightforward to verify the commutativity with $\hat{A}_{\mathbb{Z}^x_2}(\rho_i)$:
\begin{eqnarray}
    &&\hat{A}_{\mathbb{Z}^x_2}(\rho_i)\hat{M}^c_{1,1}(p_s)\ket{a,h_{i+1},v_{i+1/2}} \nonumber\\
    =&& \hat{M}^c_{1,1}(p_s) \hat{A}_{\mathbb{Z}^x_2}(\rho_i) \ket{a,h_{i+1},v_{i+1/2}}\nonumber\\
    =&& \frac{\delta(a\in \mathbb{Z}^x_2 )}{2}\ket{a,[\mathbb{Z}^x_2]h_{i+1},[\mathbb{Z}^x_2]v_{i+1/2}}
\end{eqnarray}
Commutativity with $\hat{A}_{\mathbb{Z}^x_2}(\tau_i)$ can be checked analogously. These indicates that the internal DOF $\ket{c;1}$ can freely condense onto the boundary. This provides a concrete instance of the general analysis in \S\ref{subsec:GroundDynamics}.

\begin{turnpage}
\begin{table*}
\caption{Local boundary terms and condensable internal DOFs corresponding to four gapped boundaries of $S_3$ quantum double.}
\label{tab:S3bdyterms}
\begin{ruledtabular}
\begin{tabular}{l l l c}
Lagrangian Algebra & $(H,\omega)$ & Condensable Internal DOFs & Boundary Term \\
\hline
$A_1 = \mathbb{1}\oplus a_1\oplus 2a_2$ 
& $(\{e\},e)$ 
& { $\displaystyle \begin{aligned}[t]
\hat{M}_{1,1}^{\mathbb{1}} &= \hat{Y}^{e}+\hat{Y}^{r}+\hat{Y}^{r^{2}}+\hat{Y}^{x}+\hat{Y}^{xr}+\hat{Y}^{xr^2}\\
\hat{M}_{1,1}^{a_1} &= \hat{Y}^{e}+\hat{Y}^{r}+\hat{Y}^{r^{2}}-\hat{Y}^{x}-\hat{Y}^{xr}-\hat{Y}^{xr^2}\\
\hat{M}_{1,1}^{a_2} &= \hat{Z}^e(\hat{Y}^{e}+e^{i{2\pi}/{3}}\hat{Y}^{r}+e^{-i{2\pi}/{3}}\hat{Y}^{r^{2}})\\
\hat{M}_{2,2}^{a_2} &= \hat{Z}^e(\hat{Y}^{e}+e^{-i{2\pi}/{3}}\hat{Y}^{r}+e^{i{2\pi}/{3}}\hat{Y}^{r^{2}})
 \end{aligned}$ }
& $\hat{Z}^{(1,1)}\hat{Y}^{(1,1)}$ \\[60pt]

$A_2 = \mathbb{1}\oplus a_1\oplus 2b$  
& $(\mathbb{Z}_3^{r},e)$ 
& { $\displaystyle \begin{aligned}[t]
\hat{M}_{1,1}^{\mathbb{1}} &= \hat{Y}^{e}+\hat{Y}^{r}+\hat{Y}^{r^{2}}+\hat{Y}^{x}+\hat{Y}^{xr}+\hat{Y}^{xr^2}\\
\hat{M}_{1,1}^{a_1} &= \hat{Y}^{e}+\hat{Y}^{r}+\hat{Y}^{r^{2}}-\hat{Y}^{x}-\hat{Y}^{xr}-\hat{Y}^{xr^2}\\
\hat{M}_{1,1}^{b} &= \hat{Z}^r(\hat{Y}^{e}+\hat{Y}^{r}+\hat{Y}^{r^{2}})\\
\hat{M}_{2,2}^{b} &= \hat{Z}^{r^2}(\hat{Y}^{e}+\hat{Y}^{r}+\hat{Y}^{r^{2}})
 \end{aligned}$ }
& $\displaystyle \frac{1}{3}\left[(\hat{Z}^{e}+\hat{Z}^{r}+\hat{Z}^{r^{2}})(\hat{Y}^e+\hat{Y}^r+\hat{Y}^{r^2})\right]$ \\[60pt]

$A_3 = \mathbb{1}\oplus a_2\oplus c$  
& $(\mathbb{Z}_2^{x},e)$ 
& { $\displaystyle \begin{aligned}[t]
\hat{M}_{1,1}^{\mathbb{1}} &= \hat{Y}^{e}+\hat{Y}^{r}+\hat{Y}^{r^{2}}+\hat{Y}^{x}+\hat{Y}^{xr}+\hat{Y}^{xr^2}\\
\hat{M}^{a_2}_{\frac{1}{\sqrt{2}}(1+2),\frac{1}{\sqrt{2}}(1+2)} &= \frac{1}{2}\hat{Z}^e(2\hat{Y}^{e}-\hat{Y}^{r}-\hat{Y}^{r^{2}}+2\hat{Y}^{x}-\hat{Y}^{xr}-\hat{Y}^{xr^2})\\
\hat{M}_{11}^{c} &= \hat{Z}^x(\hat{Y}^e + \hat{Y}^x)
 \end{aligned}$ }
& $\displaystyle \frac{1}{2}\left[(\hat{Z}^{e}+\hat{Z}^{x})(\hat{Y}^e+\hat{Y}^x)\right]$  \\[60pt]

$A_4 = \mathbb{1}\oplus b\oplus c$  
& $(S_3,e)$ 
& { $\displaystyle \begin{aligned}[t]
\hat{M}_{1,1}^{\mathbb{1}} &= \hat{Y}^{e}+\hat{Y}^{r}+\hat{Y}^{r^{2}}+\hat{Y}^{x}+\hat{Y}^{xr}+\hat{Y}^{xr^2}\\
\hat{M}^{b}_{\frac{1}{\sqrt{2}}(1+2),\frac{1}{\sqrt{2}}(1+2)} &= \frac{1}{2}(\hat{Z}^r+\hat{Z}^{r^2})(\hat{Y}^{e}+\hat{Y}^{r}+\hat{Y}^{r^{2}}+\hat{Y}^{x}+\hat{Y}^{xr}+\hat{Y}^{xr^2}) \\
\hat{M}^{c}_{\frac{1}{\sqrt{3}}(1+2+3),\frac{1}{\sqrt{3}}(1+2+3)} &= \frac{1}{3}(\hat{Z}^{xr}+\hat{Z}^{xr^2})(\hat{Y}^{e}+\hat{Y}^{r}+\hat{Y}^{r^{2}}+\hat{Y}^{x}+\hat{Y}^{xr}+\hat{Y}^{xr^2})
 \end{aligned}$ }
& { $\displaystyle \begin{aligned}[t] \frac{1}{6}\bigl[ &(\hat{Z}^{e}+\hat{Z}^{r}+\hat{Z}^{r^{2}}+\hat{Z}^{x}+\hat{Z}^{xr}+\hat{Z}^{xr^2}) \\ &(\hat{Y}^{e}+\hat{Y}^{r}+\hat{Y}^{r^{2}}+\hat{Y}^{x}+\hat{Y}^{xr}+\hat{Y}^{xr^2})\bigr] \end{aligned}$ } \\
\end{tabular}
\end{ruledtabular}
\end{table*}
\end{turnpage}

\section{Summary and Outlook}
\label{sec:Summary}
In this work, we have developed a systematic framework for constructing all 1+1D gapped boundaries of Kitaev's quantum double models based on the macroscopic categorical formulation of Lagrangian algebras.

The core results in Section~\ref{sec:QDbdy} report three major achievements:
\begin{enumerate}
    \item We systematically derive the boundary interaction terms (Theorem~\ref{thmph:bdyTerms}) by ensuring consistency between anyon-creating/probing processes and the axioms of Lagrangian algebras.
    \item Theorem~\ref{thm:SSBterms} and Theorem~\ref{thm:AbelianBoundaryTerms} furnish explicit expressions for two classes of boundary terms that satisfy the conditions in Theorem~\ref{thmph:bdyTerms}.
    \item We provide a microscopic characterization of bulk-to-boundary anyon condensation dynamics via the action of ribbon operators in \S\ref{subsec:GroundDynamics}.
\end{enumerate}

This work paves the way for several promising extensions:
\begin{itemize}
    \item \textbf{Generalization to $C^*$-Hopf Algebra Quantum Doubles:} Given the broad applicability of Lagrangian algebras, our construction can be extended to describe gapped boundaries of more general $C^*$-Hopf algebra quantum double models.
    
    \item \textbf{Boundary Phase Transitions:} The framework developed in this work provides a solid foundation for studying pure boundary phase transitions, offering insights into their underlying mechanisms and properties.

    \item \textbf{Topological Wick Rotation:} By dualizing the effective Hilbert space $\mathcal{H}^{\mathrm{Zig-Zag}}_{\mathrm{bdy}}$ to an appropriate Hilbert space of a 1+1D chain, we can explore the microscopic correspondence between gapped boundaries of 2+1D topological orders and 1+1D gapped quantum phases with symmetries. This correspondence will facilitate a detailed investigation of topological Wick rotation.
\end{itemize}

\begin{acknowledgments}
    We thank Liang Kong for valuable discussions. M.L. would also to thank Zhi-Hao Zhang for helpful discussions. M.L. is supported by the National Natural Science Foundation of China (NSFC, Grant No. 12574175) and the Guangdong Basic and Applied Basic Research Foundation (Grant No. 2020B1515120100). X.-H.Y. and X.-Y.D. are supported by the Innovation Program for Quantum Science and Technology (Grant No. 2021ZD0301900) and the National Natural Science Foundation of China (Grant No. 12504175).
\end{acknowledgments}

\newpage
\appendix

\begin{widetext}
\section{Algebraic computations}\label{apdx:AlgCompute}

Proof of the orthogonality of probing operators acting on the same path:
\begin{eqnarray}
\hat{P}^{[C_1,R_1]}_{m_1p_1}\hat{P}^{[C_2,R_2]}_{m_2p_2}&=& \frac{|C_1|d_{R_1}|C_2|d_{R_2}}{|G|^2}\sum_{z_1,z_2}\delta_{p_1r_{C_1}p_1^{-1},p_2r_{C_2}p_2^{-1}}\bar{\rho}^{R_1}_{m_1m_1}(z_1)\bar{\rho}^{R_2}_{m_2m_2}(z_2){\Ribbon}^{p_1r_{C_1}p_1^{-1},p_1z_1p_1^{-1}p_2z_2p_2^{-1}}\nonumber \\
  &=& \frac{|C_1|d_{R_1}|C_2|d_{R_2}}{|G|^2}\delta_{C_1,C_2}\delta_{p_1,p_2}\sum_{z_1,z_2}\bar{\rho}^{R_1}_{m_1m_1}(z_1)\bar{\rho}^{R_2}_{m_2m_2}(z_2){\Ribbon}^{p_1r_{C_1}p_1^{-1},p_1z_1z_2p_1^{-1}} \nonumber\\
  &=& \frac{|C_1|d_{R_1}|C_2|d_{R_2}}{|G|^2}\delta_{C_1,C_2}\delta_{p_1,p_2}\sum_{z_1,z}\bar{\rho}^{R_1}_{m_1m_1}(z_1)\bar{\rho}^{R_2}_{m_2m_2}(z_1^{-1}z){\Ribbon}^{p_1r_{C_1}p_1^{-1},p_1z p_1^{-1}}\nonumber \\
  &=& \frac{|C_1|d_{R_1}}{|G|}\delta_{C_1,C_2}\delta_{p_1,p_2}\delta_{R_1R_2}\delta_{m_1m_2}\sum_{z}\bar{\rho}^{R_1}_{m_1m_1}(z){\Ribbon}^{p_1r_{C_1}p_1^{-1},p_1z p_1^{-1}}\nonumber \\
  &=& \delta_{C_1,C_2}\delta_{R_1R_2}\delta_{p_1,p_2}\delta_{m_1m_2}\hat{P}^{[C_1,R_1]}_{m_1p_1}.
\end{eqnarray}

Proof of the normalization of probing operators acting on the same path:
\begin{eqnarray}
  \sum_{C,R,m,p} \hat{P}^{[C,R]}_{mp}
  &=& \sum_{C}\sum_{R}\frac{|C|d_R}{|G|}\sum_{p}\sum_{m}\sum_{z\in Z(r_C)}\bar{\rho}^{R}_{mm}(z){\Ribbon}^{pr_Cp^{-1},pzp^{-1}}\nonumber\\
  &=& \sum_{C}\sum_{R}\frac{|C|d_R}{|G|}\sum_{p}\sum_{z\in Z(r_C)}\bar{\chi}^{R}(z){\Ribbon}^{pr_Cp^{-1},pzp^{-1}}\nonumber\\
  &=& \sum_{C}\frac{|C|}{|G|}\sum_{p}\sum_{z\in Z(r_C)} \sum_{R}d_R\bar{\chi}^{R}(z){\Ribbon}^{pr_Cp^{-1},pzp^{-1}}\nonumber\\
  &=& \sum_{C}\sum_{p}\sum_{z\in Z(r_C)} \delta_{z,e}{\Ribbon}^{pr_Cp^{-1},pzp^{-1}}\nonumber\\
  &=& \sum_{C}\sum_{p}{\Ribbon}^{pr_Cp^{-1},e}
  = \sum_{g\in G}{\Ribbon}^{g,e}
  = \mathrm{Id}.
\end{eqnarray}

Proof of the Eq.~(\ref{eq:OmegaStrand}):

\begin{eqnarray}
\frac{1}{|G|}\sum_{C,R} {\frac{|C|d_R}{|G|}\hat{M}^{[C,R]}}
&=& \frac{1}{|G|^2}\sum_{C} \sum_{R}{|C|d_R} \sum_{p}\sum_{z\in Z(r_{C})}\chi^{R}(z)\hat{F}^{pz p^{-1},p r_{C} p^{-1}}\nonumber\\
&=& \frac{1}{|G|^2}\sum_{C} \sum_{p}\sum_{z\in Z(r_{C})} |C|\sum_{R} d_R\chi^{R}(z)\hat{F}^{pz p^{-1},p r_{C} p^{-1}}\nonumber\\
&=& \frac{1}{|G|^2}\sum_{C} \sum_{p}\sum_{z\in Z(r_{C})} |C|\delta_{e,z}|Z(r_c)|\hat{F}^{pz p^{-1},p r_{C} p^{-1}}\nonumber\\
&=& \frac{1}{|G|}\sum_{C} \sum_{p}\hat{F}^{e,p r_{C} p^{-1}}\nonumber\\
&=& \frac{1}{|G|}\sum_{g\in G}\hat{F}^{e,g}
= \left(\frac{1}{|G|}\sum_{g\in G}\hat{Z}^g\right)\hat{Y}^e.
\end{eqnarray}

Modular invariant of $\operatorname{alt}_\mu(h_1, h_2)$ for $h_1h_2=h_2h_1$ can be proved utilizing Eq.~(\ref{eq:defining2cocycle}):
\begin{equation}
    \frac{\mu(a, b)}{\mu(b, c)}  = \frac{\mu(a, bc)} {\mu(ab, c)}
\end{equation}
\begin{eqnarray}
    &&\frac{\operatorname{alt}_\mu(h_1, h_2)}{S[\operatorname{alt}_\mu(h_1, h_2)]} = \frac{\mu(h_1, h_2)}{\mu(h_2, h_1)}\frac{\mu(h_1^{-1}, h_2)}{\mu(h_2, h_1^{-1})}=\frac{\mu(h_1, h_2)}{\mu(h_2, h_1^{-1})}\frac{\mu(h_1^{-1}, h_2)}{\mu(h_2, h_1)}\\
    &=& \frac{\mu(h_1, h_2h_1^{-1})}{\mu(h_1h_2, h_1^{-1})}\frac{\mu(h_1^{-1}, h_2)}{\mu(h_2, h_1)} =  \frac{\mu(h_1h_2, h_2h_1^{-1})}{\mu(h_2, h_2)}\frac{\mu(h_2, h_2)}{\mu(h_1h_2, h_1^{-1}h_2)} =1
\end{eqnarray}

\begin{equation}
        {\operatorname{alt}_\mu(h_1, h_2)} = \frac{\mu(h_1, h_2)}{\mu(h_2, h_1)}
    = \frac{\mu(h_1, h_2h_1)}{\mu(h_1h_2, h_1)} = {\operatorname{alt}_\mu(h_1, h_1h_2)}
    = T[\operatorname{alt}_\mu(h_1, h_2)]
\end{equation}

\end{widetext}

\section{Anyon basis on torus}\label{apdx:Anyon_Basis}
Consider the set of states $\{|g,h\rangle,\forall g,h\in G\}$ defined on the simplest lattice decomposition of the torus as in Eq.~(\ref{eq:Torus_State}). Two different types of bases of it, labeled by $L_1$ and $L_2$, respectively, can be defined as follows: 
\begin{eqnarray}\label{basis L1}
\left\{|C,R;nq,mp\rangle_{L_1}\equiv \sum_{z\in Z(r_C)}\rho^R_{nm}(z)\ket{qzp^{-1},pr_Cp^{-1}}\right\}, \nonumber\\
\end{eqnarray}
\begin{eqnarray}\label{basis L2}
\left\{ \ket{C,R;mp,nq}_{L_2} \equiv \sum_{z\in Z(r_C)}\bar{\rho}^R_{mn}(z)\ket{pr_Cp^{-1},pzq^{-1}} \right\}.\nonumber\\
\end{eqnarray}
The labels $L_1$ and $L_2$ also denote two non-contractible loops on the torus as in Fig.~\ref{fig:TorusDecomposition}. The reason that we use the same labels here will be clear later in this section. The orthogonality between basis states in each set can be verified directly using the orthogonality of group representations. The number of states in each set is:
\begin{eqnarray}
\sum_{C}\sum_R |C|^2d_R^2 = \sum_{C}|C|^2|Z(r_C)|= |G|(\sum_{C}|C|) = |G|^2,\nonumber\\
\end{eqnarray}
which is equal to the number of states in $\{|g,h\rangle\}$.

In the set of bases labeled by $L_1$, if we choose the trivial conjugate class $C_e=\{e\}$ of $G$, and the trivial representation $1$ of $Z(e) = G$, whose element and character are both $1$, we observe that $\ket{C_e,1}_{L_1}$ is a ground state on the torus:
\begin{eqnarray}
    \ket{G.S.}_{L_1}&\equiv&\ket{C_e,1}_{L_1}\equiv\ket{C_e,1;1,1}_{L_1}=\sum_{z\in G}\rho^{1}(z)\ket{z,e}\nonumber\\
    &=&\sum_{z\in G}\chi^{1}(z)\ket{z,e}=\sum_{z\in G}\ket{z,e}.
\end{eqnarray}

Acting ribbon operator $\hat{F}^{g,h}(L_1)$ along the loop $L_1$ on the ground state $\ket{G.S.}_{L_1}$, we obtain the state $\ket{g,h}$:
\begin{eqnarray}\label{eq:corres_g_h}
\hat{F}^{g,h}(L_1)\ket{G.S.}_{L_1}&=&\sum_{z\in G}\hat{F}^{g,h}(L_1)\vcenter{\hbox{\includegraphics{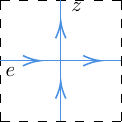}}}\nonumber\\
&=&\vcenter{\hbox{\includegraphics{TorusState.pdf}}} = \ket{g,h}.
\end{eqnarray}
This relationship sets up a one-to-one correspondence between $\hat{F}^{g,h}(L_1)$ and $\ket{g,h}$. Therefore, Eq.~(\ref{eq:corres_g_h}) establishes a linear isomorphism between two linear spaces:
\begin{equation}
    \mathfrak{C}_{L_1} \simeq \left\{\ket{g,h}\right\}_{\mathrm{Torus}}.
\end{equation}
As discussed in \S\ref{subsec:LocalProbing}, the $\mathfrak{C}$-algebras defined on different paths are isomorphic. Thus, we can omit the subscript $L_1$ of $\mathfrak{C}_{L_1}$, and the operator-state correspondence holds in general on any fixed path of the ribbon operators:
\begin{equation}
    \mathfrak{C} \simeq \left\{\ket{g,h}\right\}_{\mathrm{Torus}}.
\end{equation}

Acting $\hat{M}^{[C,R]}_{nq,mp}(L_1)$ on $\ket{G.S.}_{L_1}$, we obtain another basis state $\ket{C,R;nq,mp}_{L_1}$ in the basis set labeled by $L_1$:
\begin{eqnarray}\label{eq:L1Anyonloop}
&&\hat{M}^{[C,R]}_{nq,mp}(L_1)\ket{G.S.}_{L_1} \nonumber\\
&&=\sum_{z\in Z(r_C)}\rho^R_{nm}(z)\hat{F}^{qzp^{-1},pr_Cp^{-1}}(L_1)\sum_{z'\in G}\ket{z',e}\nonumber\\
&&=\sum_{z\in Z(r_C)}\rho^R_{nm}(z)\ket{qzp^{-1},pr_Cp^{-1}}\nonumber\\
&&=\ket{C,R;nq,mp}_{L_1}.
\end{eqnarray}
Notice that these states are not normalized:
\begin{eqnarray}
&&{}_{L_1}\bra{C,R;nq,mp}\ket{C,R;nq,mp}_{L_1}\nonumber\\
&&= \sum_{z\in Z(r_C)}\bar{\rho}^R_{nm}(z)\rho^R_{nm}(z) \nonumber\\
&&= \frac{|Z(r_C)|}{d_R}=\frac{|G|}{\mathsf{dim}([C,R])}.\label{eq:TorusNormal1}
\end{eqnarray}

The operator $\hat{M}^{[C,R]}_{nq,mp}(L_1)$ creates a pair of anyon (with internal DOF $nq$) and its dual anyon (with internal DOF $mp$), and moves one of them around the loop $L_1$. Since in general $nq$ and $mp$ are not the same, these two anyons cannot fuse into the vacuum. Thus, $\ket{C,R;nq,mp}_{L_1}$ is not a ground state. This relationship sets up a one-to-one correspondence between the anyon-creating operator $\hat{M}^{[C,R]}_{nq,mp}(L_1)$ and the basis states $\ket{C,R;nq,mp}_{L_1}$. 

Acting $\hat{M}^{[C,R]}(L_1)$, which is the trace of $\hat{M}^{[C,R]}_{nq,mp}(L_1)$, on the $\ket{G.S.}_{L_1}$, a pair of anyons are created, and then annihilated after moving one of them around loop $L_1$. The system returns to the ground state: 
\begin{eqnarray}\label{eq:L1GSAnyonloop}
&&\hat{M}^{[C,R]}(L_1)\ket{G.S.}_{L_1} \nonumber\\
&&=\sum_{p\in\{p\}^C}\sum_{z\in Z(r_C)}\chi^R(z)\hat{F}^{pzp^{-1},pr_Cp^{-1}}\sum_{z'\in G}\ket{z',e}\nonumber\\
&&=\sum_{p\in\{p\}^C}\sum_{z\in Z(r_C)}\chi^R(z)\ket{pzp^{-1},pr_Cp^{-1}}\nonumber\\
&&\equiv\ket{C,R}_{L_1}.
\end{eqnarray}
Therefore, starting from one particular ground state $\ket{G.S.}_{L_1}$, all the other degenerate ground states, in the form $\ket{C,R}_{L_1}$, can be obtained through the action of $\hat{M}^{[C,R]}(L_1)$ along the loop $L_1$. 

Similarly, in the basis set labeled by $L_2$, if we choose $C=C_e$ and $R=1$, we could also obtain a ground state on the torus: 
\begin{eqnarray}
    \ket{G.S.}_{L_2}&\equiv&\ket{C_e,1}_{L_2}\equiv\ket{C_e,1;1,1}_{L_2}\nonumber\\
    &=&\sum_{z\in G}\bar{\chi}^{1}(z)\ket{e,z}=\sum_{z\in G}\ket{e,z}.
\end{eqnarray}
Acting ribbon operator $\hat{F}^{g,h}(L_2)$ along the loop $L_2$ on the ground state $\ket{G.S.}_{L_2}$, we obtain the state $\ket{h,g^{-1}}$:
\begin{equation}
    \Ribbon^{g,h}(L_2)\ket{G.S.}_{L_2}=\ket{h,g^{-1}}
\end{equation}
Acting $\hat{M}^{[C,R]}_{nq,mp}(L_2)$ along the loop $L_2$ on $\ket{G.S.}_{L_2}$, we obtain another basis state $\ket{C,R;mp,nq}_{L_2}$ in the basis set labeled by $L_2$:
\begin{eqnarray}\label{eq:L2GSAnyonloop}
&&\hat{M}^{[C,R]}_{nq,mp}(L_2)\ket{G.S.}_{L_2} \nonumber\\
&&=\sum_{z\in Z(r_C)}\rho^R_{nm}(z)\hat{F}^{qzp^{-1},pr_Cp^{-1}}\sum_{z'\in G}\ket{e,z'}\nonumber\\
&&=\sum_{z\in Z(r_C)}\rho^R_{nm}(z)\ket{pr_Cp^{-1},pz^{-1}q^{-1}}\nonumber\\
&&=\sum_{z\in Z(r_C)}\bar{\rho}^R_{mn}(z)\ket{pr_Cp^{-1},pzq^{-1}}\nonumber\\
&&=\ket{C,R;mp,nq}_{L_2}.
\end{eqnarray}
The inner products are also:
\begin{equation}
{}_{L_2}\bra{C,R;nq,mp}\ket{C,R;nq,mp}_{L_2}=\frac{|G|}{\mathsf{dim}([C,R])}.\label{eq:TorusNormal2}
\end{equation}
Due to these intuitive pictures, we call the basis $|C,R;nq,mp\rangle_{L_1}$ and $\ket{C,R;mp,nq}_{L_2}$ as anyon basis. 

Now, we consider the $S$-transiformation of anyon-creating operator $\hat{M}^{[C,R]}_{nq,mp}(L_1)$ as defined in Eq.~(\ref{eq:Def_S_hat}), and act it on $\ket{G.S.}_{L_1}$:
\begin{eqnarray}
&&\Strans{\hat{M}^{[C,R]}_{nq,mp}(L_1)}\ket{G.S.}_{L_1} \nonumber\\
&&=\sum_{z\in Z(r_C)}\bar{\rho}^{R}_{mn}(z){\Ribbon}^{pr_Cp^{-1},pzq^{-1}}\sum_{z'\in G}\ket{z',e}\nonumber\\
&&=\sum_{z\in Z(r_C)}\bar{\rho}^R_{mn}(z)\ket{pr_Cp^{-1},pzq^{-1}}\nonumber\\
&&=\ket{C,R;mp,nq}_{L_2}.
\end{eqnarray}
And taking the trace gives:
\begin{eqnarray}
&&\Strans{\hat{M}^{[C,R]}(L_1)}\ket{G.S.}_{L_1}\nonumber\\
&&=\sum_{p\in\{p\}^C}\sum_{z\in Z(r_C)}\bar{\chi}^R(z)\hat{F}^{pr_Cp^{-1}, pz^{-1}p^{-1}}\sum_{z'\in G}\ket{z',e}\nonumber\\
&&=\sum_{p\in\{p\}^C}\sum_{z\in Z(r_C)}\bar{\chi}^R(z)\ket{pr_Cp^{-1}, pz^{-1}p^{-1}}\nonumber\\
&&=\ket{C,R}_{L_2}.
\end{eqnarray}

We also introduce a unitary operator $\hat{S}$ acting on the torus states as:
\begin{equation}
\hat{S}\ket{g,h} \equiv \ket{h,g^{-1}} = \Strans{\hat{F}^{g,h}(L_1)}\ket{G.S.}_{L_1},
\end{equation}
\begin{equation*}
    \vcenter{\hbox{\includegraphics{TorusState.pdf}}} \xrightarrow{\circlearrowleft} \vcenter{\hbox{\includegraphics{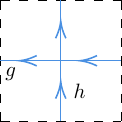}}}.
\end{equation*}
The operator $\hat{S}$ relates the basis states with label $L_1$ to that with label $L_2$:
\begin{equation}
\hat{S}\ket{C,R;nq,mp}_{L_1} = \ket{C,R;mp,nq}_{L_2}.
\end{equation}
Therefore, we obtain the relation:
\begin{eqnarray}
\hat{S} \left(\hat{M}^{[C,R]}_{nq,mp}(L_1)\ket{G.S.}_{L_1}\right) = \Strans{\hat{M}^{[C,R]}_{nq,mp}(L_1)}\ket{G.S.}_{L_1}\nonumber\\
\end{eqnarray}
This relationship sets up the correspondence between the $S$-transformation of ribbon operators and the unitary operator $\hat{S}$ on the torus states.

In a completely analogous manner, we also introduce a torus operator $\hat{T}$, defined as:
\begin{equation}
    \hat{T}\ket{g,h} = \ket{g,gh} = \Ttrans{\hat{F}^{g,h}(L_1)}\ket{G.S.}_{L_1}.
\end{equation}
We find that:
\begin{eqnarray}
&&\hat{F}^{g,h}(L_t)\ket{G.S.}_{L_2}=\sum_{z\in G}\hat{F}^{g,h}(L_t)\ket{e,z}\nonumber\\
&&= \vcenter{\hbox{\includegraphics{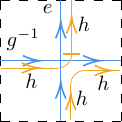}}} = \vcenter{\hbox{\includegraphics{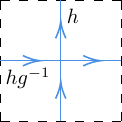}}}\nonumber\\[10pt] 
&&= \ket{h,hg^{-1}}=\hat{T}\left[\Ribbon^{g,h}(L_2)\ket{G.S.}_{L_2}\right].
\end{eqnarray}
This result indicates the correspondence between $L_2$ anyon loops and $L_t$ anyon loops:
\begin{eqnarray}
        \ket{C,R;nq,mp}_{L_t} &\equiv& {\hat{M}^{[C,R]}_{nq,mp}(L_t)}\ket{G.S.}_{L_2}\nonumber\\
        &=& \hat{T} \ket{C,R;nq,mp}_{L_2} \label{eq:ThreeTyptT}
\end{eqnarray}

\section{S-matrix of Kitaev's quantum double}\label{apdx:Smatrix}

For Kitaev's quantum double model, the following expressions for S-matrix elements can be extracted from~\cite{dijkgraaf1991quasi} or~\cite{coste2000finite}. 
\begin{eqnarray}
S_{[CR],[C'R']} &=& \frac{1}{|G|}
\smashoperator{\sum_{\substack{a\in\{a\}^{C}\\ b\in\{b\}^{C'}}}}
 \delta\Big(ar_C a^{-1}br_{C'}b^{-1},\ br_{C'}b^{-1}ar_C a^{-1}\Big) \nonumber\\
 &&\times \bar{\chi}^{R}\big(a^{-1}br_{C'}b^{-1}a\big)\bar{\chi}^{R'}\big(b^{-1}ar_{C}a^{-1}b\big).
\end{eqnarray}
Noticed that for only $a^{-1}br_{C'}b^{-1}a=z\in Z(r_C)$, the $\delta$ coefficient not equal to $0$. Then, $b^{-1}ar_{C}a^{-1}b$ can be viewed as an element in $C^{Z(z)}_{r_C}$, which is the conjugate class of $r_C$ in the subgroup $Z(z)$.

It was proved in~\cite{coste2000finite} that the S-matrix can be expressed as:
\begin{equation}
\begin{aligned}
    S_{[CR],[C'R']} =& \sum_{a\in\{a\}^{C}}\sum_{z\in Z(r_C)} \delta(C',C_z)\bar{\chi}^R(z)\bar{\chi}^{R'}(C^{Z(z)}_{r_C})\\
    =& \frac{|C|}{|G|} \sum_{z\in Z(r_C)}\delta(C',C_z)\bar{\chi}^R(z)\bar{\chi}^{R'}(C^{Z(z)}_{r_C}).
\end{aligned}
\end{equation}

Two lemmas that we're going to use are shown below.
\begin{lemma}
    Consider the definition of anyon-probe operator:
    \begin{equation}
    \begin{aligned}
            \hat{P}^{[C,R]} = \frac{d_R|C|}{|G|}\sum_{z\in Z(r_C)}\sum_{q\in\{q\}^C}\bar{\chi}^{R}(z) {\Ribbon}^{qr_Cq^{-1},qzq^{-1}}\\
            = \frac{d_R|C|}{|G|}\sum_{D}\bar{\chi}^{R}(D)\left[\sum_{q\in\{q\}^C}\sum_{d\in D} {\Ribbon}^{qr_Cq^{-1},qdq^{-1}}\right].
    \end{aligned}
    \end{equation}
    Here $D$ is a conjugate class in $Z(r_C)$.

    Given a group $H$, recall the column orthogonality of characters of representations:
    \begin{equation}
        \sum_{R}\bar{\chi}^{R}(K'){\chi}^{R}(K) = \frac{|H|}{|K|} \delta_{K,K'}.
    \end{equation}

    This deduces:
    \begin{equation}
    \frac{|C||Z(r_C)|}{|G||D|}\left[\sum_{q\in\{q\}^C}\sum_{d\in D} {\Ribbon}^{qr_Cq^{-1},qdq^{-1}}\right] = \sum_R \frac{{\chi}^{R}(D)}{d_R}\hat{P}^{[C,R]}.
    \end{equation}
\end{lemma}

\begin{lemma}
    Using the disassembly of the group by the conjugate class and the centralizer, the following formula can be proved:
    \begin{equation}
        \sum_{q\in \{q\}^{C_z}}\sum_{d\in C^{Z(z)}_{r_C}} {\Ribbon}^{qdq^{-1},qzq^{-1}} = \frac{|C^{Z(z)}_{r_C}|}{|Z(z)|}\sum_{g\in G}{\Ribbon}^{gr_Cg^{-1},gzg^{-1}}.
    \end{equation}
    Similarly:
    \begin{equation}
        \sum_{p\in \{p\}^{C}}\sum_{k\in C^{Z(r_C)}_{z}} {\Ribbon}^{qr_{C}q^{-1},qkq^{-1}} = \frac{|C^{Z(r_C)}_{z}|}{|Z(r_C)|}\sum_{g\in G}{\Ribbon}^{gr_Cg^{-1},gzg^{-1}}.
    \end{equation}
    These give:
    \begin{equation}
    \begin{aligned}
       & \sum_{q\in \{q\}^{C_z}}\sum_{d\in C^{Z(z)}_{r_C}} {\Ribbon}^{qdq^{-1},qzq^{-1}}  \\
      =& \frac{|C^{Z(z)}_{r_C}|}{|Z(z)|}\frac{|Z(r_C)|}{|C^{Z(r_C)}_{z}|}\sum_{p\in \{p\}^{C}}\sum_{k\in C^{Z(r_C)}_{z}} {\Ribbon}^{qr_{C}q^{-1},qkq^{-1}}.
    \end{aligned}
    \end{equation}

\end{lemma}

\begin{widetext}
  Now, we are ready to calculate:
  \begin{align}
      & |G|\sum_{[C',R']} \bar{S}_{[C,R],[C',R']}\frac{1}{|C'|d_{R'}}\hat{P}^{[C',R']}\\
      =& |G|\sum_{[C',R']} \frac{|C|}{|G|} \sum_{z\in Z(r_C)}\delta(C',C_z){\chi}^R(z){\chi}^{R'}(C^{Z(z)}_{r_C})\frac{1}{|C'|d_{R'}}\hat{P}^{[C',R']} \\
      =& |C|\sum_{R'} \sum_{z\in Z(r_C)}{\chi}^R(z){\chi}^{R'}(C^{Z(z)}_{r_C})\frac{1}{|C_z|d_{R'}}\hat{P}^{[C_z,R']} \\
      =& |C|\sum_{z\in Z(r_C)}\frac{{\chi}^R(z)}{|C_z|}\sum_{R'} \left[\frac{{\chi}^{R'}(C^{Z(z)}_{r_C})}{d_{R'}}\hat{P}^{[C_z,R']}\right] \\
      \overset{\text{lem1}}{=}& \frac{|C|}{|G|}\sum_{z\in Z(r_C)}\frac{{\chi}^R(z)}{|C_z|} \frac{|C_z||Z(z)|}{|C^{Z(z)}_{r_C}|}\left[\sum_{p\in\{p\}^{C_z}}\sum_{d\in C^{Z(z)}_{r_C}} {\Ribbon}^{qr_Cq^{-1},qdq^{-1}}\right] \\
      \overset{\text{lem2}}{=}& \sum_{z\in Z(r_C)}{{\chi}^R(z)}\frac{1}{|C^{Z(r_C)}_{z}|}\sum_{q\in \{q\}^{C}}\sum_{k\in C^{Z(r_C)}_{z}} {\Ribbon}^{qr_{C}q^{-1},qkq^{-1}} \\
      =& \sum_{q\in \{q\}^{C}}\sum_{D}\sum_{d\in D}{{\chi}^R(r_D)}\frac{1}{|C^{Z(r_C)}_{r_D}|}\sum_{k\in C^{Z(r_C)}_{r_D}} {\Ribbon}^{qr_{C}q^{-1},qkq^{-1}} \\
      =& \sum_{q\in \{q\}^{C}}\sum_{D}{{\chi}^R(r_D)}\sum_{k\in C^{Z(r_C)}_{r_D}} {\Ribbon}^{qr_{C}q^{-1},qkq^{-1}} \\
      =& \sum_{q\in \{q\}^{C}}\sum_{z\in Z(r_C)}{{\chi}^R(z)}{\Ribbon}^{qr_{C}q^{-1},qzq^{-1}}.
  \end{align}
  This is exactly what we expected.
\end{widetext}

\section{Global boundary degeneracy}\label{apdx:SSB_GSD} 
We now turn to the discussion of the global GSD on the boundary under open boundary conditions with infinite extent in the other direction. To unify notation and facilitate reference, we carry out the subsequent analysis on the boundary lattice illustrated below, and we restate here the definitions of Eq.~(\ref{eq:BdyRibbon1}) and (\ref{eq:BdyRibbon2}).
\begin{equation}\label{eq:localbdyRegion}
    \vcenter{\hbox{\includegraphics{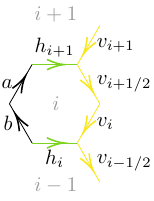}}}
\end{equation}

\begin{eqnarray}
&&\Ribbon^{k,g}(\rho_i) \left[\vcenter{\hbox{\includegraphics{BdyRibbonState1.pdf}}} \right]\nonumber\\
&&= \delta_{k,v_{i}^{-1}v_{i+\frac{1}{2}}}\left[\vcenter{\hbox{\includegraphics{BdyRibbonState3.pdf}}} \right].\label{eq:BdyRibbon1_apdx}
\end{eqnarray}
\begin{eqnarray}
&&\Ribbon^{k,g}(\tau_i) \left[\vcenter{\hbox{\includegraphics{BdyRibbonState4.pdf}}} \right]\nonumber\\
&&= \delta_{k,h_{i+1}^{-1}}\left[\vcenter{\hbox{\includegraphics{BdyRibbonState6.pdf}}} \right].\label{eq:BdyRibbon2_apdx}
\end{eqnarray}

Next, consider an arbitrary finite group $G$ and its arbitrary subgroup $K$. We adopt the convention that the sum over the subgroup:
\begin{equation}
    [K] \equiv \sum_{k\in K} k,
\end{equation}
which serves as the label on the edge, and similarly, the superscripts of the operators $\hat{Y}$ and $\hat{Z}$. Recalling Theorem~\ref{thm:SSBterms}, we have the following SSB boundary terms:
\begin{equation}
    \hat{A}_{K} = \frac{1}{|K|} \sum_{k_1, k_2 \in H} \Ribbon^{k_1, k_2} = \frac{1}{|K|}\hat{Y}^{[K]}\hat{Z}^{[K]^{\prime}},
\end{equation}
where $[K]$ and $[K]^{\prime}$ are summed independently. The actions of these SSB terms on the $\rho$ path and the $\tau$ path are given by:
\begin{eqnarray}
&& \hat{A}_{K}(\rho_i) \left[\vcenter{\hbox{\includegraphics{BdyRibbonState1.pdf}}} \right]\nonumber\\
&&= \frac{\delta(v_{i+\frac{1}{2}}\in v_{i}K) }{|K|} \left[\vcenter{\hbox{\includegraphics{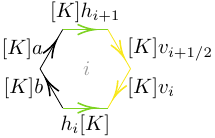}}} \right]\label{eq:AK_rhoi}.
\end{eqnarray}
\begin{eqnarray}
&&\hat{A}_{K}(\tau_i) \left[\vcenter{\hbox{\includegraphics{BdyRibbonState4.pdf}}} \right]\nonumber\\
&&= \frac{\delta(h_{i+1}\in K)}{|K|}  \left[\vcenter{\hbox{\includegraphics{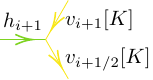}}} \right]\label{eq:AK_taui}.
\end{eqnarray}
We first examine the common eigenstates of $\hat{A}(\tau_i)$ and the vertex operator at the same location. Clearly, these eigenstates are labeled by elements $k_1 \in K$ and right cosets.
\begin{gather}
    \ket{k_1,v_{i+1/2}} \equiv \left[ \vcenter{\hbox{\includegraphics{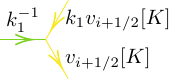}}} \right]\\
    \ket{k_1,v_{i+1/2}} = \ket{k_1,v^{\prime}_{i+1/2}} \Leftrightarrow v^{\prime}_{i+1/2} \in v_{i+1/2}K
\end{gather}
Similarly, the eigenstates of $\hat{A}(\tau_{i-1})$ and the vertex operator are given by:
\begin{equation}
    \vcenter{\hbox{\includegraphics{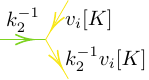}}}
\end{equation}
We now consider the action of the $\hat{Y}(\rho_i)$ component in $\hat{A}(\rho_i)$. It is straightforward to see that this requires $v_i$ and $v_{i+1/2}$ to lie within the same right coset of $K$, such that:
\begin{equation}
    v_iK = v_{i+1/2}K \Rightarrow v_i[K] = v_{i+1/2}[K]
\end{equation}
Next, considering the action of $\hat{Z}(\rho_i)$, it is readily seen that the common eigenstates of all boundary operators in the region depicted in Eq.~\ref{eq:localbdyRegion} take the following form:
\begin{equation}
    \vcenter{\hbox{\includegraphics{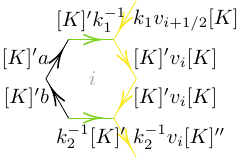}}}.
\end{equation}
Upon extending the chain, it becomes evident that the boundary eigenstates are distinguished by the distinct double cosets $[K]^{\prime}v[K]$—the double cosets of the subgroup $K$. Thus, there exists a global degeneracy given by the number of double cosets:
\begin{equation}
    d_K = |K\backslash G/K|.
\end{equation}
Finally, we examine the remaining portion in the bulk:
\begin{equation}
    \hat{A}_{K}(\rho_i)[\sim] = \frac{1}{|K|} \left[\vcenter{\hbox{\includegraphics{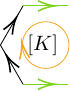}}}\right],
\end{equation}
\begin{equation}
    \hat{A}_{K}(\tau_i)[\sim] = \frac{\delta(h_{i+1}\in K)}{|K|} \left[\vcenter{\hbox{\includegraphics{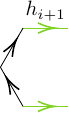}}}\right].
\end{equation}
This is precisely consistent with the results in Ref.~\cite{beigi_quantum_2011}, and does not introduce additional degeneracy.

\newpage 
\bibliography{reference}

\end{document}